\documentclass[universe,article,accept,oneauthor,pdftex]{Definitions/mdpi} 

\usepackage{subfigure}
\makeatletter
\renewcommand{\@thesubfigure}{\normalsize(\textbf{\alph{subfigure}})}
\makeatother
\usepackage{wrapfig}
\usepackage{multicol}
\usepackage{wasysym}
\usepackage{amssymb}

\firstpage{1} 
\pubvolume{1}
\issuenum{1}
\articlenumber{0}
\pubyear{2021}
\copyrightyear{2021}
\externaleditor{Ulisses Barres de Almeida and Michele Doro} 
\datereceived{17 September 2021}   
\dateaccepted{28 October 2021} 
\datepublished{5 November 2021} 
\hreflink{https://doi.org/10.3390/universe7110421} 

\Title{The Making of Catalogues of Very-High-Energy \gray\ Sources}
\TitleCitation{The Making of Catalogues of Very-High-Energy \gray\ Sources}

\Author{Mathieu de Naurois \orcidA{}}
\AuthorNames{Mathieu de Naurois}
\AuthorCitation{de Naurois, M.} 
\address[1]{%
LLR Ecole Polytechnique; denauroi@in2p3.fr} 

\abstract{Thirty years after the discovery of the first very-high-energy \gray\ source by the Whipple telescope,
the field experienced a revolution mainly driven by the third generation of imaging atmospheric Cherenkov telescopes (IACTs).
The combined use of large mirrors and the invention of the imaging technique at the Whipple telescope, stereoscopic observations, 
developed by the HEGRA array and the fine-grained camera, pioneered by the CAT telescope, led to a jump by a factor of more than ten
in sensitivity. The advent of advanced analysis techniques led to a vast improvement in background rejection, as well as
in angular and energy resolutions. Recent instruments already have to deal with a very large amount of data
(petabytes), containing a large number of sources often very extended (at least within the Galactic plane) and
overlapping each other, and the situation will become even more dramatic with future instruments. The first large
catalogues of sources have emerged during the last decade, which required numerous, dedicated observations and developments, but also 
made the first population studies possible. This paper is an attempt to summarize the evolution of the field towards 
the building up of the source catalogues, to describe the first population studies already made possible, and to  give
some perspectives in the context of the upcoming, new generation of instruments.}


\def\apj{Astrophys. J.}%
\def\aspp{Astropart. Phys.}
\def\apj{Astrophys. J.}%
\def\aap{Astron. Astrophys.}%
\def\prd{Phys.~Rev.~D}%
\def\prl{Phys.~Rev.~Lett.}%
\def\nat{Nature}%
\def\gray{$\gamma$-ray}
\def\grays{$\gamma$-rays}
\def\glike{$\gamma$-like}
\def\hess{H.E.S.S.}
\def\deg{^\circ}

\begin{document}


\section*{Introduction}

Soon after the discovery of cosmic rays by Victor Hess in 2012~\citep{Hess262750}, it was realised that very-high-energy $\gamma$ rays could allow 
the identification of their sources, mainly because, in contrast to charged cosmic rays, neutral $\gamma$ rays are unaffected by extragalactic and galactic
magnetic fields, and therefore travel undeflected in space.
Direct observation of high-energy $\gamma$ rays from space is, however, limited to energies $\lesssim 100\,\mathrm{GeV}$ due to the 
steeply falling source flux as a function of increasing energy. 
At the same time, due to the overall thickness of the atmosphere ($\approx$1$\,\mathrm{kg}\,\mathrm{cm}^{-2}$), high-energy particles ($\gamma$ rays or charged nuclei) 
entering the atmosphere do not reach the ground, but interact at high altitudes and trigger the development of a so-called ``{\it extended air shower} (EAS)'' of particles.
These showers contain numerous ultra-relativistic electrons and positrons, travelling faster than light in the air and consequently emitting ultrashort (nanosecond) flashes
of Cherenkov light~\citep{Cherenkov1934a}. After an initial suggestion from Blackett~\cite{1948esns.conf...34B}, the first attempts to detect the Cherenkov light emitted by 
atmospheric showers dates back to 1953~\cite{1953Natur.171..349G}.
It took, however, several decades before the emergence of ground-based very-high-energy gamma-ray astronomy. The Whipple collaboration 
established the imaging atmospheric Cherenkov technique~\citep{1989ApJ...342..379W}, whereby large telescopes, equipped with an ultra-fast 
camera, capture the Cherenkov light emitted by ultrarelativistic electrons and positrons in the atmospheric showers, and form the image of the latter.
A detailed analysis of the shower image allows the reconstruction of the parameters of the incoming particle: direction of arrival, impact on the ground, energy, 
and, on a statistical basis, allows for the discrimination of $\gamma$ rays from the much more numerous charged cosmic rays.

During the last decades, the field of very-high-energy (VHE) \gray\ energies over 100 GeV evolved from the observation of isolated, 
well defined sources to very large projects,
spanning several years, and covering a large fraction of the sky. These projects resulted in very large and inhomogeneous
data sets, with deep exposures on specific regions of interest (ROI) and much shallower exposures on other ones. 
The resulting large exposure gradients are tricky to handle in analysis pipelines and imposed the development of new acceptance
determination and background subtraction techniques.
In addition, these large data sets are acquired across several years, resulting in very diverse observational conditions,
in terms of array configuration (number of operational telescopes, trigger settings, etc.), zenith angle, night sky brightness (NSB), etc. 
Dedicated analysis techniques have been developed to permit the consistent analysis of such data sets which are key 
ingredients for the build-up of catalogues. The first catalogues, elaborated in the last decade, revolutionised our view
on the VHE sky and initiated the statistical analysis of populations of the same type, revealing some of their
evolution scheme. The next generation of instruments, and in particular, the upcoming Cherenkov telescope array (CTA),
will sample the sky with unprecedented sensitivity and is expected to make quantitative studies on source populations
a major activity, thus pushing forward our understanding of particle acceleration and \gray\ production in VHE
sources.

This paper is divided into three sections. The first is dedicated to the technical aspects of catalogue construction.
In a second part, the existing major surveys are described, together with the first population studies that they made possible. The third
and last section presents the upcoming projects and some personal perspectives.

\section{Technical Aspect of Survey and Catalogue Constructions }

In the VHE \gray\ domain, the construction of catalogues arises essentially from two different observational strategies.
On the one hand, observations were historically mainly conducted on sources of particular interest, identified from observations
at different wavelengths ({targeted observations}). This mode of observation is still valid for extragalactic observations,
where the density of sources of sufficient brightness is not high enough to undertake systematic surveys.
Such targeted observations result in sparse and incomplete catalogues with very heterogeneous depth. On the other
hand, a few large-scale surveys ({survey observations}) have been conducted, essentially in the Galactic plane (see
Section~\ref{sec:Surveys}), allowing for (partially) unbiased samples of sources.
These two observational strategies have numerous implications, first on the way the array of telescopes is operated,
but also on the way in which the analysis pipeline is constructed and run. 
In this section, we review the technical aspects of the catalogue construction. The important steps towards a source
catalogue are:

\begin{itemize} 
  \item {\bfseries Array operation and observational strategy:} the way in which the array of telescopes is operated and optimised for a given
  physics goal (optimised on sensitivity or on field-of-view (FoV) width for instance).
  \item {\bfseries Event reconstruction and classification:} separation of \glike\ events from the much more numerous
  background-like events and construction of events classes.
  \item {\bfseries Background model:} determination of the expected background in the field of view, taking into
  account the instrument response.
  \item {\bfseries Excluded region determination:} identification of regions which are
  potentially contaminated by genuine \gray\ signal. These regions should not be used to estimate the remaining background in the subsequent
  background subtraction procedure, so as to avoid signal over-subtraction. 
  \item {\bfseries Background subtraction:} comparison of the number of events in a region of interest with the
  expected number of events (coming from the background model), to assess the potential existence of a
  localised excess.
  \item {\bfseries Automated catalogue pipeline:}  separation of regions of significant \gray\ emission into
  individual source components and extraction of their physical characteristics (flux, energy spectrum, morphology,
  temporal variability, \dots).
\end{itemize}

\subsection{Observational Strategies
\label{sec:Observations}}

Observations of IACTs are usually divided into chunks of $\sim$30~$\mathrm{min}$, called {runs}, which could correspond
to an {exposure time} in different domains of astronomy. This typical duration results from a trade-off between 
opposite constraints:
on the one hand, it takes some time to slew the telescopes to a different target, to configure the system and to start
the observations, so a run should not be too short (at least a few minutes). Since the instrument trigger rate ranges from a
few hundreds of $\mathrm{Hz}$ to a few $\mathrm{kHz}$, it takes at least a few minutes to collect enough events to be able to
assess the instrument performance and stability (and to be able to estimate the background, see
Section~\ref{sec:BackgroundSubtraction}). On the other hand, the instrument response function varies strongly with the
observational conditions (and in particular with the zenith angle and meteorological conditions, both on time scales of
a few minutes), making very long runs more prone to systematics and more complicated to~analyse. 

Since any astronomical source can only be observed for a few hours every night, and only during certain
periods of the year, and given the very low flux of very high energy $\gamma$ rays, even from the strongest known sources,
many runs, spread over days, months and even years have to be combined in a consistent manner in the analysis procedure
to produce a {\it stacked} data-set. This observation procedure also requires the performance of the instrument 
and the atmosphere to be
monitored precisely over very long periods of time. 
The current generation of IACTs carry out two main modes of pointing corresponding 
to the targeted and survey modes of observation:

\begin{figure}[H]
 
\includegraphics[height=3.5 cm]{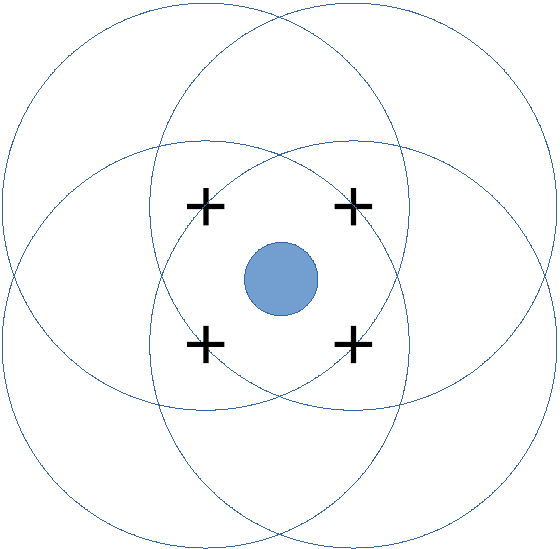} \hspace{1cm}
\includegraphics[height=3.5 cm]{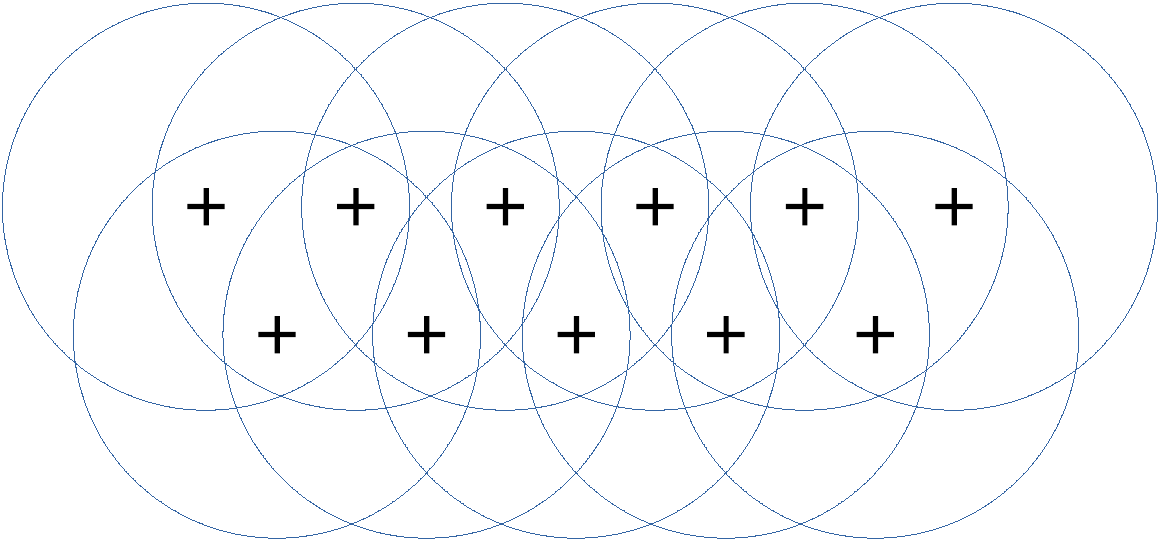}\\
\caption{\label{fig:Pointing}({\bf Left:}) Classical {Wobble} pointing mode, where the source is offset in the field
of view.
({\bf Right:}) Survey pointing mode, with observations overlapping with each other. The black ``+'' mark denotes the
pointing direction for the various runs, and the blue circles the instrument field of view.}
\end{figure}   

\begin{itemize}
\item {\it Wobble mode of observation} (Figure~\ref{fig:Pointing}, left), where several observations are taken with different
pointing directions around the source of interest. The source is displaced with respect to the centre of the field
of view, to allow for proper background determination (Section~\ref{sec:ReflectedBackground}).
This mode of observation is appropriate for point-like or moderately large sources of known position, and in particular for
targeted observations. Historically, the pointing positions were taken with a shift in the right ascension (RA) equal to
the temporal separation between runs, in order to reproduce the exact same trajectory of the telescopes on the sky for each
pair of runs.  By doing so, no correction for the variation of telescope response had to be applied, simplifying a lot
the analysis.  Recent IACTs, using more elaborate background models (Section~\ref{sec:Acceptance}), dropped this observational
constraint and combined observations  with wobble offset in any direction (right ascension, declination or any other coordinate).
\item {\it Survey mode of observation} (Figure~\ref{fig:Pointing},
right), where a large region of the sky is scanned with observations overlapping each other (in order to minimise the
background gradients). Several rows can be conducted in parallel or one after the other, and different spacing between
pointing positions can be used. This mode of observation is usually optimised to maximise the sky coverage, while minimising
the acceptance variations across the surveyed region.
\end{itemize}

\clearpage
\end{paracol}
\nointerlineskip
\begin{figure}[H]
 \widefigure
 \centering
\subfigure[]{\includegraphics[height=3.5 cm]{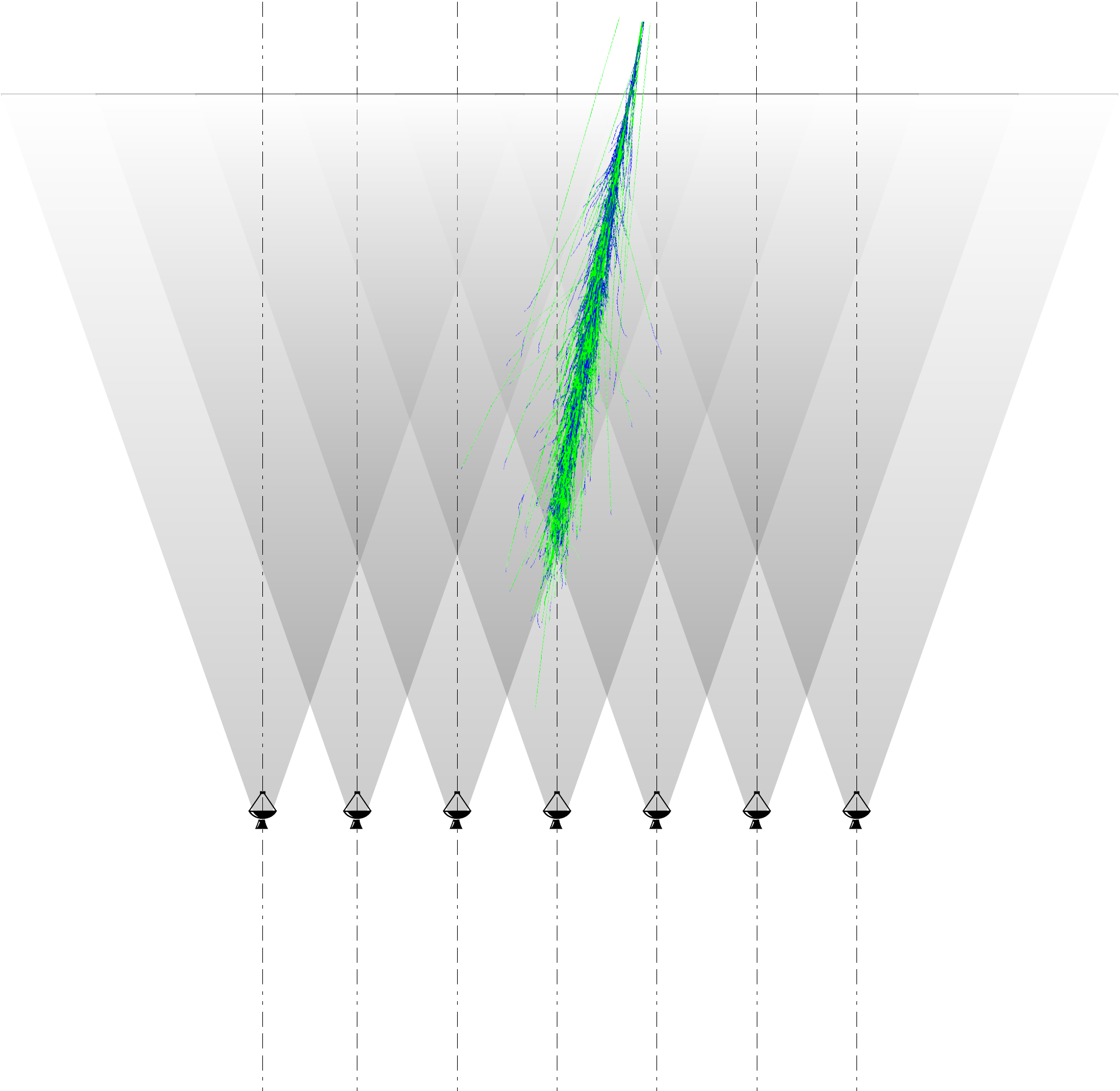}} \hspace{0.5em}
\subfigure[]{\includegraphics[height=3.5 cm]{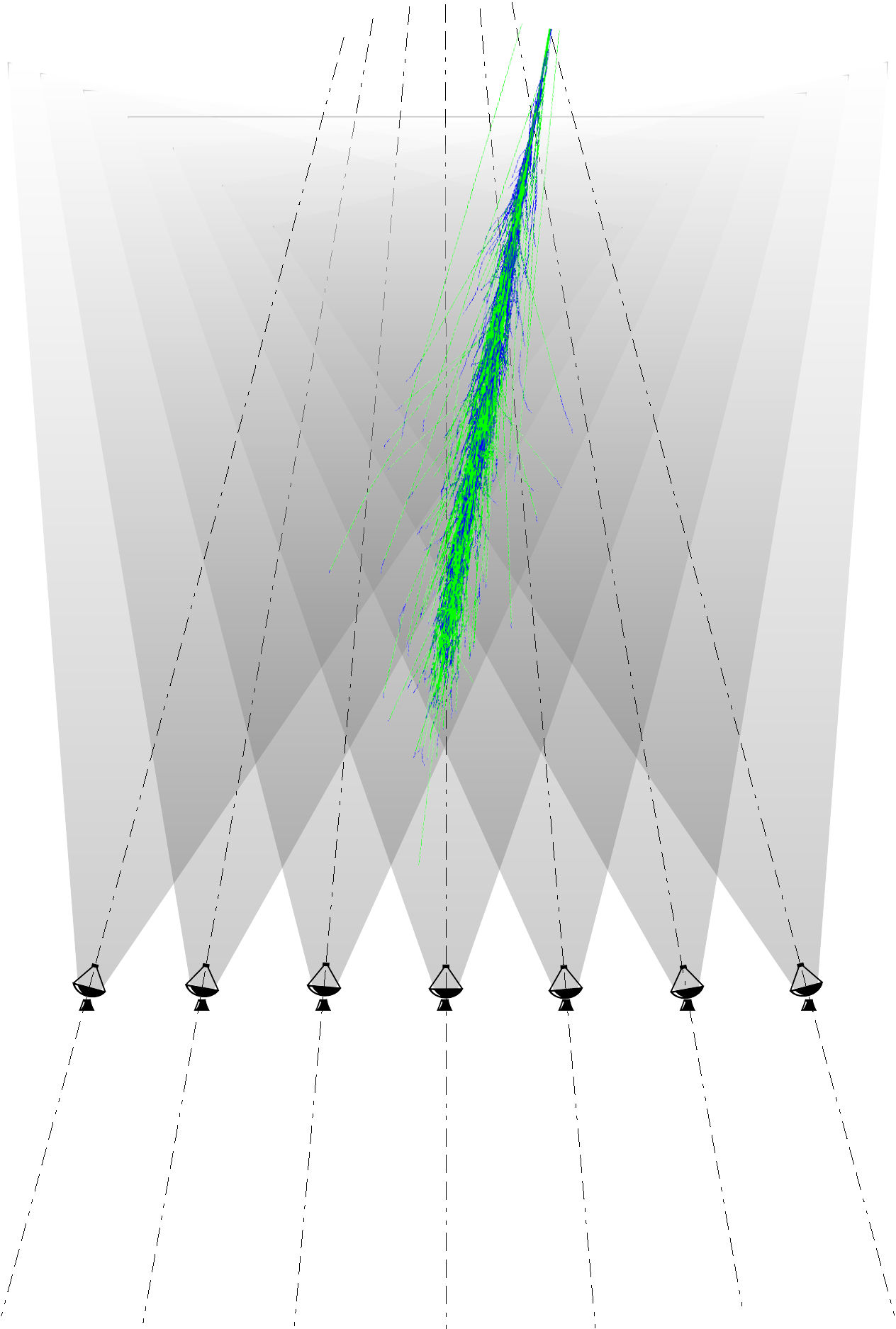}} \hspace{0.5em}
\subfigure[]{\includegraphics[height=3.5 cm]{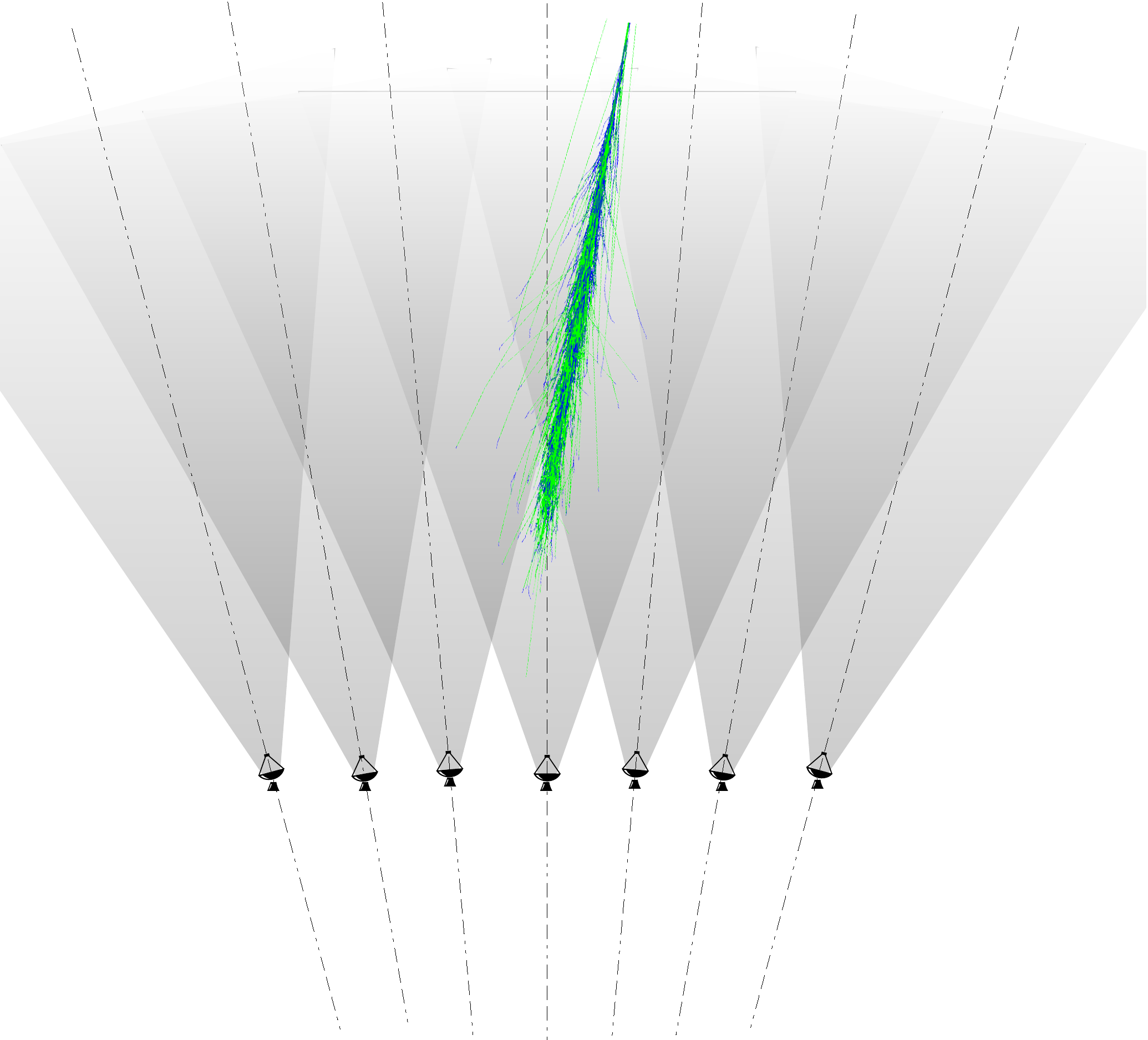}} \hspace{0.5em}
\subfigure[]{\includegraphics[height=3.5 cm]{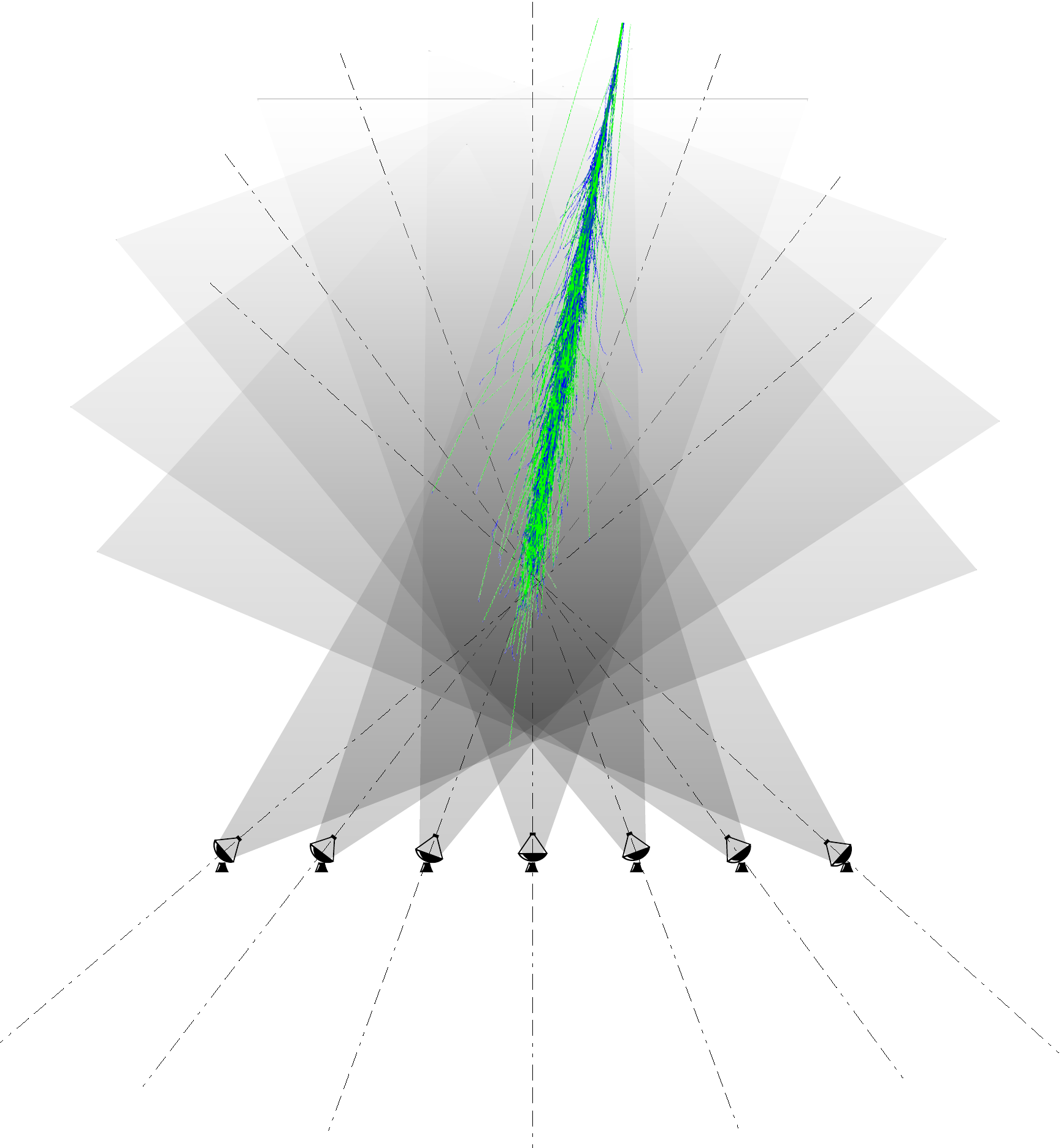}}\\
\caption{\label{fig:ParallelePointing}From {\bf left} to {\bf right}: ({\bf a})~parallel pointing, ({\bf b})~convergent
pointing at high altitude, ({\bf c})~divergent pointing, ({\bf d})~convergent pointing at low altitude, also denoted {skewed pointing}.}
\end{figure} 
\begin{paracol}{2}
\switchcolumn


These two modes of observation correspond also to different possible pointing optimisation schemes. In the wobble observation
mode, one usually wants to reach the best possible sensitivity. To achieve that, all telescopes are pointed in the same direction 
({\it parallel} pointing) (Figure~\ref{fig:ParallelePointing}{a}), or even pointed at the altitude of the maximum of
development of the showers ({convergent pointing}---Figure~\ref{fig:ParallelePointing}{b}) to maximise the
collection of light. In contrast, in the survey mode of observation, one might want to increase the sky coverage 3 at the
expense of point-like sensitivity. 
This can be achieved by splitting the array in several groups of telescopes pointing at different directions,  or even,
although this was not yet effectively used, by implementing a {divergent} pointing mode
(Figure~\ref{fig:ParallelePointing}{c})  where telescopes point on directions slightly 
offset from each other to increase the effective field of view. 
Technically, divergent pointing can easily be implemented as a convergent pointing to a negative altitude.  
Convergent pointing at very low altitude (a few km above the ground, Figure~\ref{fig:ParallelePointing}{d}),  also
denoted {\it skewed pointing} here, can also be used, and is technically not more difficult to implement.

Depending on the telescope angular separation, divergent pointing can result in a non-flat exposure across the sky,
which can significantly complicate the subsequent steps of the analysis. To investigate the merits of each telescope pointing
strategy, we performed a simulation of an array of 37 \hess-I telescopes ($5\deg \diameter\ \mathrm{FoV}$ each) placed on a
square grid with lines of 3,5,7,7,7,5,3 telescopes at the altitude of the \hess\ site (1800\,m a.s.l)  (\mbox{Figure~\ref{fig:CTADArray}}), and separated by $120~\mathrm{m}$ each (for
a total array size of $720\times 720~\mathrm{m}^\mathrm{2}$). Pointing altitudes ranging from
$3~\mathrm{km}$ to $10~\mathrm{km}$ above site level were used in both convergent and divergent (negative altitude) modes, and parallel
pointing was also included for reference. Diffuse \grays\ between $100~\mathrm{GeV}$ and $10~\mathrm{TeV}$
($20\deg\diameter$ cone opening angle) were simulated on a circle of $700~\mathrm{m}$ (enclosed in the array),
using the \texttt{kaskade}/\texttt{Smash} simulation chain developed for \hess~\cite{TheseJulienGuy}.
Data were analysed using the \texttt{Model++}~\citep{denaurois2009} within the \hess\ software framework.
Results of this simulation are presented in Figures~\ref{fig:PointingComparison} and \ref{fig:PointingApperture}.

\begin{wrapfigure}{r}{5cm}
\includegraphics[width=4.9 cm]{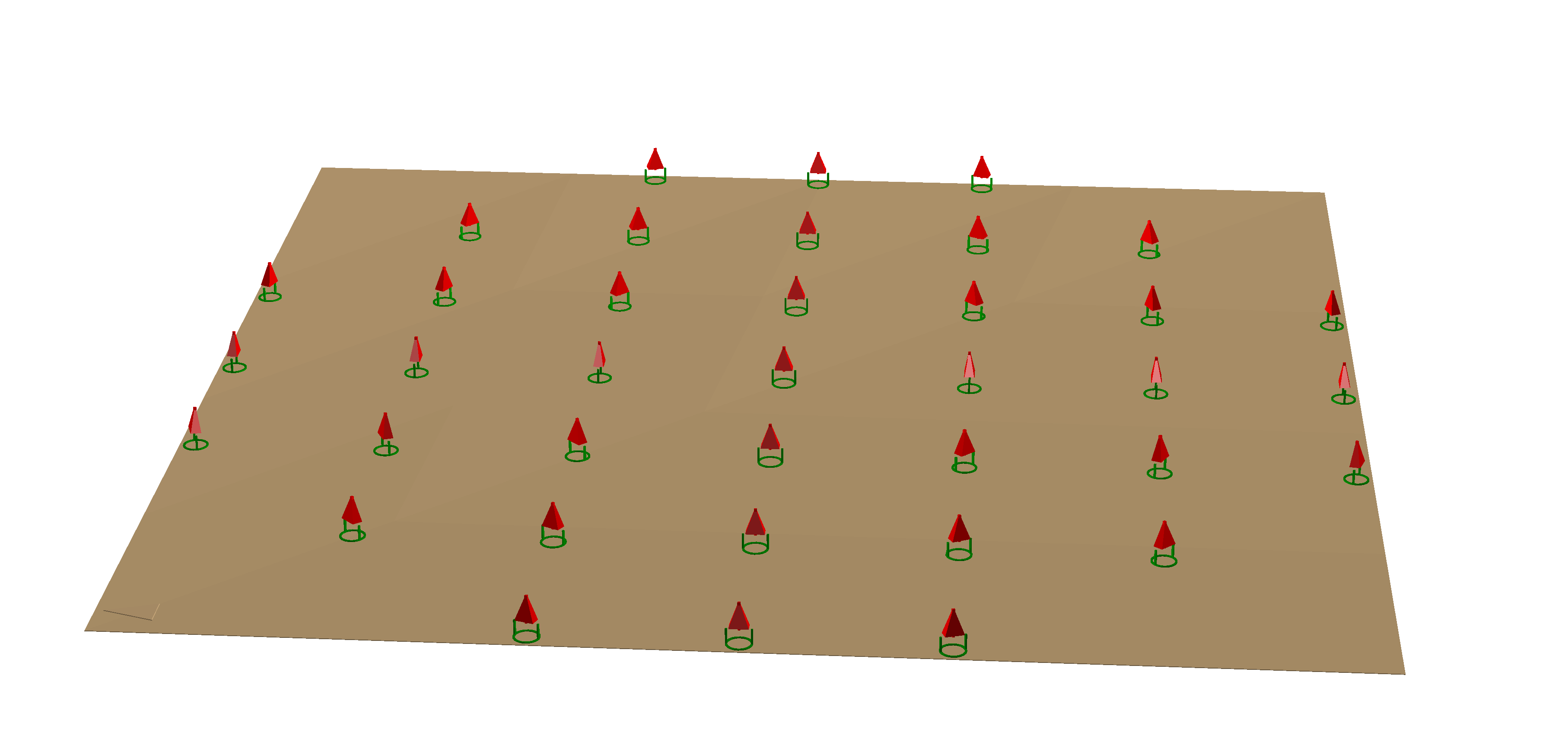}
\caption{\label{fig:CTADArray}Array of 37 \hess-I telescopes used in the simulation of various pointings.}
\end{wrapfigure}

Figure~\ref{fig:PointingComparison}, {\bfseries a} shows that convergent pointing at high altitude maximises the event
multiplicity (number of triggered telescopes). As a consequence, this mode of observation also maximises the precision of the
reconstruction (angular and energy resolution in particular), as shown in panel {\bfseries d}. 
In contrast, this corresponds to a rather modest size of the effective field of view, as measured by the squared angular
distance of the observed events to the optical axis (panel {\bfseries b}).

As expected, the largest effective fields of view are obtained by pointing at low altitude, either in divergent or
convergent (skewed) modes, as shown in panel {\bfseries b}, with rather similar and quite flat distributions.
Panel {\bfseries c} shows the distribution of squared impact distance with respect to the centre of the array, which is
used to derive the effective area of the array. Low altitude pointings (convergent or divergent) tend to select mostly
events close to the array centre, whereas convergent pointing at high altitude tends to maximise the effective area.

\begin{figure}[H]
 
\subfigure[]{\includegraphics[height=5 cm]{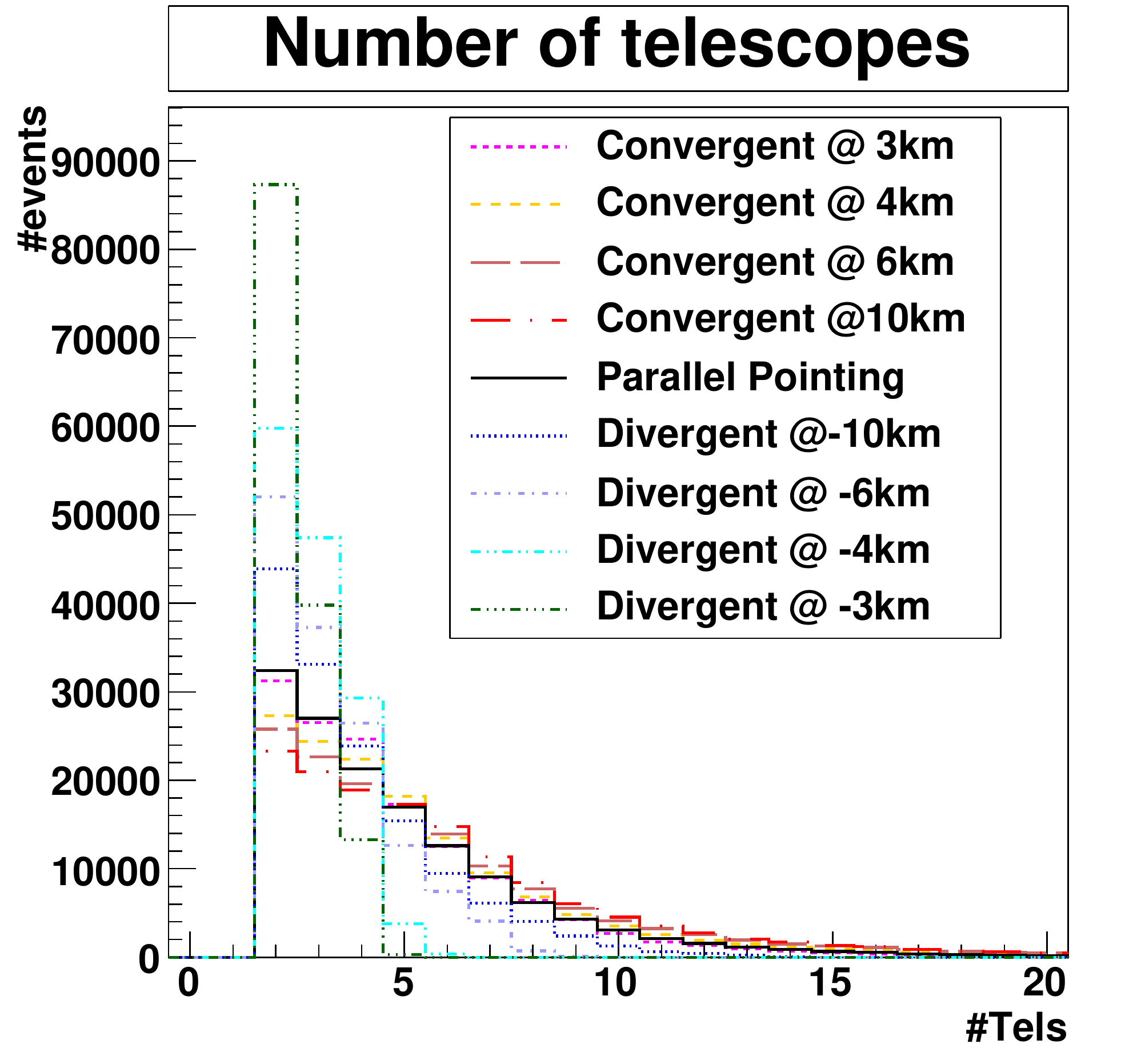}} 
\subfigure[]{\includegraphics[height=5 cm]{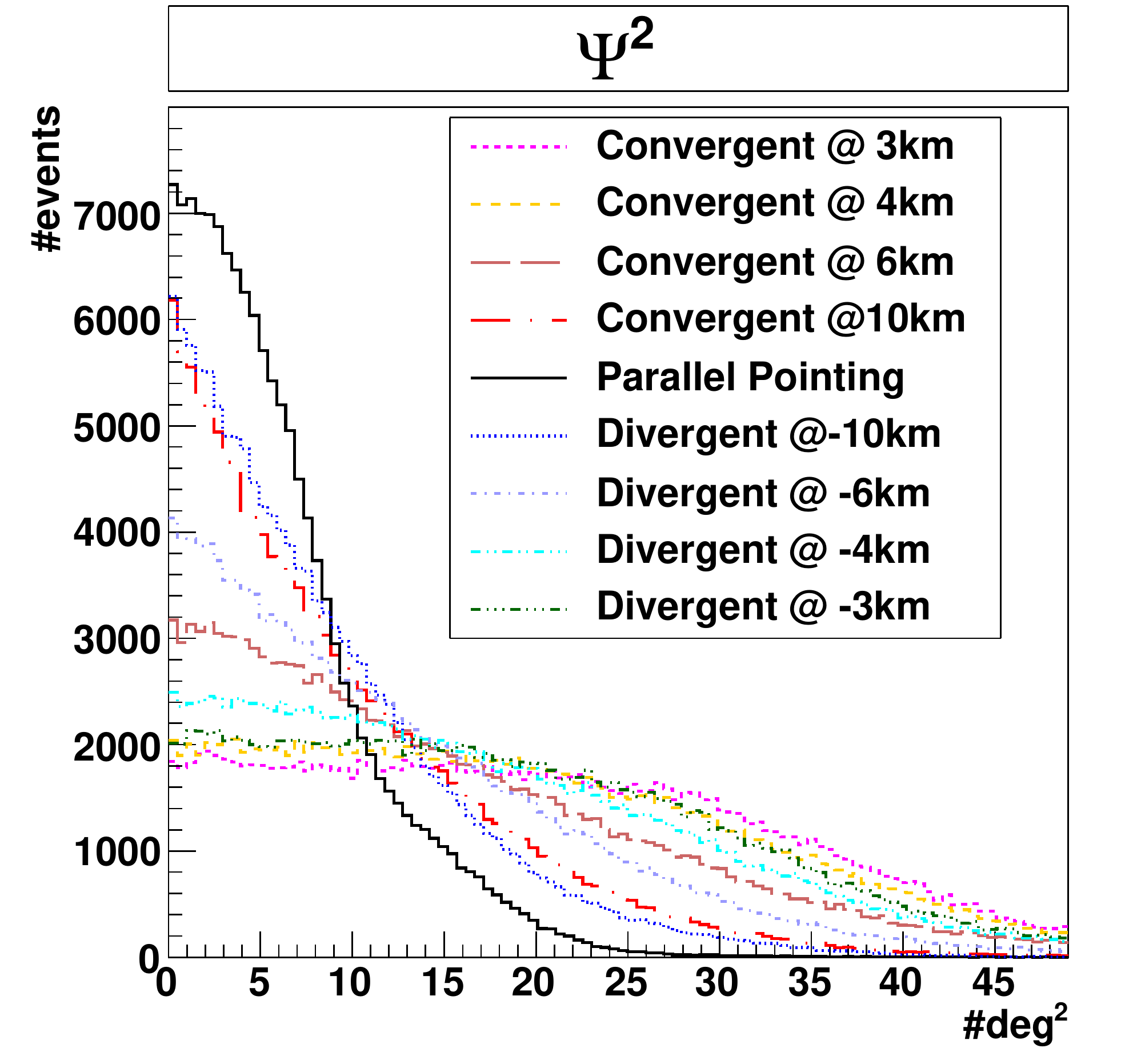}}  \\
\subfigure[]{\includegraphics[height=5 cm]{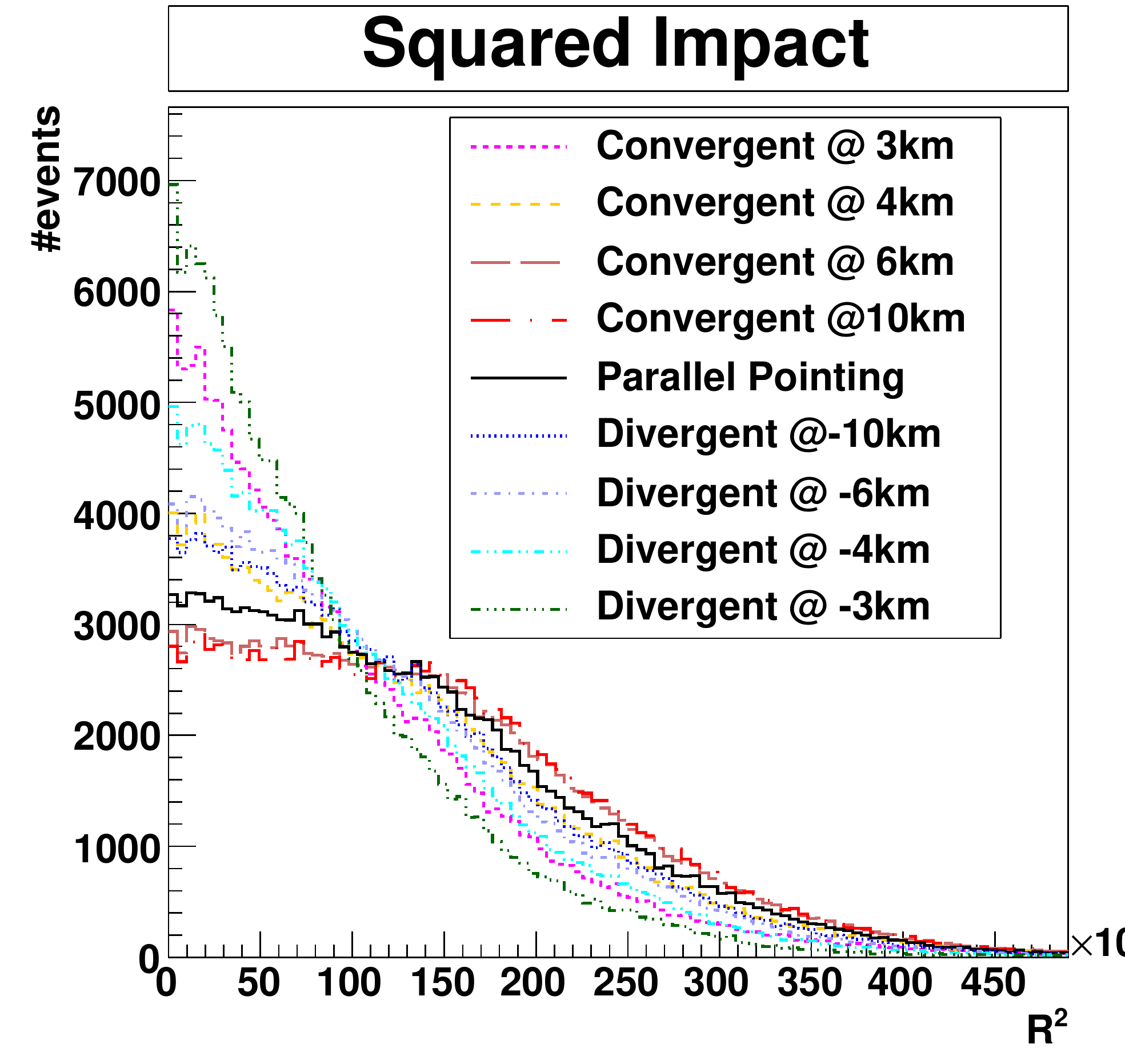}} 
\subfigure[]{\includegraphics[height=5 cm]{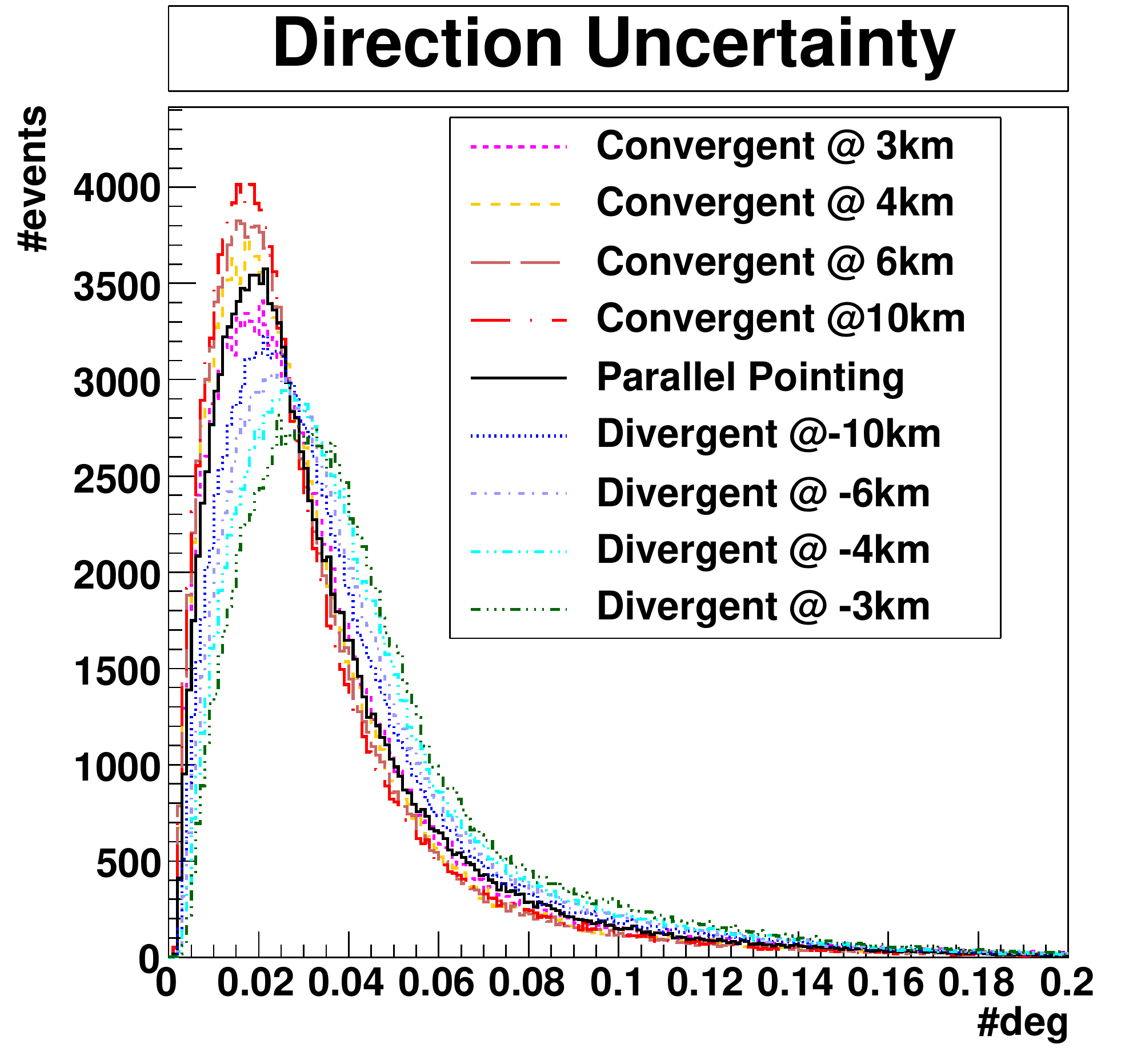}}\\
\caption{\label{fig:PointingComparison}Comparison of performances of various pointing strategies. From {\bfseries left}
to {\bfseries right}: ({\bfseries a})~ Event multiplicity,  ({\bfseries b})~ Squared angular distance to optical axis, 
({\bfseries c})~ Squared impact distance to the centre of the array,  ({\bfseries d})~ Event reconstruction precision,
measured as the fit uncertainty on the event direction.}
\end{figure}   
\vspace{-9pt}

\begin{wrapfigure}{r}{5cm}
\includegraphics[width=4.9 cm]{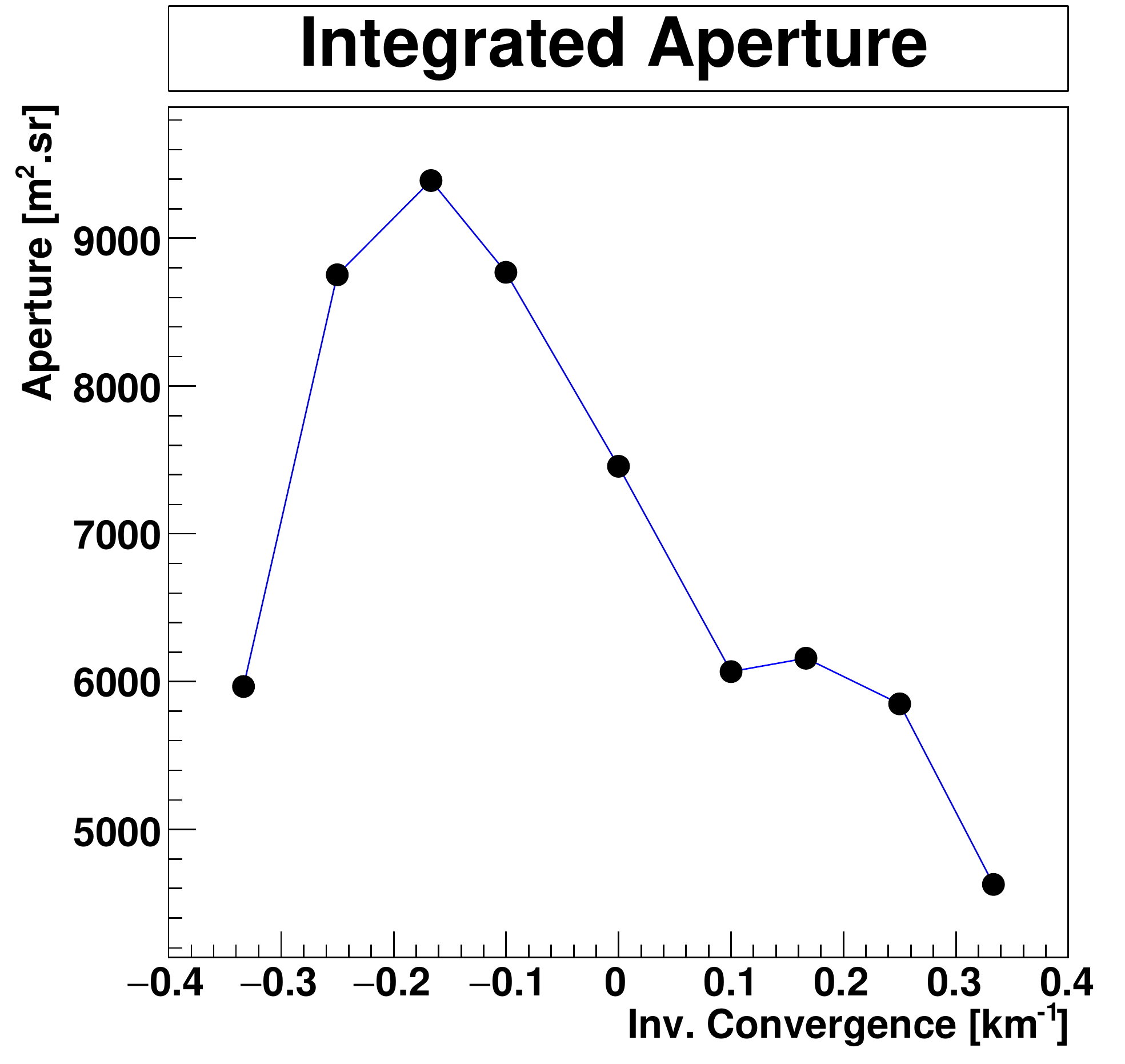}
\caption{\label{fig:PointingApperture}Integrated aperture for the various pointing
strategies.} 
\end{wrapfigure}

From the squared angular distance and squared impact distance distributions (panel {\bfseries b} and {\bfseries c}), an
{\it integrated aperture} can be derived, which corresponds, to a normalisation factor, to the rate of detected \grays.
The integrated aperture for the presented simulation is displayed in Figure~\ref{fig:PointingApperture} as a function of the
inverse of the pointing altitude (such that parallel pointing is at the origin of the X axis, negative values
correspond to divergent pointings, and positive ones to convergent pointings).
It turns out that for this particular simulated array configuration, the detection rate is maximised for moderate
divergent pointing (at an altitude of $-6~\mathrm{km}$), thus confirming the potential of the divergent observation
mode.
In particular, for a uniform distribution of sources with sufficient density, as expected in the extragalactic sky, divergent
pointing might indeed be the most effective mode of observation. On the other hand, distant sources are affected by 
\gray\ absorption by pair creation on the EBL. Limiting this absorption requires the reduction of the energy threshold to its
minimum possible, which is better achieved in convergent pointing mode.
These preliminary results, although confirming the
findings of other authors~(e.g., \citep{2015APh....67...33S}), need to be confirmed by a full scale simulation using 
realistic, next generation arrays (CTA) and investigating not only the integrated aperture,
but also the event reconstruction, $\gamma$-hadron separation and background subtraction.  The question of background
modelling and subtraction might become complicated to handle (due to possible non-trivial variations across the FoV), and
will certainly require further studies before such alternate pointing strategies can be used in large-scale surveys.

\subsection{Event Separation and Classification
\label{sec:Classification}}

Genuine \grays\ represent a tiny fraction ($\approx$0.01\%) of the events recorded by IACTs, the vast majority being
charged cosmic rays, composed of mainly protons and nuclei, but also a small fraction of cosmic electrons.
The details of event reconstruction and $\gamma$-hadron separation are covered in an extensive bibliography and could be the subject 
of a review on their own, and will not be covered here. A very large spectrum of techniques is indeed used in the
field, ranging from simple image parametrisation to template fitting, and even image deep learning techniques.
Whatever method is used to reconstruct the events, one or several {discriminating} parameters are constructed to
separate \gray\ events from the charged cosmic rays.
The probability density functions (PDFs) for the \gray\ and charged cosmic-ray events always overlap, rendering a
perfect separation impossible. In particular, a small ($\sim$$10^{-3}$) fraction of protons generate a $\pi^0$ high in the
atmosphere, which initiates the development of an electromagnetic shower which is very similar to that induced by \grays. 
Similarly, electrons also initiate electromagnetic showers and are therefore almost indistinguishable from
genuine \grays.
The discriminating parameters can be
used to construct several, well separated event {classes} used in the subsequent steps of the analysis. Two main
event classes are usually used:

\begin{itemize}
\item \glike\ events: these events are very likely (probability depending on the analysis strategy) to
originate from a genuine $\gamma$ ray.
\item background-like events: these events have a marginal, tiny probability of originating from a genuine $\gamma$ ray,
and most likely come from a charged cosmic ray.
\end{itemize}

Due to the overlap of the PDFs, this classification is incomplete, with many events falling between the two
cases. The separation also remains imperfect, as some remaining background events always survive the selection.
Alternatively, one can make use of the full PDFs to derive a {\it ``gammaness''}  or {\it ``hadroness''} parameter~({e.g}., \citep{PhysRevD.103.123001}),  
giving the probability for the event to respectively originate from   a \gray\ or a charged cosmic ray. So far, the
subsequent steps of the analysis, and in particular the background subtraction, have not really been adapted to the
use of continuous probability distributions, so the use of event classes remains the state-of-the-art for what concerns IACTs. 
 
\begin{wrapfigure}{r}{5cm}
\includegraphics[width=4.9 cm]{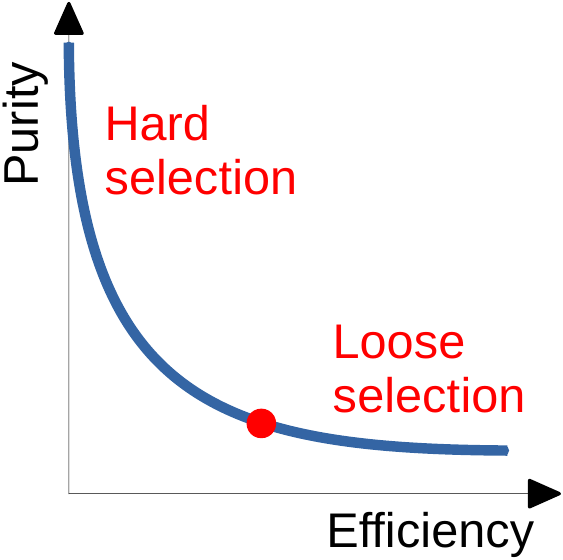}
\caption{\label{fig:EfficiencyPurity}Efficiency-purity plot.}
\end{wrapfigure} 
  
Different event selection strategies can be used, which can be visualised in an {efficiency} vs. {purity} plane
(Figure~\ref{fig:EfficiencyPurity}), where the {efficiency} denotes the fraction of $\gamma$ rays that are retained in the
analysis, while the {purity} is the relative fraction of $\gamma$ rays in the selected sample, i.e., one minus the
background contamination fraction. It is, in general, possible to achieve a very high \gray\ efficiency (retain almost all $\gamma$'s) but
at the price of large background contamination (bottom-right in the plot). It is also possible to have a rather large
purity of the sample (almost no background) but at the price of a very bad efficiency to $\gamma$ rays (top-left). 
One usually denotes as ``{\it loose selection}'' a selection corresponding to first case, while ``{\it hard selection}'' is used for
a high purity, low efficiency selection strategy.
In general, low-energy showers are subject to more statistical fluctuations, and are therefore more difficult to distinguish 
from hadronic showers. As a consequence, hard selections usually lead to a higher energy threshold than loose selections.

The question of optimal selection strategy is fully non-trivial, as it is intimately linked to the questions of {background subtraction} (Section~\ref{sec:BackgroundStatistics}) and {background systematics}
(Section~\ref{sec:Systematics}). In terms of pure statistics, a theoretical optimal selection exists along the curve (red
point), which maximises the statistical significance of the detection of a given \gray\ source. This optimal
point however differs for each and every source, as it depends on the source intensity and spectral shape. If it is
possible to adapt the selection to the source characteristics in the case of individual, targeted observation, large
scale surveys used in catalogue construction require, in contrast, a homogeneous selection to be used consistently
throughout the whole data set.
One general trend that can guide the choice is the fact that, due to the steeper energy spectrum of cosmic rays compared
to that of galactic \gray\ sources, the background is reduced faster than the signal when moving towards harder selection cuts. In order to
maximise the detection potential of faint sources, rather hard selection cuts were used in most surveys so far, with the
drawback of reduced efficiency at low energies. Hard cuts also have the advantage of significantly mitigating  the problems
arising from the imperfect modelling of the acceptance and uncontrolled background systematics.

Since the population of VHE sources might actually vary with the energy domain, future surveys might be optimised also
towards low energy, imposing the use of looser cuts. It is also possible to release several sub-versions of the
same catalogue, corresponding to different selection schemes, as has already been done in other experiments such as {Fermi}-LAT or HAWC.

\subsection{Acceptance---Background Model
\label{sec:Acceptance}}

The term  {background model} or {acceptance} denotes the shape of the distribution of events across the field
of view {\bf in the absence of genuine \gray\ sources}.
It can be determined for the various event classes (Section~\ref{sec:Classification}), and needs to be determined in
particular for \glike\ events prior to background subtraction (Section~\ref{sec:BackgroundSubtraction}). 
For genuine \grays\, it is usually determined from Monte Carlo simulation, whereas for cosmic ray events, 
it is usually determined directly from the data, either from the considered data set, or from a different, control data set.
The background distribution across the field of view  depends on multiple parameters, and 
must be derived for each and every analysis configuration. It depends of course on the array geometry (number of 
telescopes and position), on the reconstruction method and on the event selection, but also on the observational conditions
(zenith angle) and on the energy.
The deeper the observations, the more accurate the background models needs to be to avoid uncontrolled systematics
across the field of view. 

The background model is usually determined on a run-by-run basis, and is then reprojected onto the sky to compute the
background model for the full data set, as shown in Figure~\ref{fig:AcceptanceStacking}; left: background models are
determined for every run, and then stacked together. Several algorithms have been developed for the computation of
acceptance:
\begin{figure}[H]
 
\begin{tabular}{cc}
\includegraphics[height=5 cm]{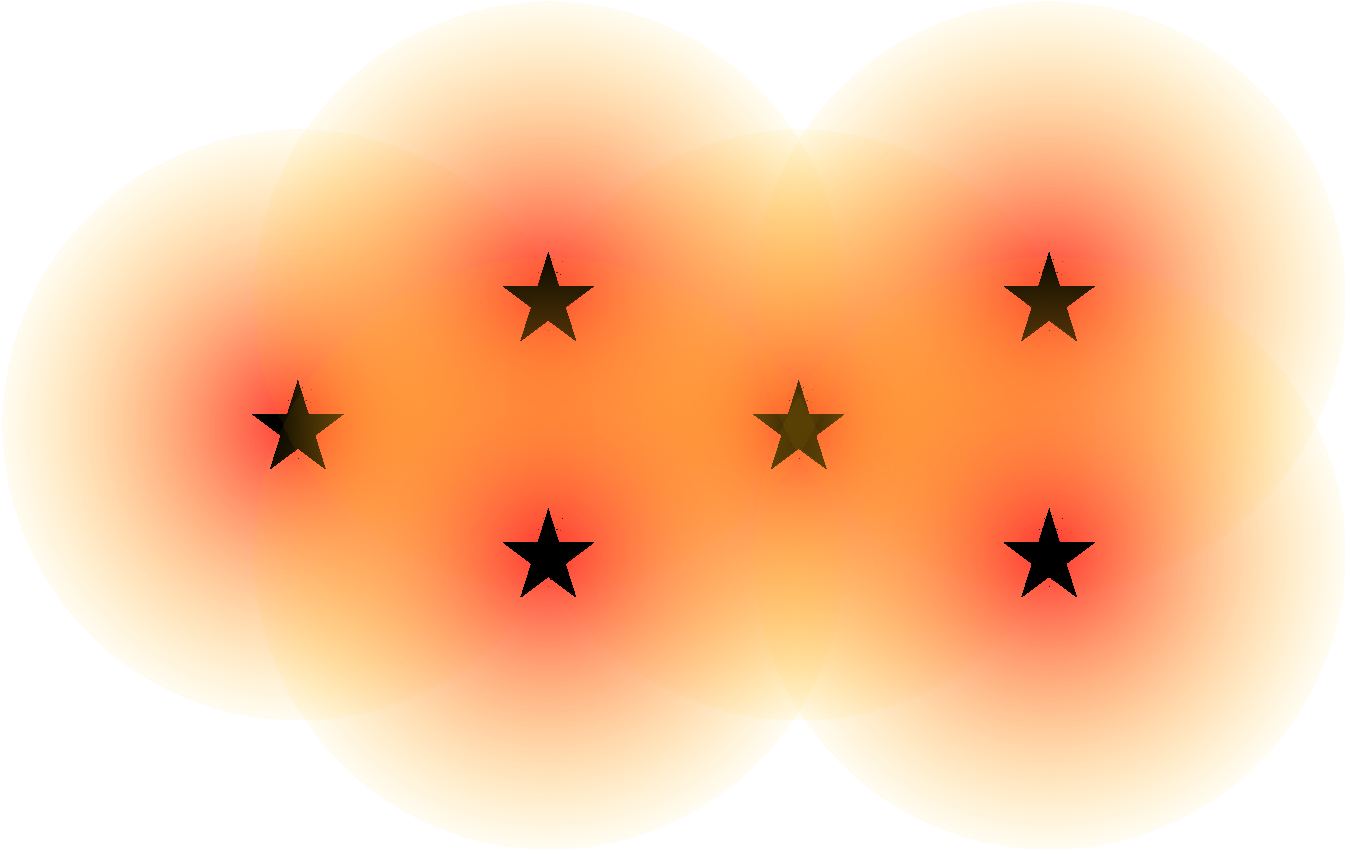} & \includegraphics[height=4
cm]{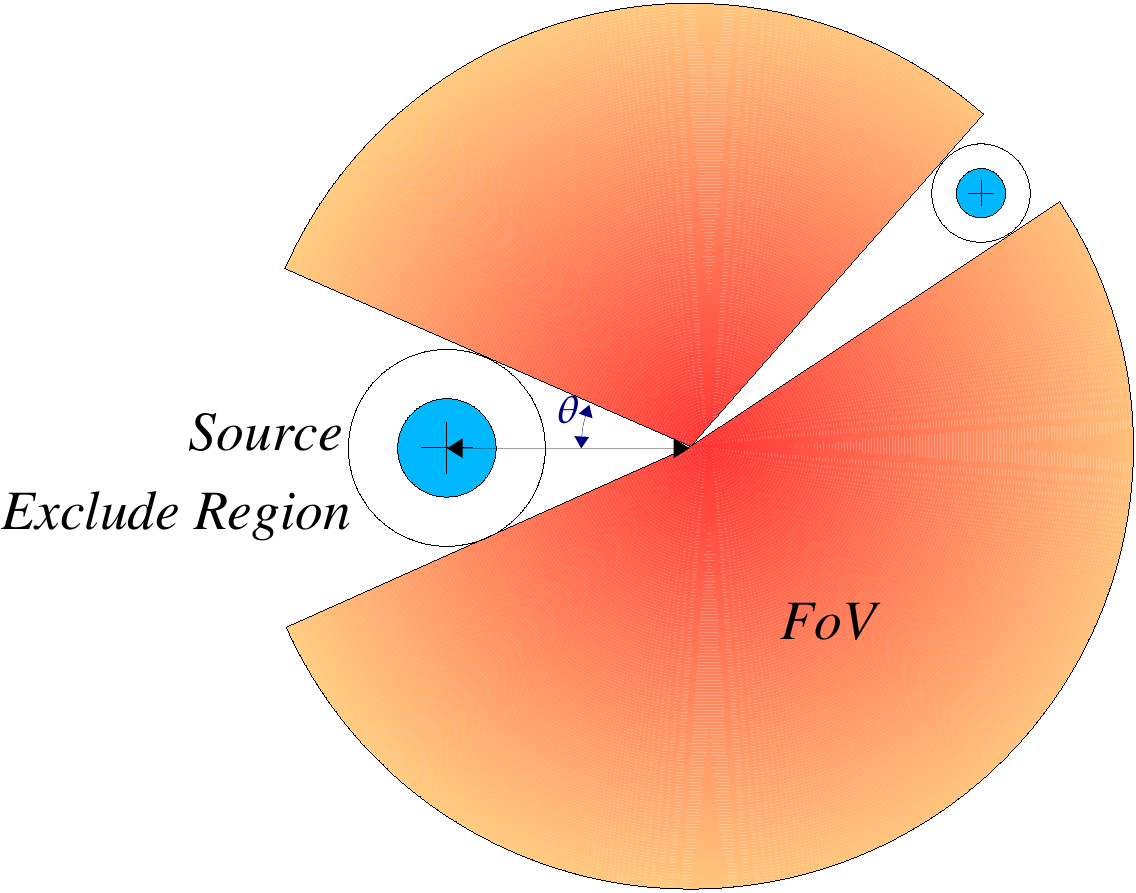} \\
\end{tabular}
\caption{\label{fig:AcceptanceStacking} ({\bf Left}) Stacking of background models for a data set with different
pointing. ({\bf Right}) Radial acceptance determination in the presence of known or putative \gray\
sources.}
\end{figure}

\subsubsection{\label{sec:RadialAcceptance}Radial Acceptance}

The {radial acceptance} model is the simplest acceptance model, and the easiest to implement. It assumes a rotational
symmetry of the instrument response around the pointing direction, which is an acceptable assumption for not-too-deep
data sets.
Thanks to its simplicity, it can be determined easily in different energy slices, thus providing the input for a 3D
analysis.
Radial acceptance curves usually depend also on the zenith angle range. The incorporation of both zenith angle bands
and energy slices results in a 3-dimensional model which represents the current state-of-the-art.

To avoid contamination of the acceptance, known or putative \gray\ sources can be excluded from the acceptance
determination, if not overlapping with the centre of the FoV, by excluding a sector from the radial  acceptance
determination (Figure~\ref{fig:AcceptanceStacking}, right). 
Additional gradients, due in particular to the variation of zenith angle across the field of view, can also be taken into account.

The evolution of the radial acceptance curves with zenith angle (left) and energy (right) is shown in
Figure~\ref{fig:RadialAcceptanceEvolution} for the \hess-I array of 4 telescopes, and for a given reconstruction
(Model++ Std). For a different reconstruction and/or a different set of cuts, the curves will be different but will exhibit a
similar trend. As can be seen from the figure, differences of more than 20\% between different bands can easily exist,
stressing the fact that the use of zenith angle bands is mandatory to avoid systematic effects. 

The advantages and drawbacks of the radial acceptance model are the following:

\noindent{\bfseries Advantages} 
\begin{itemize}
  \item Conceptually easy
  \item Known sources (if not overlapping with the neater of the FoV) can be excluded easily
  \item Acceptance can be determined from the actual data set or from an alternate one
  \item Can be computed in energy slices and in zenith angle bands
  \item Simple gradients (zenith angle gradient) can be taken into account rather easily
\end{itemize}

\noindent{\bfseries Drawbacks}
\begin{itemize}
  \item Does not take into account the non-symmetrical response of the camera, nor the actual array shape
  \item Does not take into account inhomogeneities of response 
  \item Does not take into account varying conditions (NSB, etc.) 
  \item Requires a significant amount of data to be already taken with the corresponding array configuration
\end{itemize}
 
\end{paracol}
\nointerlineskip
\begin{figure}[H]
\widefigure
\centering
\includegraphics[width=0.45\textwidth]{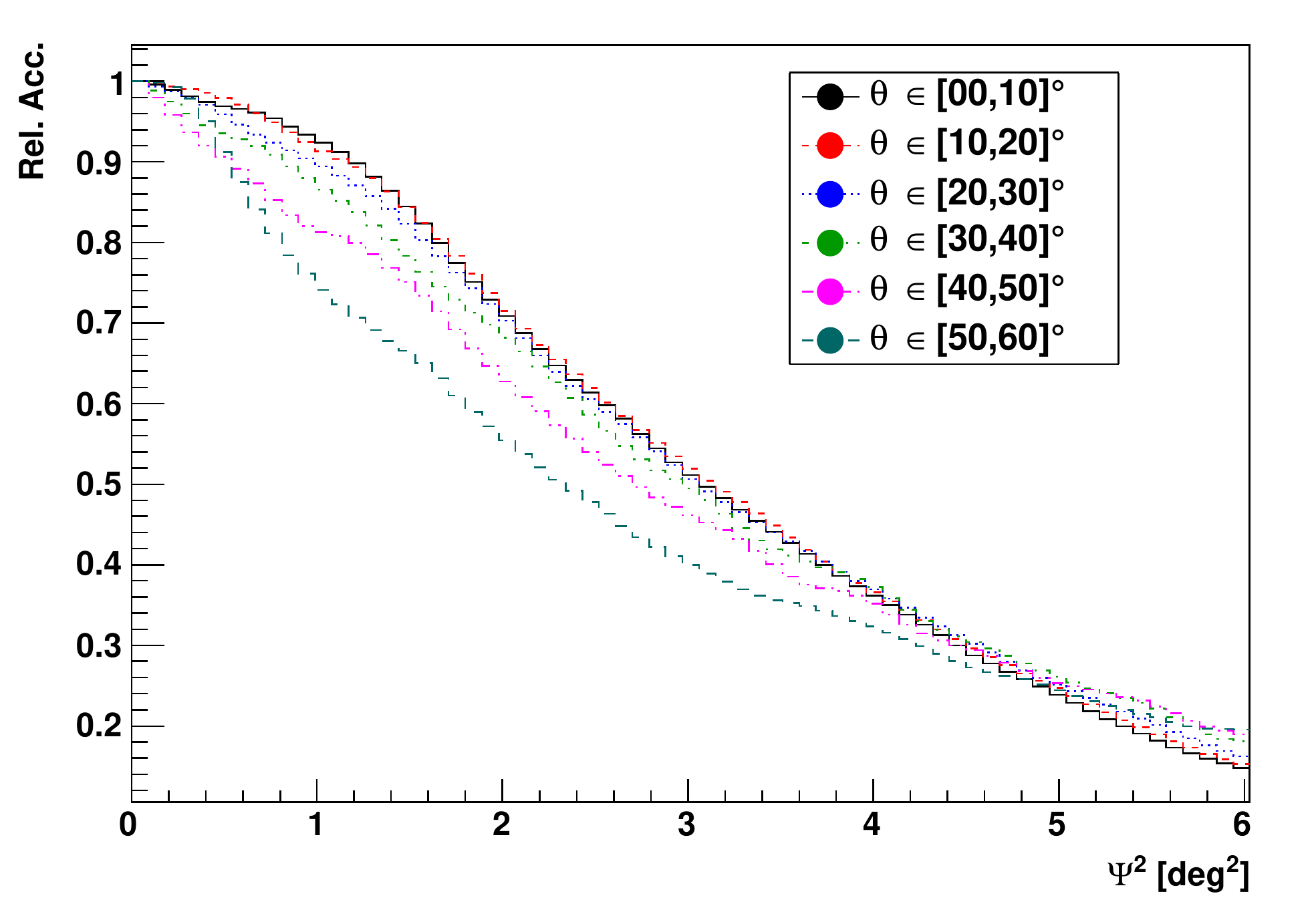} \hspace{1cm}
\includegraphics[width=0.45\textwidth]{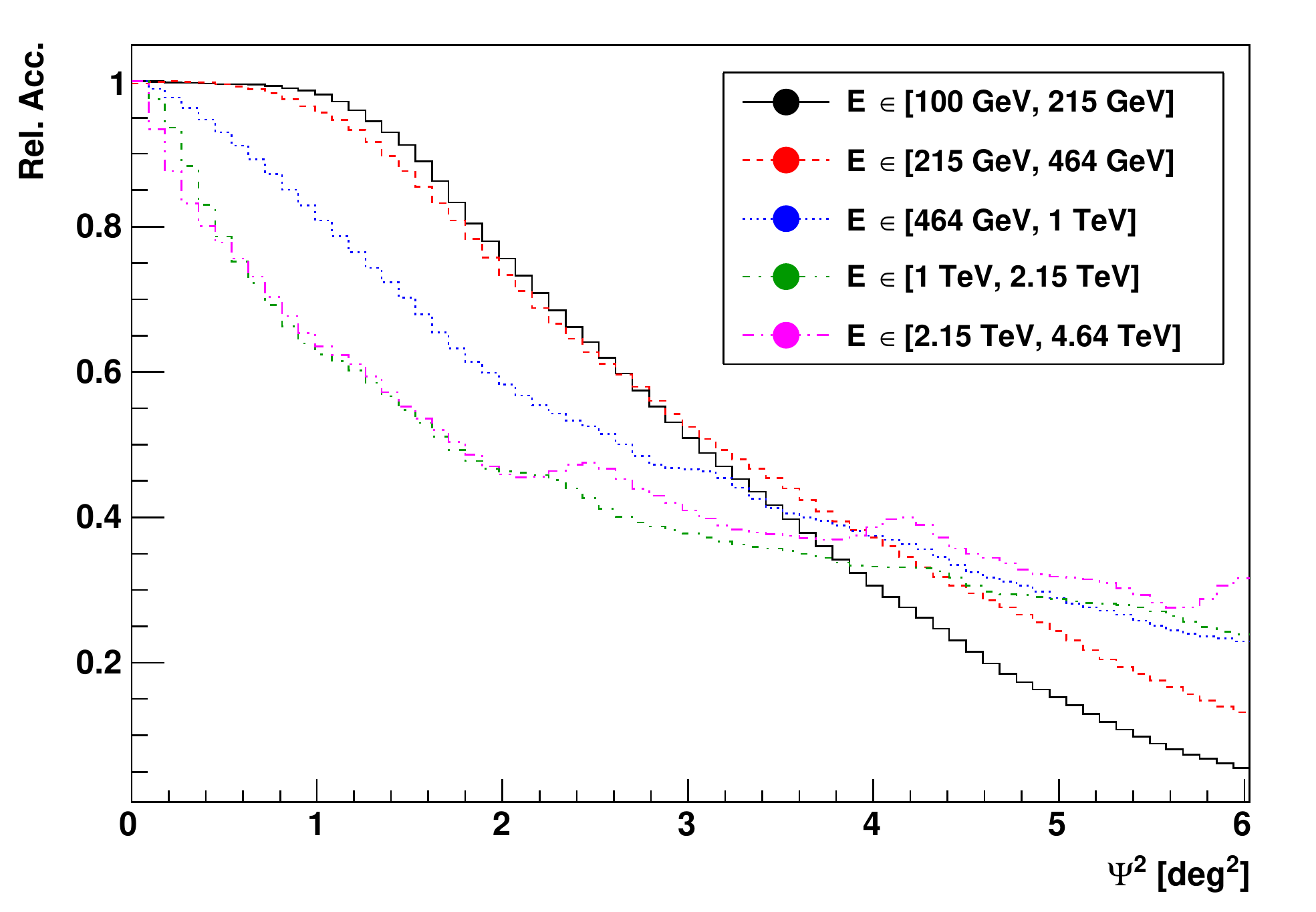}\\
\caption{\label{fig:RadialAcceptanceEvolution}({\bf Left}) Radial acceptance curves for different zenith angle bands.
({\bf Right}) Radial acceptance curves for different energy bands.}
\end{figure}
\begin{paracol}{2}
\switchcolumn

\subsubsection{\label{sec:2DAcceptance}2D Acceptance}
{Bi-dimensional} acceptance (or ``{\it 2D}'' acceptance) is relatively similar in principle to radial acceptance,
but without the assumption of radial symmetry. The response of the array is computed in the nominal frame (i.e., in the 
frame attached to the pointing direction) for every run, and then reprojected onto the celestial coordinates. 
Instead of a radial description of the instrument response, a 2D representation
is used. Since the input statistics are spread over a wider phase space, 2D acceptance needs more data than radial
acceptance to be produced with a similar level of precision.

The exclusion of known and/or putative \gray\ sources is also more complicated than for the radial acceptance, because
sources move in the field of view during the observations. One working algorithm is depicted in
Figure~\ref{fig:2DAcceptance}: throughout the observations, an exposure map is computed by counting the faction of time in which
each pixel is not within one excluded region (top left). The exposure maps of each run are stacked together (top
right), with a weight corresponding to the total number of events per run.
An event map is computed at the same time for each run, excluding the events in the corresponding region (bottom left).
The event maps of all runs are summed up. The final acceptance map is then computed by taking the ratio of the
stacked event map to the exposure map (bottom right). The whole procedure can be performed in parallel for different
event classes (\glike, hadron-like), for different zenith angle bands, or for different energy slice bands. An implementation of 
the 2D acceptance model has recently been made available in \texttt{gammapy}~\citep{2019A&A...632A..72M}.

\begin{figure}[H]
 
\includegraphics[width=10 cm]{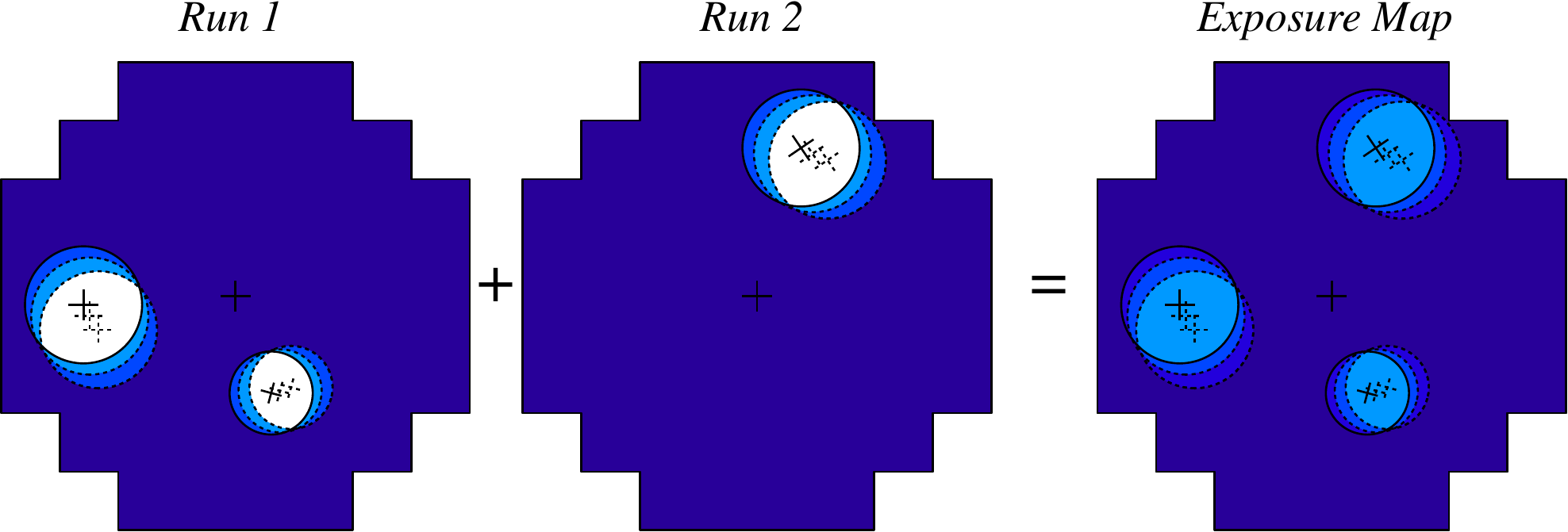} \\
\vspace{1em}
\includegraphics[width=10 cm]{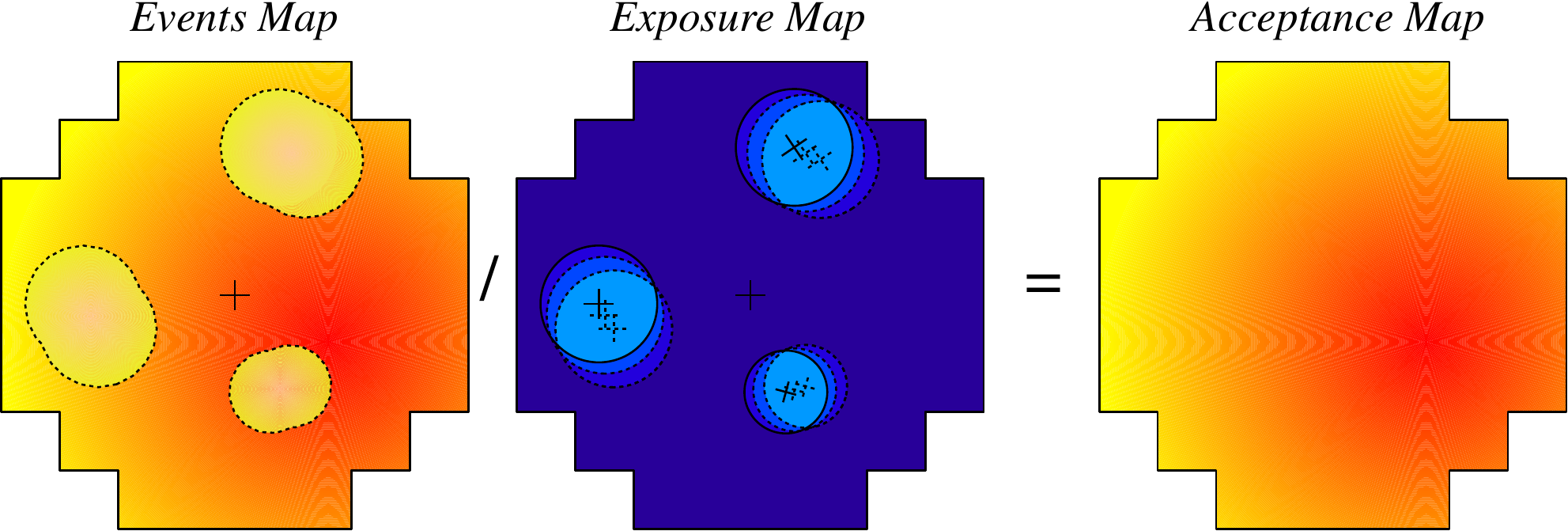} \\
\caption{\label{fig:2DAcceptance}2D acceptance determination. ({\bfseries Top}) Determination of the exposure map by
stacking of maps from individual runs.
({\bfseries Bottom}) Determination of the final acceptance map from the ratio of the event maps to the exposure map.
Reproduced from \cite{HDRMathieu}.}  
\end{figure}

The advantages and drawbacks of the 2D acceptance model are the following:

\noindent {\bfseries Advantages}
\begin{itemize}
  \item Takes into account actual camera shape and inhomogeneities of response
  \item Known sources can be excluded as soon as several different pointing positions are used in the data set (one needs
  to make sure, however, that no part of the FoV is excluded in {\bf all} pointing positions)
  \item Acceptance can be determined from the actual data set or an alternate one (i.e., extragalactic observations)
  \item Can be computed in energy slices
  \item Simple gradients (zenith angle gradient) can be taken into account
\end{itemize}
 
\noindent {\bfseries Drawbacks}
\begin{itemize}
  \item Technically more complicated
  \item Requires a minimum number of runs with sources at different locations
  \item Does not take into account varying conditions (NSB, optical efficiency, \ldots) 
  \item Requires a significant amount of data to be already taken with the corresponding array configuration
\end{itemize}

Both the radial and the 2D acceptance models assume some underlying symmetry. In particular, they assume that the
distribution of events in the field of view does not vary with the azimuth angle of the observation (for a given zenith angle band).
This assumption appears in practice reasonable for arrays with a sufficient number of telescopes and high degree of
symmetry. For very sparse or very asymmetric arrays (or when some telescopes are non-operational), this becomes a
limitation. For instance, in the case of a two-telescope system such as MAGIC, the acceptance exhibits an elongated,
altitude/azimuth-dependant shape, which can be partially corrected by Monte Carlo simulations~(and references
therein, \citep{2012Klepser}).
In addition, the asymmetry caused by the direction of the magnetic field and the induced
asymmetric broadening of showers can induce some additional acceptance systematics, particularly at low energy. 
Generating acceptance models for different array sub-configurations and for different azimuth bands can quickly 
become prohibitive, as it further increases the amount of required data.
In very deep observations, the imperfection of the acceptance models can be readily observed~\citep{HDRMathieu}. 


\subsubsection{\label{sec:RWSAcceptance}Simulated Acceptance}

Since the advent of so-called {\it RunWise simulations}~\citep{2020RWS}, the possibility of generating an acceptance
model exclusively from simulations has been investigated~\citep{HollerRWS}. While theoretically possible, the simulation
of cosmic ray background is in practice prohibitive in terms of computing time, due to the extremely large phase space
and rather low triggering efficiency. Moreover, we are interested mostly in the \glike\ acceptance, corresponding to the
tiny fraction of background events surviving selection cuts. It was instead assumed that \glike\ acceptance would be
rather close to genuine \gray\ acceptance, and could be derived from \gray\ simulations.
For this purpose, diffuse \gray\ simulations over the field of
view are generated for each individual run, using settings as close as possible to the real observations. Actual
calibration coefficients per pixel are used (gains, flat-fielding, non-operational pixels, level of NSB,
pixel threshold, \ldots) and the evolution of the pointing direction during the run (due to the rotation of the
sky) is reproduced in the simulation. It has already been shown in~\citep{2020RWS} that the RunWise simulation offers
a more precise modelling of the instrument response than classical simulations performed on specific grid points of the
phase space. Now, it appears that RunWise simulations can also be used to generate more precise acceptance models,
by taking into account properly any inhomogeneity of response, as well as varying, atmospheric conditions.

One important point to address in this scheme is the aforementioned difference between the cosmic ray \glike\ and \gray\
events, which might exhibit a different distribution across the field of view.
It has been shown however that diffuse \gray\ simulations reproduce fairly well the \glike\ hadronic background, and
that a radial correction, obtained by comparing simulations and actual data from fields free of \gray\ emission, can account
for the difference and lead to a usable background model.
The advantages and drawbacks of the simulated acceptance model are the following:

\noindent{\bfseries Advantages}
\begin{itemize}
  \item Conceptually rather simple
  \item Takes into account the actual array configuration for each individual run
  \item Takes into account varying conditions (NSB, high voltage gradients, pixel gains, \ldots) across the field of
  view
  \item Reproduces naturally the zenith angle gradients (no correction needs to be applied afterwards)
  \item One model per run,  no need to
  generate zenith angle slices or whatsoever, nor to use a multidimensional interpolation scheme
  \item No need to exclude known or putative \gray\ sources, no risk of contamination by large scale diffuse emission
  \item Can be derived as soon as observations are made; no need for a large, pre-existing data set
\end{itemize}
\noindent{\bfseries Drawbacks}
\begin{itemize}
  \item Computationally more intensive (in order to produce enough statistics)
  \item Needs to be produced for every run
  \item Requires some radial corrections due to the difference between cosmic-ray \glike\ events and real gammas
\end{itemize}

\subsubsection{Comparison Elements and Limits
\label{sec:BackgroundComparison}}

For most moderately deep observations, the radial and 2D models usually perform similarly well.
Figure~\ref{fig:AcceptanceComparison} shows a comparison between the radial (top) and 2D acceptance (middle) models for a
very large data set of more than 5000 runs (2500 h of observations) in the
inner galactic plane ($l \in [-50,50]\deg$). The two models agreed within $\approx$1\% (bottom panel), which is
generally sufficient for the standard analyses. This value is similar to what is quoted in~\citep{2007HESS-Backgrounds}, where a typical
detector acceptance inhomogeneity of the order of $3\%$ is also mentioned, with possible larger values in specific
fields that have large NSB variations and/or large zenith angles.


Analysis of deep fields with the current generation of instruments is, however, already dominated by background
systematics arising from, amongst others, an imperfect determination of the acceptance. The actual layout of the
telescopes has an altitude/azimuth dependant imprint on the acceptance, which is not fully predicted by neither the radial
nor the 2D acceptance model. Improving the precision of the acceptance model is a major, but mandatory challenge for the
next generation instruments.
CTA, with a factor of ten larger effective area, will require the acceptance to be determined with a sub-percent level.
This will require significant efforts to include the various sources of systematic differences, arising in particular 
from the actual array layout or the variation of NSB across the
field of~view.

\end{paracol}
\nointerlineskip
\begin{figure}[H]
\widefigure
\centering
\includegraphics[width=\textwidth]{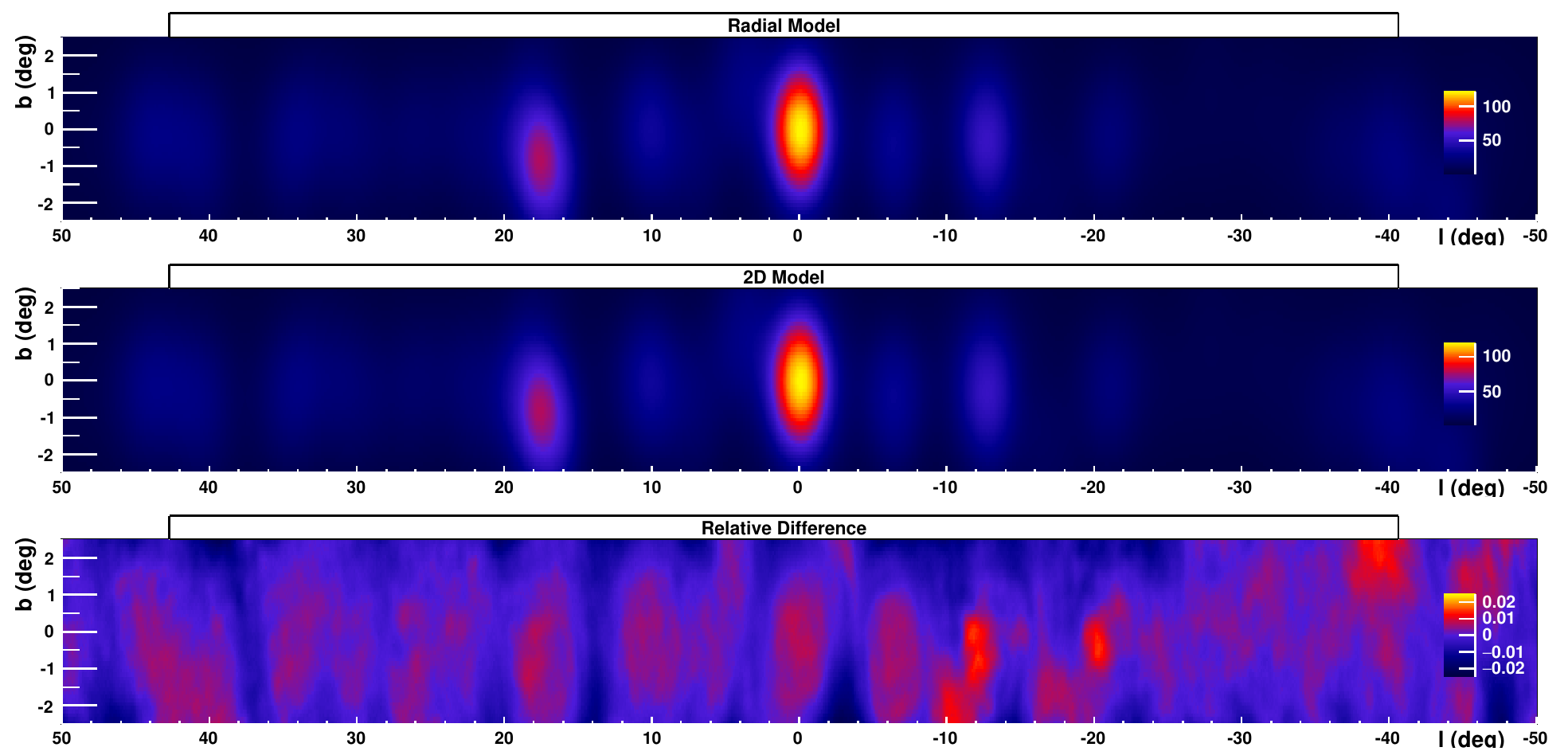}
\caption{\label{fig:AcceptanceComparison}Comparison of radial and 2D acceptances  determination.}  
\end{figure}
\begin{paracol}{2}
\switchcolumn


\subsection{Background Subtraction
\label{sec:BackgroundSubtraction}}

The next step in data analysis corresponds to the comparison of the recorded number of \glike\ events in a region of
interest with an expected number of background events, in order to assess the putative presence of a significant
excess signalling the presence of a \gray\ source.
The evaluation of the expected number of background events can arise from different origins: \glike\ events in different
parts of the field of view or in different regions of the sky, with various reprojection techniques, hadron-like events
at the same location, or Monte Carlo simulations. 
Throughout the history of VHE \gray\ astronomy, a variety of techniques have been developed; some of them
suitable for source detection and morphology determination, some of them also used to derive the energy spectrum
of the sources.
This dichotomy arises because the detector response varies with observational
conditions, and, in particular, depends strongly on the zenith angle: to be able to determine the energy spectrum of
the source, the background subtraction needs to be performed in different energy slices (``{\it Cube}'' analysis).
Some background subtraction techniques are done on a run-by-run basis; some use the complete stacked data set. The main
algorithms used in the field are:

\begin{itemize}
  \item {Reflected background}, using \glike\ events in regions at identical distances from the centre of the field
  of view, on a run-by-run basis,
  \item {On-Off background}, using \glike\ events in identical regions of different, usually consecutive (but not always)
  observations,
  \item {Ring background}, using \glike\ events in a ring around the ROI or around the centre of the field of view,
  \item {Template background}, using hadron-like events at the test position, 
  \item {Field-of-view background}, using calculated acceptance as background,
  \item {RunWise Simulated background}, using completely simulated background.
\end{itemize}

\subsubsection{\label{sec:BackgroundStatistics}Basics of Background Statistics}

When subtracting some background estimate from the number of recorded \glike\ events in an ROI, one needs
to assess the significance of the resulting excess (or deficit). The computation of this significance depends on
the way in which background is estimated.

Whether the background is estimated from the number of events in a different region of the phase space (i.e.,
from a different direction, or from a different event class), the number of background events is subject to Poisson
fluctuations, just like the number of \glike\ events in the ROI. In that case, the Li and Ma statistics~\cite{liandma}
apply. Considering $\mathrm{N_{on}}$, the number of \glike\ events in the ROI and $\mathrm{N_{off}}$, the number
of background events with a normalisation ratio $\alpha$, the significance of an excess $\mathrm{N_{on}} - \alpha\times
\mathrm{N_{off}}$ is given by $S = \sqrt {-2\ln \lambda } $, where $\lambda$ is the {likelihood ratio} between the
null (background only) and the (signal+background) hypotheses:

\begin{equation}
\label{eq:LiMaStatistics}
\lambda = \frac{ P_0(\mathrm{N_{on}},\mathrm{N_{off}} | \overline B_0)}{ P(\mathrm{N_{on}},\mathrm{N_{off}} | \overline S, \overline B)} = \
\left[ \frac {\alpha}{1+\alpha} \left( \frac {\mathrm{N_{on}} + \mathrm{N_{off}}}{\mathrm{N_{on}}}\right)\right]^\mathrm{N_{on}} \times \
\left[ \frac {1}{1+\alpha} \left( \frac {\mathrm{N_{on}} + \mathrm{N_{off}}}{\mathrm{N_{off}}}\right)\right]^\mathrm{N_{off}} 
\end{equation}

This method applies to the reflected, on-off, ring and template backgrounds (see \mbox{Sections~\ref{sec:ReflectedBackground}}
to \ref{sec:Template}). In contrast, when the background is estimated from a model, and not subject to
Poisson fluctuations (as in the field-of-view background), one should use the so-called {cash statistics}~\citep{1979ApJ...228..939C}, from which a similar formula can be derived: 


\begin{equation}
\label{eq:CashStatistics}
\lambda = \frac{ P_0(\mathrm{N_{on}} | \overline B_0)}{ P(\mathrm{N_{on}} | \overline S, \overline B)} = \
\left( \frac{\overline B} {\mathrm{N_{on}}}\right)^{\mathrm{N_{on}}} \exp\left(\mathrm{N_{on}} - 
\overline B\right)
\end{equation}

This method, however, assumes perfect knowledge of the background model, which is, in practice, incorrect. Some ways in which to
take into account the uncertainty in the background model are discussed in Section~\ref{sec:Systematics}. 

\subsubsection{\label{sec:ExcludedRegions}Excluded Regions}

When subtracting some background estimate from the number of events in the ROI, it is of prime importance to ensure that
the background estimate used is not itself contaminated by $\gamma$ rays. This issue is relevant when the background is estimated 
from the population of \glike\ events in different regions of the sky, as for the reflected or ring background in
particular. 
This concern also applies to the case when some normalisation factor between the number of \gray\ candidates and hadron-like events is required, 
which is the case for the template or field-of-view background. 
All in all, the definition of
proper ``{\it excluded regions}'', possibly contaminated by genuine \gray\ events from an astrophysical source, appears
more or less mandatory.

The definition of excluded regions is usually done manually, at least for targeted observations or for modest regions of
the sky.
In some cases however, such as when constructing a complete catalogue over a large region of the sky populated with many
sources, an automatic procedure is required to avoid biases, and becomes mandatory.
Such an iterative procedure was used in the H.E.S.S. Galactic plane survey~\citep{2018HESS-HGPS3}, see
Section~\ref{sec:Surveys}, by excluding all regions with a statistical significance $> 5\,\sigma$ augmented by a margin of
$0.3\deg$ around them.


\subsubsection{\label{sec:ReflectedBackground}Reflected Regions}

The {reflected background} uses \glike\ events from the {\bf same} observation, and from regions located at the
{\bf same angular distance} from the centre of the field of view (\mbox{Figure~\ref{fig:ReflectedBackground}}). 
For each observation (pointing direction displayed as black star in Figure~\ref{fig:ReflectedBackground}), the ROI (red
circle) is located at a different angular distance from the pointing direction. OFF regions  (blue circles) of identical
shape are spaced evenly in the field of view, at the same angular distance from the pointing direction. Regions which
intersect one or several excluded regions (grey regions) are then eliminated from the background estimate.

\begin{wrapfigure}{r}{7cm}
\includegraphics[width=6.9 cm]{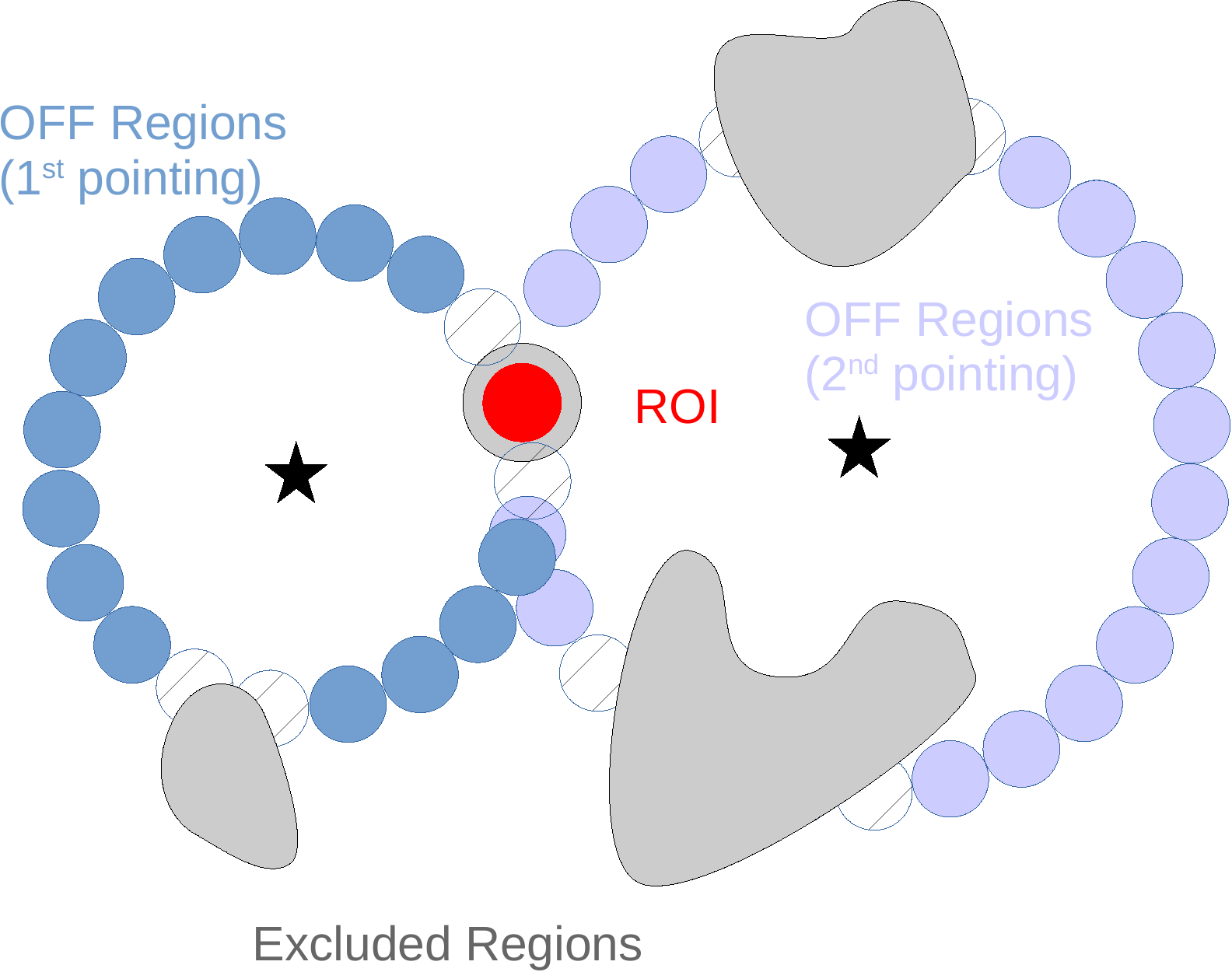}
\caption{\label{fig:ReflectedBackground}Illustration of the reflected regions background, for two different observation positions
(shown as black stars). The positions of the selected OFF regions are shown as filled circles. The excluded regions, which have non-empty overlap 
with excluded regions (displayed in grey) are shown as dashed circles.}
\end{wrapfigure}

Besides its technical simplicity, the main advantage of the reflected background resides in the fact that, with all
regions being located at the same distance from the pointing direction, no radial dependence of the acceptance  has to be
taken into account. 
Only gradients caused by the variation of the zenith angle across the field of view need to be accounted for in the
$\alpha$ normalisation factor. Moreover, with the acceptance being essentially the same in all regions (with identical energy
dependence), the reflected background is very well suited to the determination of the energy spectrum of the source.

In contrast, since the background regions differ for every test position, and are different for each run, the
determination of sky maps using this technique appears non-trivial {(The author is not aware of any implementation of the
reflected background algorithm suitable to the production of sky maps)}. Note that there is one case in which the reflected
background cannot be used: when the ROI overlaps with the pointing direction, no OFF regions can be found with this
algorithm. This imposes the need for the careful planning of observations. 

\subsubsection{\label{sec:OnOffBackground}On-Off Background}

The {On-Off background} is somewhat similar in spirit to the {reflected background}. It also uses \glike\ events
in the field of view, but instead of taking the control (OFF) regions from different positions in the {\bf same}
run, it uses {\bf pairs} of runs with the same observing conditions. This was one of the first methods used in the
field~\cite{1989ApJ...342..379W}, as it is particularly robust to systematics. Observations were paired in right
ascension, such that the telescope trajectory on the sky was completely identical in both runs, thus cancelling the effect
of the varying zenith angle. 

The On-Off background also allows the energy spectra to be derived, and is suitable for very extended sources, but presents
two main disadvantages:
first, the amount of data needed is at least doubled, since for every ON run, a paired OFF run is needed. 
Using a single OFF run for each ON run gives $\alpha = 1$ in Equation~(\ref{eq:LiMaStatistics}), and means that the fluctuations in the background
are dominant in the calculation of the significance. To limit the effect of the fluctuations in the background, one might need
5--10 OFF runs per ON run, which further increases the amount of data, and leads to very poor efficiency.
Second, it requires the
OFF run to be clear of $\gamma$ rays. With the large increase of known \gray\ sources in recent decades, this becomes
tricky, if not impossible, in crowded regions such as the galactic plane. Nowadays, the On-Off background is barely used
anymore. It is still used in very specific projects concerning very extended sources (covering most of the field of view),
for which other methods fail, e.g.,~\citep{2012A&A...548A..38A}. OFF runs are no longer taken from dedicated
observations, but from archival observations of extragalactic fields taken under similar conditions that are empty of \gray\ sources.
This method, where archival data are used instead of paired observation, is also called {matched run background}.

\subsubsection{\label{sec:RingBackground}Ring Background}

The {ring background}~\citep{2007HESS-Backgrounds} also only uses \glike\ events,
but from a different part of the phase space. The overall idea is to compute the expected number of background events 
in the ROI using a ring around its position (see Figure~\ref{fig:RingBackground}). 
The radius and thickness of the ring have a direct influence on the
normalisation ratio $\alpha$, and thus on the final statistics. 
In general, the size (area) of the ring should be set
to a value that is large compared to the size of the ROI (to limit the statistical fluctuations in the OFF regions), but
should not exceed a significant fraction of the size of the field of view, to avoid introducing additional systematics.

 
\end{paracol}
\nointerlineskip
\begin{figure}[H]
\widefigure
\centering
\includegraphics[height=4cm]{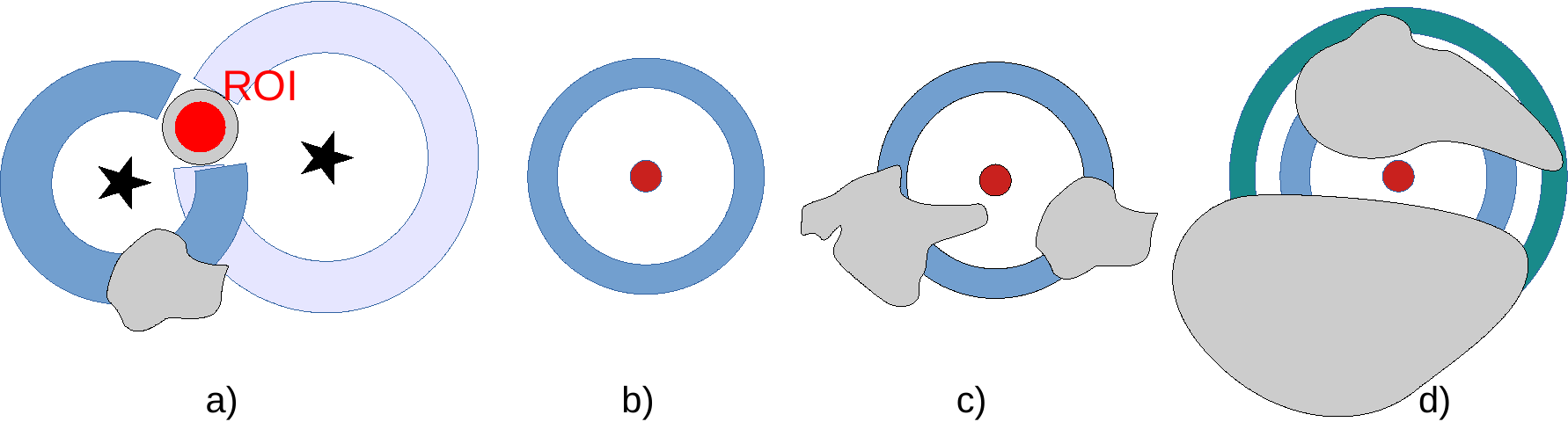} 
\caption{\label{fig:RingBackground}Principle of Ring- and Adaptative Ring Backgrounds. From left to right: ({\bf a})
ring background in camera frame, ({\bf b}) ring background in astronomical frame, ({\bf c}) ring background in
astronomical frame with excluded regions ({\bf d}) adaptative ring background.}
\end{figure} 
\begin{paracol}{2}
\switchcolumn

Two different versions of the ring background currently exist, corresponding to different use cases:

\begin{itemize}
  \item The ring can be constructed around the pointing direction in the camera frame (\mbox{Figure~\ref{fig:RingBackground}{a}}), and then differ from run to run.
  This algorithm is then very similar to that of the reflected regions, and shares the same general properties (spectral
  reconstruction capabilities, \ldots)
  \item The ring can be constructed around the ROI in the astronomical frame (equatorial, galactic, \ldots,
  Figure~\ref{fig:RingBackground}{b}--{d}), and then uses the stacked data set, instead of individual runs to
  generate sky maps.
  The determination of the energy spectrum of the source is, however, very challenging in this version, because the ring
  around the ROI encompasses many different runs, corresponding to different observational conditions which need different
  response functions. The ring background can, however, be performed in energy slices (thus requiring the acceptance
  to also be determined in energy slices).
\end{itemize}

By averaging the background over a large region around the ROI, the ring background is rather robust against
localised background systematics in the OFF region caused, in particular, by small-scale variations of the NSB (bright stars, \dots).
It also permits large values of $\alpha$ in Equation~(\ref{eq:LiMaStatistics}), thus reducing the effect of the background fluctuations and
improving the statistical power of the analysis.
The drawback is that it tends to remove any large structure of \gray\ emission, such as the large-scale galactic diffuse
emission. 

\subsubsection{\label{sec:AdaptativeRingBackground}Adaptative Ring Background}

In very crowded regions, such as the Galactic plane, the presence of (very) extended sources can make large
fractions of the ring unusable, as shown in Figure~\ref{fig:RingBackground}{d}. The normalisation ratio $\alpha$ then takes
very different values depending on the position of the ROI, leading to inhomogeneous performances.
In some cases, the full ring would be excluded, leading to holes in the significance map. 
For that reason, the concept of {adaptative} ring background was
introduced in~\citep{2018HESS-HGPS3}: for a given test position, the size of the ring is increased progressively until
the acceptance integrated within the ring (and outside excluded regions) reaches at least four times the acceptance
integrated in the ROI.

\subsubsection{\label{sec:Template}Template Background}

The {template background}~\cite{2003Rowell-Template} differs completely from the previous models. It makes use of the
fact that only a small fraction of events in the ROI are \glike\ events, the vast majority being cosmic ray events, 
also denoted as hadron-like events, which can be used to estimate the background.
It assumes that the rate of \glike\ and cosmic-ray events are, to some predictable factor, proportional. The ratio
between the two is estimated from the ratio of the relative acceptances to \glike\ and hadron-like events calculated
previously.

Until relatively recently, the template background was only used to derive the morphology of \gray\ sources. 
Spectrum determination appears very challenging, since the population of events is made from the superposition of 3
categories, for which the response functions have to be determined, either from Monte Carlo simulation or from OFF data:

\begin{itemize}
  \item \glike\ events in the ROI (entering the ON sample), mostly made of hadronic and electronic cosmic rays within
  the \glike\ selection
  \item true \gray\ events, corresponding to signal being sought (also entering the ON sample)
  \item hadron-like events in the ROI (entering the OFF sample)
\end{itemize}

A method was proposed in \citep{2014A&A...568A.117F}, in which the template background normalisation is done in
reconstructed energy bands, and various lookup corrections are made to correct for the different shape of the acceptance
for \glike\ and hadron-like events. Although it provides consistent results with classical methods and can be applied in
crowded regions where there are no \gray\ free regions (which will be a clear advantage in the context of the
upcoming CTA), its complexity might introduce new systematics which are not easy to assess. 
This is a substantial problem at low energies where the ratio of \glike\ events to hadron-like events degrades. In general,
such methods work rather well with hard selections, but are subject to large systematics when using loose selections.

\subsubsection{\label{sec:FoVBackground}Field-of-View Background}

In the {field-of-view background}~\citep{2007HESS-Backgrounds}, the acceptance is directly used as the background model,
with a normalisation factor usually derived from specific regions in the field-of-view (regions assumed to be free
from \gray\ emission, such as side bands in the case of the Galactic plane). The acceptance can be derived from the same
data set, or from OFF observations. Since much larger statistics are used to derive the acceptance at each test
position, the statistical fluctuations of the background model are usually considered negligible, and the {cash
statistics} are used (Equation~(\ref{eq:CashStatistics})). The field-of-view background can be applied to very extended sources,
or even to diffuse structures, and has the largest statistical power (as the normalisation factor $\alpha$ is null), but is
prone to systematics induced by the imperfect determination of the acceptance.

The field-of-view background was, until now, rarely used in VHE \gray\ astronomy. It has recently been used in a detailed
comparison between the \hess\ and HAWC views of the galactic plane~\citep{2021ApJ...917....6A}.

\subsubsection{\label{sec:Systematics}Assessment of Systematics}

When the background is properly modelled, the significance distribution derived from Equations (\ref{eq:LiMaStatistics})
or (\ref{eq:CashStatistics}) (depending on the algorithm used) should follow a normal distribution. 
In the presence of non-negligible systematic differences between the actual background distribution and its model,
the distribution is widened. Noting $\sigma_\mathrm{sig}$ the Gaussian width of the significance distribution, the
relative level of background systematics $f_\mathrm{syst}$ in the field of view can be estimated simply as 

\begin{equation}
\label{eq:Systematics}
f_\mathrm{syst} = \sqrt {\frac{\sigma_\mathrm{sig}^2 - 1}{\left\langle B \right\rangle}}
\end{equation}

\noindent where $\left\langle B \right\rangle$ is the average number of background events per sky bin. This simple
evaluation provides an easy-to-calculate, single number per field, but does not take into account the fact that the
number of background events varies significantly across the field of view, notably in the presence of strong acceptance
gradients. More elaborate models have been developed to quantify more precisely the level of background
systematics, e.g.,~\cite[][]{HollerSystematics}. 
State-of-the art analyses of IACT data reach background systematics of the order of $1 - 2\%$.
Background systematics can arise in particular from variation of the
night sky background across the field of view, the variation of calibration coefficients (high voltage, pixel gains, \ldots) across
the camera, changing atmospheric conditions, but also pointing direction with respect to the earth's magnetic
field direction (which affects the lateral development of showers).
When using the simulated acceptance (Section~\ref{sec:RWSAcceptance}), the systematics level should not
depend much on the FoV, because most of the predictable, field-of-view dependant effects are properly taken into account
in the acceptance.
For other acceptance models, the field-of-view effects are expected to be the dominant source of systematics.

Background systematics are already the limiting factor for very deep exposures and/or very extended sources in the current
generation of instruments, and have been identified as a major challenge for the next generation instruments, and in
particular for CTA. In this context, several strategies for the mitigation of the background systematics have already been
investigated.
In~\citep{2015Spengler}, it is proposed to take the systematics into account by adding an uncertainty to the $\alpha$
factor in Equation~(\ref{eq:LiMaStatistics}), and by modelling the resulting significance distribution. This restores the
correct statistical behaviour of the significance across the field of view (and in particular its normal distribution), but the
price to pay is a significant reduction of the sensitivity.
In~\citep{2013Dickinson}, a joint-likelihood is used to compute the total significance instead of stacking the
individual observations together. The $\alpha$ parameter is modelled as a random variable for each observation.
This solves some of the problems that occur when stacking observations with very different values of $\alpha$ for which the error
propagation appears problematic, while offering equivalent or superior sensitivity, but it implies a good knowledge of
the $\alpha$ distribution for each observation. In~\citep{2012Klepser}, no assumption on the shape of the acceptance is
made. Instead, observations are grouped by similar observational conditions (array configuration, zenith angle, \ldots),
and a generalised likelihood ratio is used to derive simultaneously the signal and the background at a given position,
assuming identical relative acceptance shapes for the observations belonging to the same group. 
It does not, however, solve the problem of field-of-view systematics which
vary from observation to observation, even within the same group.

\subsubsection{Comparison}

A comparison of three background subtraction algorithms, using the same data set as for
Figure~\ref{fig:AcceptanceComparison} ($100\deg$ of the \hess\ galactic plane survey), with a top-hat smoothing of
$0.25\deg$), is shown in Figure~\ref{fig:BackgroundSubstraction}.
The panels look overall very similar, however the template and field-of-view backgrounds
tend to produce more ``diffuse'' emission or ``bridges'' between the well identified, localised sources and exhibit
consistently larger systematics than the ring background. Note, however, that in this example, the acceptance model (here
2D acceptance) was determined using the same data set, and might therefore contain some residual contamination from large-scale galactic diffuse
emission. Moreover, the excluded regions were not optimised again for this analysis and might be undersized. This example
should therefore serve as an illustration of the sensitivity differences between different algorithms, and not as an input for a scientific discussion.

\end{paracol}
\nointerlineskip
\begin{figure}[H]
\centering
\widefigure
\includegraphics[width=\textwidth]{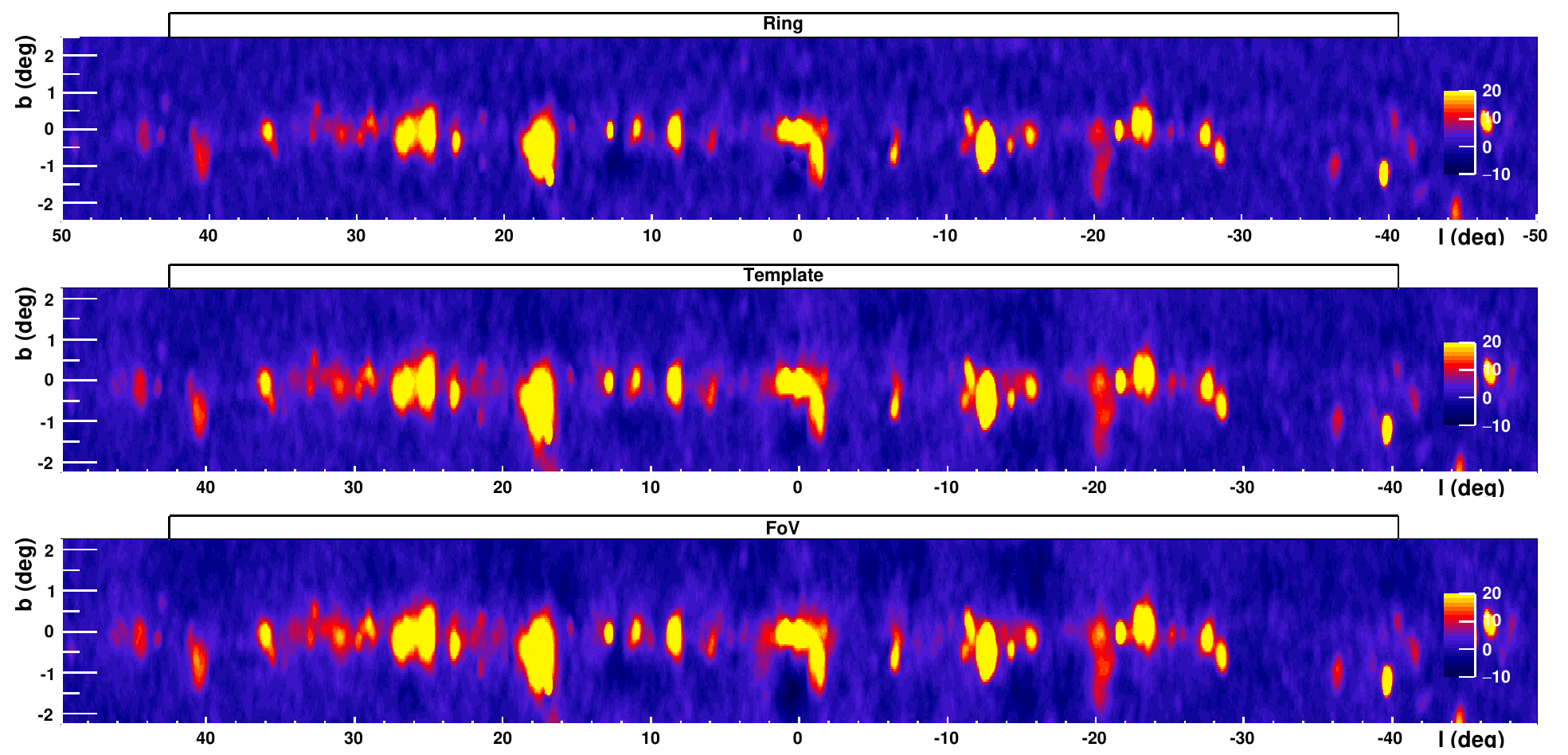}
\caption{\label{fig:BackgroundSubstraction}Comparison of three background subtraction algorithms for the inner
$100\deg$ of the \hess\ Galactic plane survey, with a source size of $0.25\deg$
({\bfseries Top}) ring background.
({\bfseries Middle}) template background. ({\bfseries Bottom}) field-of-view background.}
\end{figure} 
\vspace{-10pt}
\begin{paracol}{2}
\switchcolumn

\subsection{Toward Template Fitting
\label{sec:TemplateFitting}}

Template fitting is the state-of-the-art in high-energy \gray\ astronomy, and is the default in {Fermi}-LAT data
analysis: the counts maps or photon lists are compared in an iterative procedure to a composite model using a
Likelihood analysis.
The model describing the data is built from the following ingredients:

\begin{itemize} 
\item a model of isotropic diffuse emission, corresponding to extra-galactic diffuse $\gamma$ rays, unresolved
extra-galactic sources, and residual (misclassified) cosmic-ray emission.
\item a model of the Galactic diffuse emission, which is developed using, in particular, spectroscopic HI and CO surveys as
tracers of the interstellar gas, and diffusion codes such as
GALPROP~\citep{1998ApJ...509..212S} ({\url{https://galprop.stanford.edu/}}) to derive the inverse Compton emission
\item a source model, comprising the \gray\ source properties (morphology and energy spectrum) within the region of
interest. Characteristics of the sources (position, shape, energy spectrum and brightness) can be fixed (for
instance to the published values) or kept free, in which case they will be adjusted throughout the log-likelihood
maximisation procedure.
\end{itemize}

Additional models for large scale components, such as the {Fermi} Bubbles for instance, can be incorporated as well. The
source model is usually constructed iteratively, by adding new sources until the likelihood converges.
In contrast to the high-energy domain, template fitting is so far still in its infancy in very high-energy \gray\ data
analysis, but will certainly become one of the major, if not the default, analysis procedure in the coming years.

Building on its success in high-energy \gray\ astronomy, the MAGIC collaboration recently implemented such a
template fitting  {procedure}~\citep{2018A&A...619A...7V}.
Open-source software such as
\texttt{gammapy}~\citep{2019A&A...632A..72M} and \texttt{ctools}~\citep{2016A&A...593A...1K} already propose a template
fitting procedure. One very important difference with respect to high-energy \gray\ astronomy lies in the way in which the
background model is generated: high-energy \gray\ instruments are signal dominated, and the so called {background}
consists mostly of genuine $\gamma$ rays, but of diffuse origin. This model can be incorporated directly in the final part of
the analysis, using the standard instrument response functions.
In contrast, IACTs are background-dominated, and the
remaining background consists of mostly hadronic or electronic cosmic rays, which are much more complicated to evaluate.
The model used in template fitting analysis must, therefore, incorporate such a background model, or {acceptance},
produced by the procedure described in Section~\ref{sec:Acceptance}.

\subsection{Catalogue Pipelines
\label{sec:CataloguePipelines}}
 
In this section, the tools used in the final part of the catalogue construction are described.
 
\subsubsection{Requirements
\label{sec:CataloguePipelinesRequirements}}

Until recently, the analysis of large data sets was done in a completely supervised way, with most tasks being the
responsibility of the scientist. In particular, the excluded regions were defined manually, based on the known existing
sources and on results obtained previously. Similarly, source identification was done based on existing spatial overlap and 
similarly in shape with counterparts at other wavelengths, and was subject to human judgement. With the
increasing exposure and consequent depth of the data sets, the problem of source confusion and overlapping has also
become crucial, pushing for fully automated catalogue pipelines. The main tasks of an automated catalogue pipeline are: 

\begin{enumerate}
  \item Selection of good quality data, based on instrumental and atmospheric measurements (stability of instrument
  trigger rate, cloud monitoring, atmospheric transparency measurement, \dots).
  \item Construction of an excluded regions mask, incorporating already-known \gray\ sources, but also new sources and/or
  possible diffuse contamination within the data set under investigation.
  \item Computation of acceptance.
  \item Construction of background subtracted maps (excess and significance maps) using the appropriate
  algorithm (adaptative ring background, \dots).
  \item Determination of source components and morphologies.
\end{enumerate}
 
The whole procedure usually needs to be executed several times in an iterative way: when new sources are identified at
step 5, the excluded regions from step 2 need to be refined, and the whole loop needs to be performed again. Some
quantitative criteria are also needed to decide when to stop the iterations. 
The analysis pipeline can also incorporate additional tasks, such as automatic searches for transients events and for
source variability, as well as search for counterparts at other wavelengths, which are currently still mainly done manually,
since this requires some physics expertise.

\subsubsection{Completeness, Angular Resolution and Horizon
\label{sec:CataloguePipelinesHorizon}}

As mentioned already, IACT are background-dominated instruments. This has numerous implications for the large-scale 
surveys and for the construction of catalogues.
For sufficiently high statistics and low signal to background ratio (reasonable assumption), the significance of a
detection scales with the source intensity and observation time as:

\begin{equation}
\sigma \propto {\phi \, A} \sqrt{\frac {t}{B}}
\end{equation}

\noindent with $\phi$ the source flux (at Earth), $A$ the effective area of the array, $t$ the observation time and $B$
the background rate, which depends on the detector characteristics and, therefore, indirectly on the effective area. The
minimum detectable flux thus scales as $1/\sqrt t$, which usually limits the depth of existing surveys.

The background rate, $B$, depends on various instrumental characteristics (array geometry, background rejection power,
and angular resolution, among others), but also on the source extent. In the context where most of the galactic sources
are (very) extended, as demonstrated by the results accumulated over recent years, one can neglect the effect of
the angular resolution and assume that the background rate scales as the source solid angle ($B \sim b \times
\Omega_{\mathrm{s}}$).
Assuming a source at a distance $d$ with a physical extent $R$ and an intrinsic luminosity $L$, the scaling of the
significance for point-like and extended sources becomes, respectively:

\begin{equation}
\sigma_{\text{pt-like}} \propto \frac{L \, A}{d^2}  \sqrt{\frac {t}{B}}; ~~~~~~
\sigma_{\text{extended}} \propto \frac{L \, A}{d \, R} \sqrt{\frac {t}{b}}
\end{equation}

It follows that:

\begin{itemize}
  \item For a homogeneous population of sources of the same luminosity and size, the maximum detection distance (horizon)
  scales as $d_\mathrm{max} \propto (L/R)\sqrt{t/b}$. For point-like source, it scales as $d_\mathrm{max} \propto
  \sqrt{L} (t/B)^{1/4}$. The horizon of a given survey
  therefore depends on the type of sources that one considers. It is usually defined for point-like sources, but can
  be reduced substantially for extended sources.
  \item The reduction of apparent size $\Omega_s$ with increasing source distance $d$  partially compensates for the
  decrease in flux.
  Indeed, the minimum detectable luminosity scales as $L_\mathrm{min} \propto d\, R \sqrt{b/t}$ for extended sources vs. 
  $L_\mathrm{min} \propto d^2 \sqrt{B/t}$ for point-like ones. The survey depth depends on the source
  class considered.
  In the case of source class for which the extent varies with age (as for instance, for expanding shell-type supernova remnants),
  better flux limits can be obtained in the early ages, when the source is still rather compact, whereas the peak of the
VHE emission can occur at later stages.
  \item The horizon scales as $t^{1/2}$ for extended sources and $t^{1/4}$ for point-like ones, and is currently still
  limited to a rather small fraction of the Milky Way. It is usually more effective to increase the spatial coverage  of
  a survey (if possible) to collect more sources, rather than to increase its depth.
\end{itemize}
\section{A New View on the Milky Way\label{sec:Results}}

Over the last 20 years, major collaborations in the field have conducted several surveys of varied angular extent,
completeness and depth, which has led to the discovery of many sources, and allowed for the first population studies to be
performed.

\subsection{Existing IACT Surveys\label{sec:Surveys}}

\subsubsection{\label{sec:EarlySurveys}Early Times}

The HEGRA collaboration conducted the first systematic survey of modern TeV \gray\
astronomy~\citep{2002A&A...395..803A}. It consisted of 176 h of observations covering one quarter of the Galactic
plane ($-2\deg < l < 85\deg$) and resulted in no source detection, thus placing upper limits in the range between $0.15$
to a few Crab units, depending on the observational conditions. Source stacking on some source populations (bright GeV sources, nearby
Supernova Remnants, powerful and nearby pulsars) was used to derived more constraining, so-called {ensemble
limits}.

\subsubsection{\label{sec:GalacticSurveys}Galactic Plane Surveys}

Following on this, the \hess\ collaboration conducted the most comprehensive survey of the Milky Way so far, as part of
a decade-long observational program, dubbed HGPS. Nearly 2700 h of good quality data were accumulated between 2004 and
2013, in the longitude range ($-110\deg < l < 65\deg$), with a sensitivity better than $\leq$1.5\% Crab flux. The survey
was published in 3 successive papers~\citep{2005HESS-HGPS1,2006HESS-HGPS2,2018HESS-HGPS3}, comprising data sets of increasing size.
Whereas the first two papers used a manual source identification procedure, the last paper proposed, for the first time,
a semi-automated pipeline, similar to that described in Section~\ref{sec:CataloguePipelinesRequirements}. The resulting
flux map is shown in Figure~\ref{fig:HGPS} and comprises 78 firmly identified VHE sources. 

Out of these 78 sources, the majority (47) are associated with an energetic pulsar, and 12 of them correspond to a
firmly identified pulsar wind nebula (PWN). The second population by frequency corresponds to supernova remnants (SNR),
with 24 sources associated with a shell-type SNR (although the number of chance coincidences is non-negligible, due to the
number and the large angular extent of such objects). Six VHE sources are firmly identified as SNRs, with two additional
candidates based on their shell-type morphologies. Three binary systems finally form the only class of variable galactic 
sources at these energies (so far).

\end{paracol}
\nointerlineskip
\begin{figure}[H]
\centering
\widefigure
\includegraphics[width=.97\textwidth]{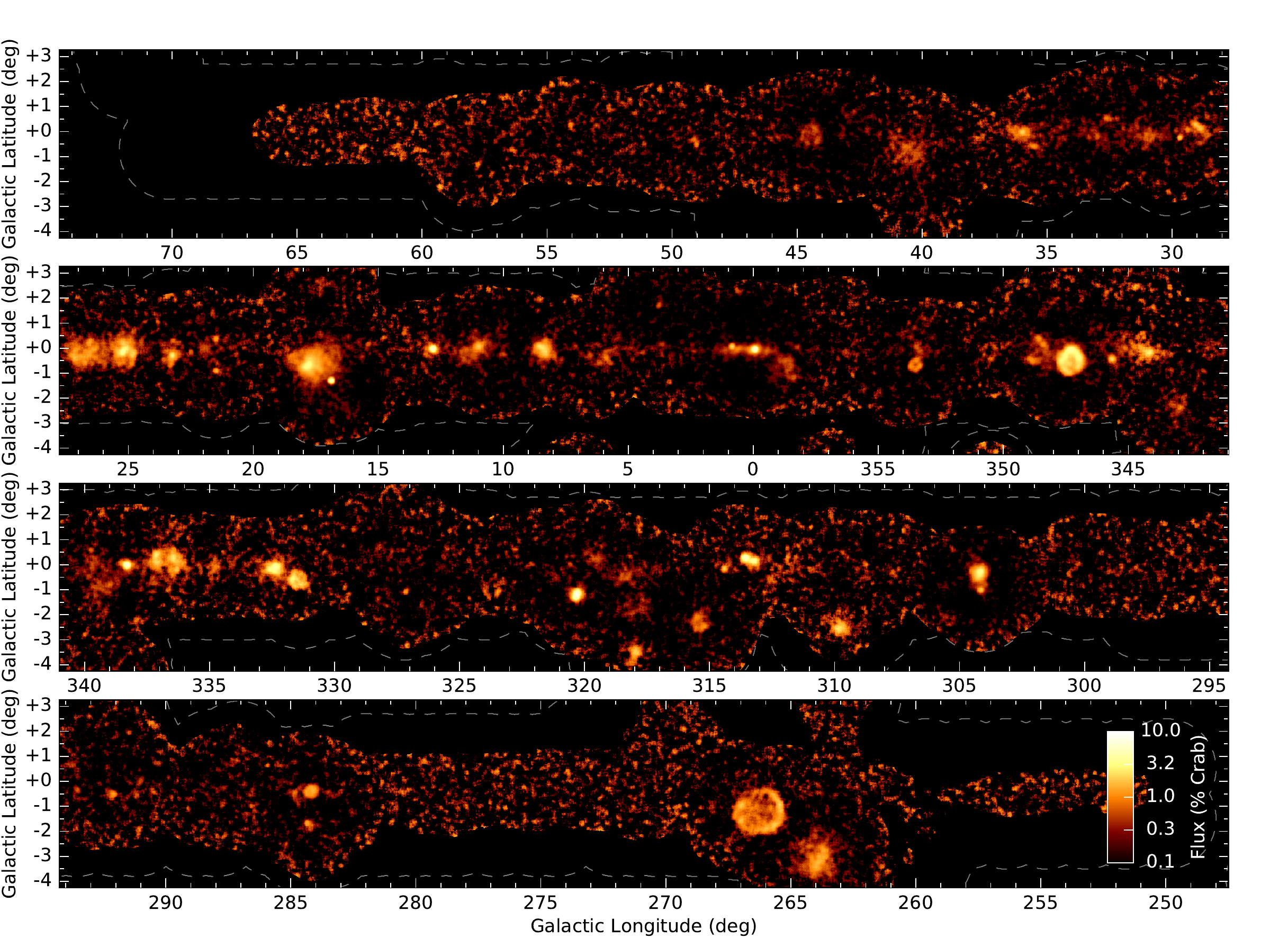}
\caption{\label{fig:HGPS}\hess\ Galactic Plane Survey: Integral flux above 1 TeV. Reproduced from
~\citep{2018HESS-HGPS3}  with permission \copyright\ ESO.} 
\end{figure}
\begin{paracol}{2}
\switchcolumn

It should be noted that a large number of sources (36) cannot be firmly identified with the rather strict
association criteria used in the process (positional evidence and, depending on the source class, energy-dependant
morphology consistent with other wavelengths, variability,\ldots). In most cases there are, however, 
plausible counterparts.
Eleven sources, denoted as ``Not associated'', did not have any plausible association at the time of publication of the
paper.


While a rather large fraction of the galaxy has been sampled to $10\%$ of the Crab flux (point-like sensitivity), a flux
limit of $1\%$ Crab can only have been reached in the solar system's neighbourhood. 
From the $\log N$--$\log S$ distribution an estimate of $\sim$600 sources in the Galaxy above $1\%$ Crab was obtained
(with a statistical error of a factor of 2). The HGPS included a large-scale emission model, accounting for both unresolved
sources and genuine, diffuse emission, due to the interaction of cosmic rays with the interstellar medium. This ``diffuse''
component, already established in~\citep{2014HESSDiffuse}, has a latitude distribution similar to that of the HGPS
sources.
Based on a source population synthesis, \citep{2020A&A...643A.137S} estimated that a significant fraction (13--32\%)
of the the \gray\ emission within the HGPS is due to yet unresolved sources. They estimate the total number of VHE
sources in the Galaxy to be in the range from 800 to 7000.

\hess\ also performed a deep survey of the large magellanic cloud~\cite{2015HESS-LMC}, with 210 h of data.
Although the LMC is located much further away than the Galactic Centre, the survey resulted in the detection 
of three sources of exceptional intrinsic luminosity: the superbubble 30 Dor C,
the energetic pulsar wind nebula N~157B, and the radio-loud supernova remnant N~132D.
Since the LMC is seen almost face-on, source confusion is not
a problem as it might be in the Milky Way. N~157B and N~132D belong to the classes of sources that are represented in 
the HGPS, but they stand out by their distinguishing characteristics. N~157B is indeed being powered by the most
energetic young pulsar known so far, while N 132D is one of the oldest VHE \gray\ emitting SNRs,
with possibly a very high cosmic-ray acceleration efficiency.

\subsubsection{\label{sec:RegionsSurveys}Particular Regions}

A few years ago, VERITAS published a survey of the Cygnus region~\citep{2018VERITAS-CYGNUS}, based on 
300 hours of data collected over 7 years. This region, where the Cygnus arm of the galaxy is observed 
tangentially, is the brightest region of diffuse \gray\ emission in the northern sky, and could also exhibit 
one of the largest density of sources of \grays. The VERITAS survey covered a region of $15\deg$ by $5\deg$
(Galactic latitude $l\in [67\deg;82\deg]$) and reached a point-like sensitivity of $\lesssim$3\% Crab.
Four already known \gray\ sources (out of which three are significantly extended) are detected in this survey. Detailed
analysis of the significance distribution did not indicate the presence of additional, sub-threshold sources.
Upper limits on a large number of potential targets were derived (including, in particular, energetic pulsars and
supernova remnants). Many {Fermi}-LAT sources visible at lower energies were not detected in VHE in this survey,
and the ratio of VHE to HE sources appears rather similar to that in the \hess\ survey region.

\subsection{\label{sec:SurveysFromSurvey}Results from Particle Array Survey Instruments}

Non-imaging particle array instruments such as Milagro and its successor, HAWC, rely on a completely different technique. 
Instead of detecting the Cherenkov light emitted by the charged particles in the atmospheric showers,
they detect the particles of these air showers that reach the ground. Various techniques have been investigated 
in the past, including very large surfaces of resistive plate chambers~\citep{2002ARGO}, plastic
scintillators~\citep{2019TibetAS}, and water Cherenkov~\citep{2003Milagro, 2017ApJ...841..100A}.
More recently, LHAASO started to operate a system consisting of three interconnected detectors, 
combining water and air Cherenkov with scintillators~\citep{2021LHAASO}.

Particle array survey instruments are confronted with very large amounts of raw data collected over many years,
which pose some specific challenges for the analysis. Data analysis techniques usually use a likelihood formalism, e.g.,~\cite[][]{2015ICRC...34..948Y},
in which a physics model (sky position of \gray\ sources, spectrum, angular extent, etc.)
is confronted with the data through a likelihood maximisation routine that takes into account the detector response.
The number of background events (hadronic events passing the selection cuts) in each sky bin is usually estimated directly from 
the data, either prior to the maximisation procedure (using off-source data), or directly in the procedure via an additional, nuisance parameter 
inserted in the log-likelihood.

Compared to IACTs, survey instruments have a much better duty cycle (close to 100\%), very large fields of view (nearly
half of the sky), but poorer hadronic rejection and reconstruction capabilities, leading to poorer angular resolution 
(of the order of $1\deg$) and limited spectral performances. They are, however, well suited to the analysis of 
extended sources in general, and to the study of large-scale diffuse emission in particular. 
HAWC, with a sensitivity improved by one order of magnitude compared to Milagro, started to provide a very complementary 
view on the Milky Way in VHE with unbiased, large-scale surveys.

While the Milagro survey only yielded two sources of $\gamma$ rays, the Crab Nebula and Mrk~421~\citep{2004Milagro},
the first HAWC catalogue~\cite{2016HAWC-1HWC}, with an incomplete array, already contained ten sources and candidate sources,
three of them being detected with significances $>5 \sigma$ (post-trials).
The two following catalogues, 2HWC~\citep{2017HAWC-2HWC} and 3HWC~\citep{2020HAWC-3HWC} contain, respectively, 39 and 65
sources (among which 17 are considered as secondary sources, being not well separated from neighbouring sources).
As the first, large-scale, unbiased catalogue, this constitutes a major contribution to the field. 
The all-sky significance map, under the assumption of point-like sources, is displayed in Figure~\ref{fig:3HWC}:  most VHE sources are,
like for \hess, concentrated along the galactic plane.

 
\end{paracol}
\nointerlineskip
\begin{figure}[H]
\centering
\includegraphics[width=0.8\textwidth]{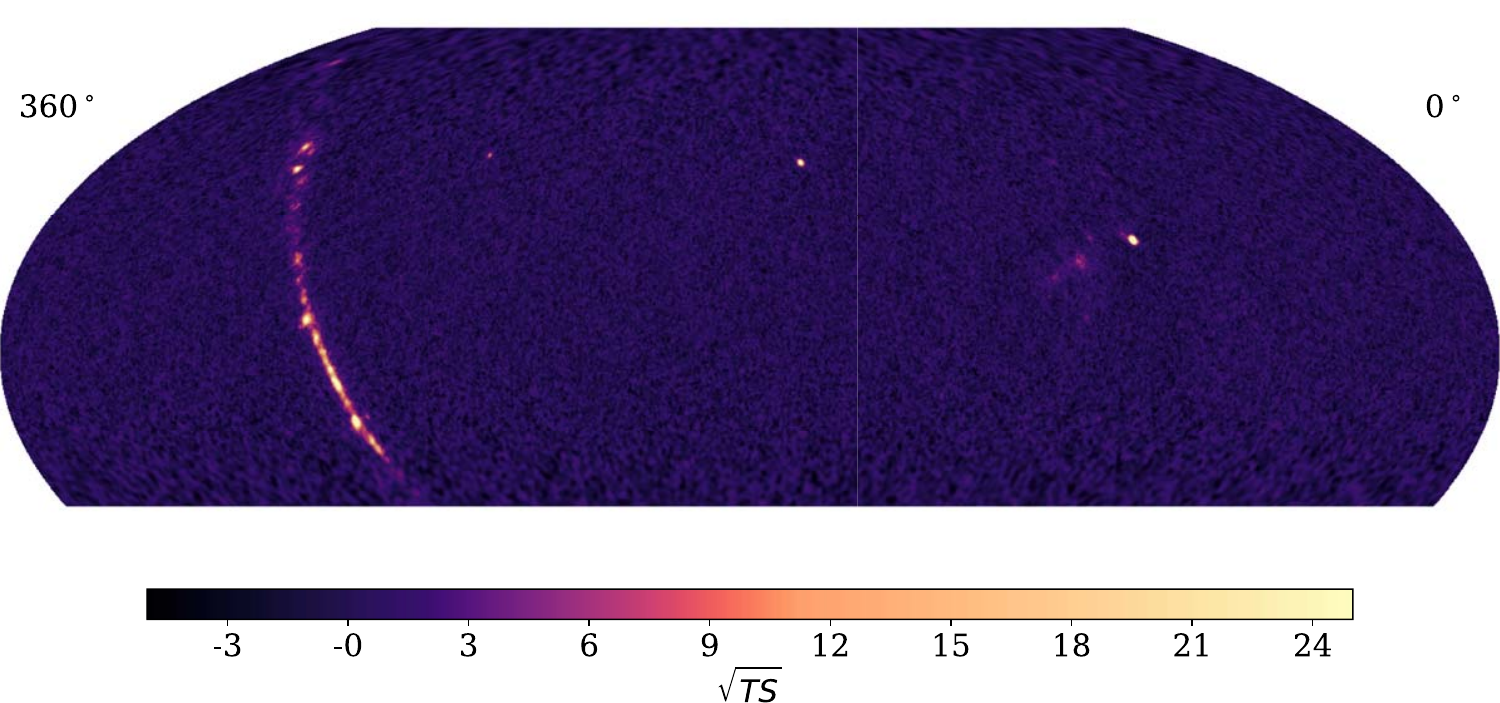}
\caption{\label{fig:3HWC}All-sky significance map of the third HAWC Catalogue of Very-high-energy Gamma-Ray Sources.
Reproduced from~\citep{2020HAWC-3HWC} with permission \copyright\ The American Astronomical Society.}
\end{figure} 
\begin{paracol}{2}
\switchcolumn

The overall analysis and catalogue construction is very different from that in use for IACTs, and is not the main
subject of this paper. In general, the shower core is reconstructed using the density of particles on the ground, while the
timing provides the shower axis, and thus the reconstruction of the direction. The homogeneity of the particle density is used to
discriminate between \grays\ and charged cosmic rays, and a likelihood ratio procedure is used to produce the
significance map.

One of the most interesting features of the HAWC data is the presence of very extended \gray\ emission
around young pulsars (and in particular Geminga and Monogem~\citep{2017Sci...358..911A}), which indicates
that such extended pulsar ``halos'' could be a rather common feature, even for old pulsars which could have already left
their SNR shell, or whose shell could have already vanished. Indeed, out of the 65 detected HAWC sources,
56 have a pulsar as a plausible counterpart. 
This could open new prospects for the quite numerous unidentified VHE sources.

Since the results from \hess\ and HAWC, in the part of the sky that is visible to both instrument, appeared to be rather
different at a first glance (due to the different instrumental performances), it appeared mandatory to compare the
results more thoroughly. This was done in~\cite{2021ApJ...917....6A}, using the field-of-view background (suitable for
very extended sources) and smoothing the \hess\ data to mimic the HAWC angular resolution. 
The results, shown in Figure~\ref{fig:HESSvsHAWC}, indicate a reasonable agreement with some remaining, intriguing,
differences.

\end{paracol}
\nointerlineskip
\begin{figure}[H]
\widefigure
\centering
\includegraphics[width=.95\textwidth]{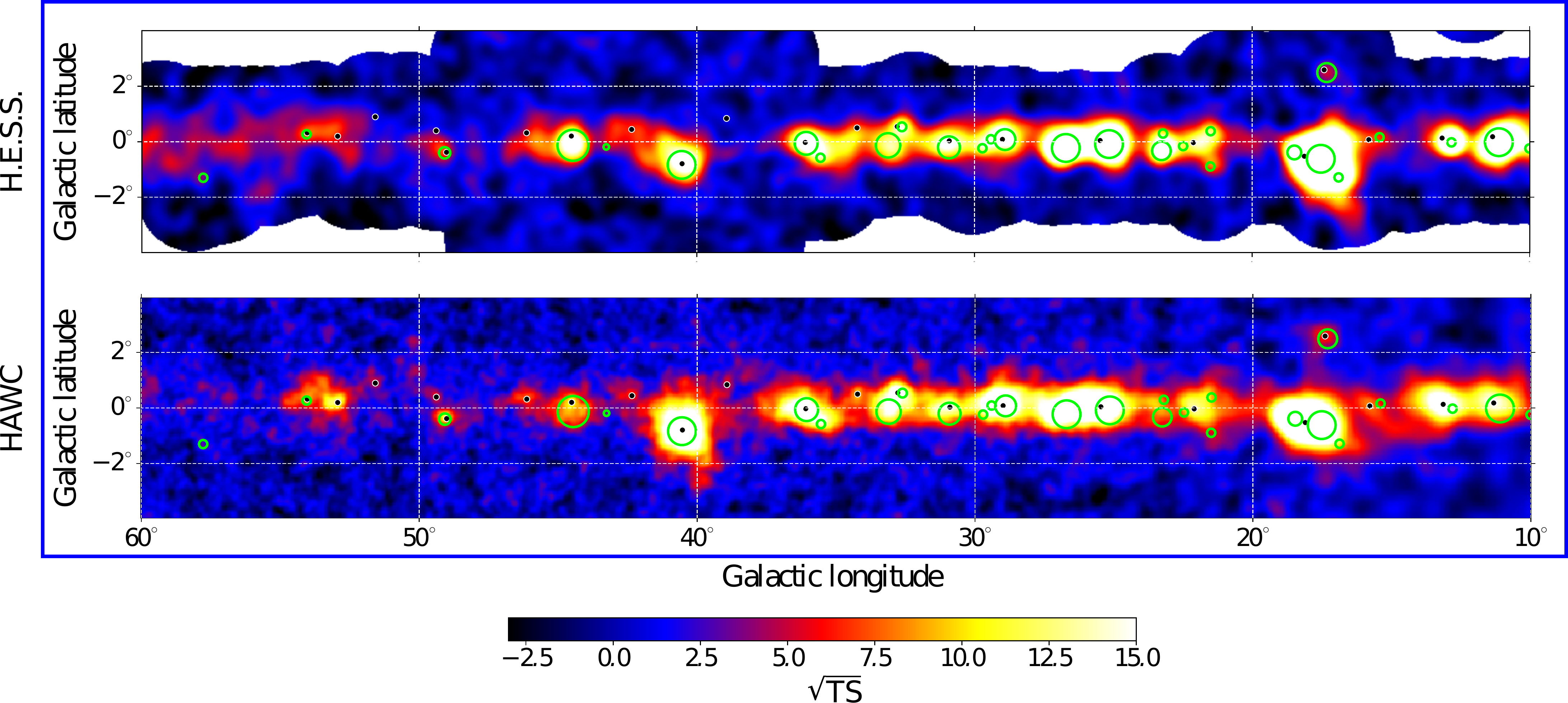} 
\caption{\label{fig:HESSvsHAWC}The galactic plane as see by \hess\ and HAWC with the same angular resolution. Adapted
from \cite{2021ApJ...917....6A}  with permission \copyright\ The American Astronomical Society.} 
\end{figure}
\vspace{-9pt}
\begin{paracol}{2}
\switchcolumn


\subsection{\label{sec:MetaCatalogues}Meta-Catalogues and Population of VHE Sources}


Meta-catalogues are online catalogues collecting the results of several instruments in a unique database. 
IACTs sometimes publish such catalogues~(e.g., \citep{2021arXiv210806424P}) summarising many years of observations.
In the field of VHE astronomy, TeVCat~\citep{2008TeVCat} is the standard tool {(\url{http://tevcat2.uchicago.edu/})}. 
Although the collected data correspond to different thresholds and uneven exposures, these catalogues are useful to perform statistical 
studies, but do not constitute unbiased and/or complete samples, and
need to be filled manually (for the time being). TeVCat only reports positive detections and not upper limits,
which could be useful to study source variability and, for instance, examine a transition from an emitting state to a non-emitting state
or vice versa.

The populations of VHE sources, as for September 2021, were extracted from TeVCat, and are displayed in Figure~\ref{fig:TeVCapPop}
for galactic sources ({\bf left}) and extragalactic sources ({\bf right}). Whereas PWNs comprise the largest population by
number in the galactic plane, followed by SNRs and binary systems, it should be noted that the majority of the sources
remain unidentified. Most galactic sources are (very) extended, and thus several plausible counterparts exist.
In contrast, the extragalactic sky is currently largely dominated by well-identified BL~Lacs (plus some other AGNs), which
might well result from a selection bias, since no systematic survey of the extragalactic sky has been conducted so far. The
identification of extragalactic sources is, except for in very rare cases, not problematic, due to their (mostly) 
point-like nature and to the lower density of possible counterparts.

\clearpage
\end{paracol}
\nointerlineskip
\begin{figure}[H]
\centering
\includegraphics[width=0.48\textwidth]{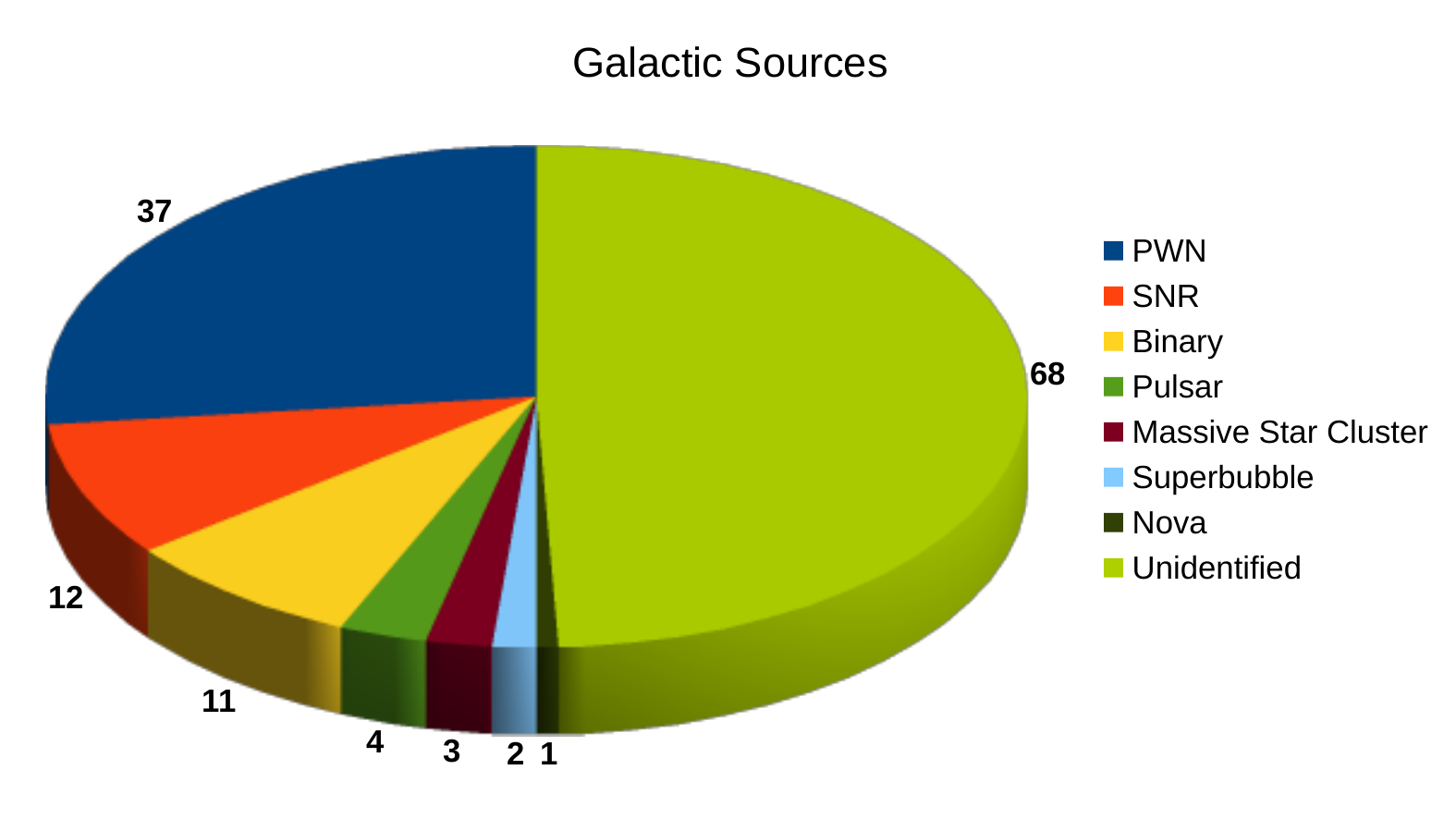} 
\includegraphics[width=0.48 \textwidth]{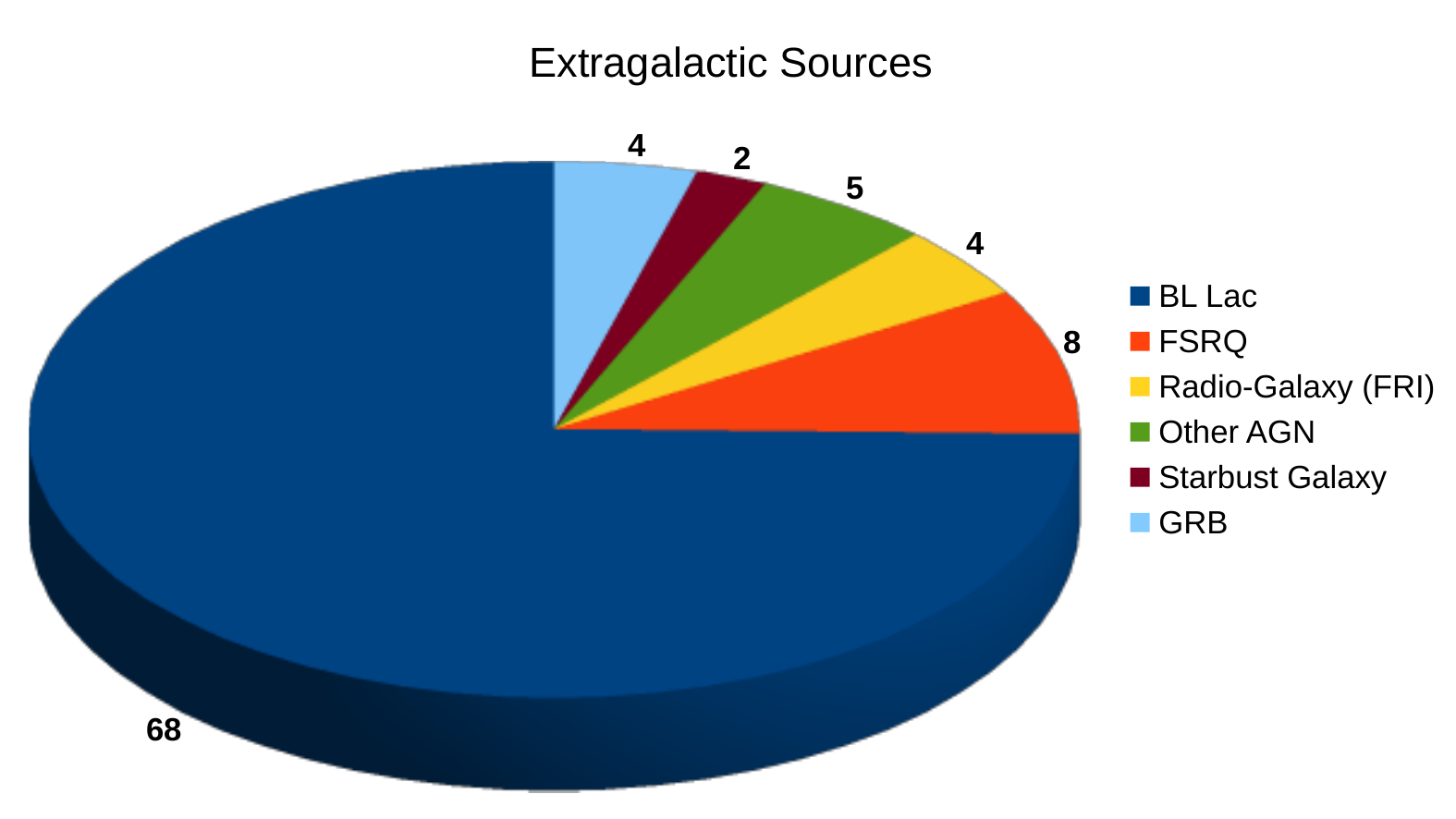} 
\caption{\label{fig:TeVCapPop}Population of established VHE sources extracted from the TeVCat~\cite{2008TeVCat} 
meta-catalogue.
({\bf Left}) Galactic sources. 
({\bf Right}) Extragalactic sources.}
\end{figure}  
\vspace{-9pt}
\begin{paracol}{2}
\switchcolumn
\subsection{\label{sec:SourcePopulations}Population of VHE Sources}

Unbiased surveys are essential tools for the analysis of source populations, which can identify global trends and possible
evolution schemes within one source class. For the first time ever, VHE \gray\ astronomy is now opening this possibility
with large scale surveys. The moderate depth and relative incompleteness of the existing surveys makes these first
studies not completely conclusive though and subject to future improvements. Two main population studies were already
performed based on the \hess\ HGPS data.

\subsubsection{Population of Pulsar Wind Nebula}

The \hess\ HGPS data have been used in a systematic population study of pulsar wind nebul\ae~\cite{2018HESS-PWNPOP}.
In addition to the 14 HPGS sources firmly identified as PWNs, 10 additional sources are found likely to be PWNs. 
Actually, most young and energetic pulsars are found to be associated with a plausible PWN candidate (Figure~\ref{fig:PWDPop}, left).
The data showed, for the first time, a correlation of the TeV surface brightness with pulsar spin-down power $\dot E$, which
can be quite well explained by a rather simple evolutionary model of PWNs, indicated by blue bands in the various plots: 
assuming a simple dipole-like radiation, the pulsar spin-down power decreases with increasing age (as measured from its
characteristic age $\tau_c = P / 2 \dot P$). The dynamical evolution of PWNs is then modelled in three distinct
phases, first the free expansion phase which lasts for a few kyr, followed by the reverse shock interaction (until some
tens of kyr), and finally the relic stage. A one-zone, time-dependant injection model is then used for the
population of electrons, from which the TeV luminosity is computed using standard radiative models.
The results of this model are reproduced in Figure~\ref{fig:PWDPop}. The extension increases quickly
in the free expansion phase (middle, $R \sim t^{1.2}$) and then slows down at later stages ($R\sim t^{0.3}$). The TeV
luminosity vs characteristic age (right) shows a rather large data scatter, still compatible with the varied model band
(blue bands in the plot). This scatter might reflect the intrinsic variability of the PWNs and their environments.

This study is a first attempt to model, in a rather comprehensive way, the TeV emission of PWNs. It suffers, however, from
several selection biases, due to the incompleteness of the survey and the difficulty in detecting very extended nebul\ae.
Going beyond this result requires the use of a population synthesis model to address these biases in a proper way. Future,
deeper surveys will also aim to improve the precision of the modelling.
\clearpage
\end{paracol}
\nointerlineskip
\begin{figure}[H]
\centering
\includegraphics[width=0.32 \textwidth]{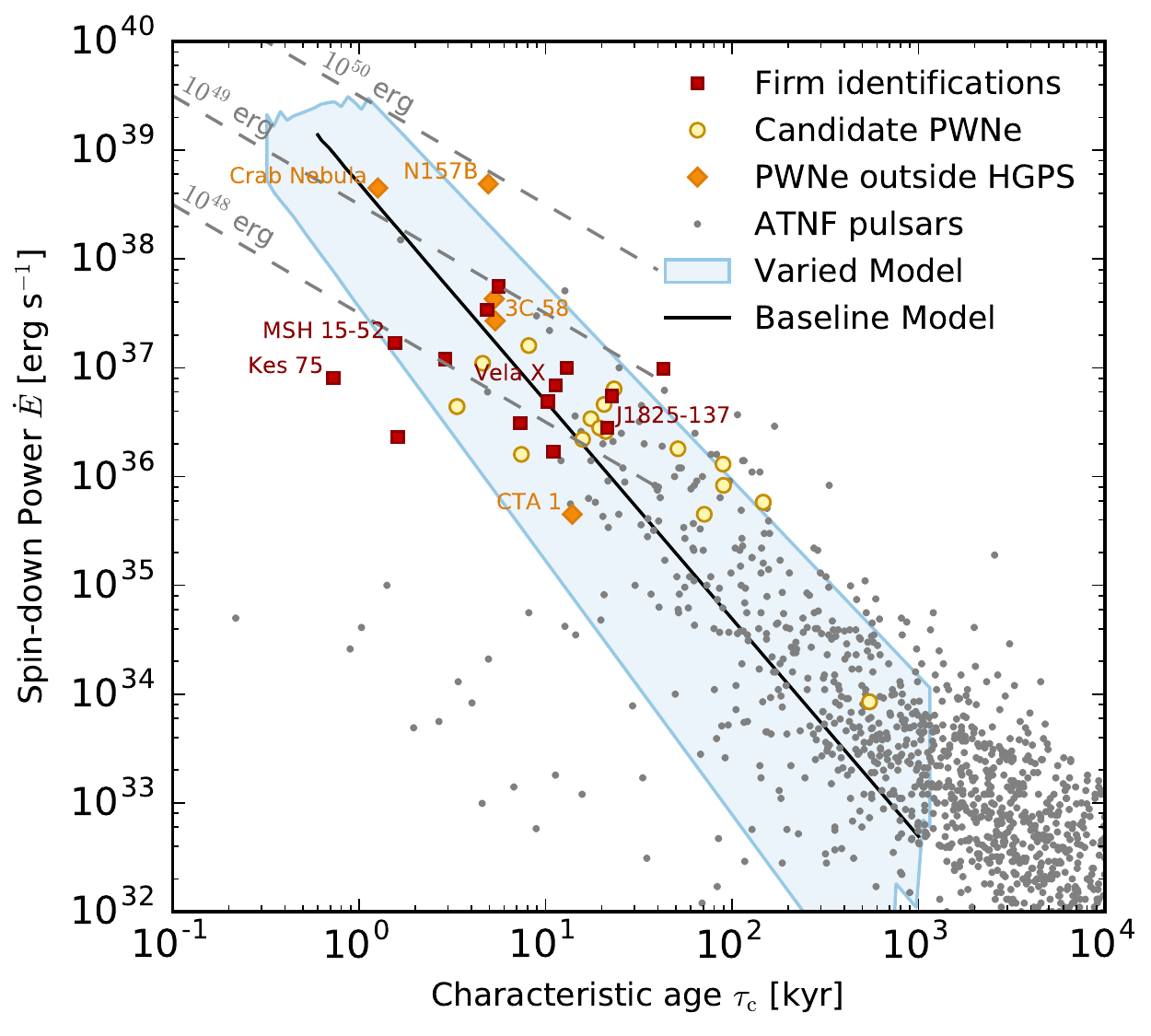}
\includegraphics[width=0.32 \textwidth]{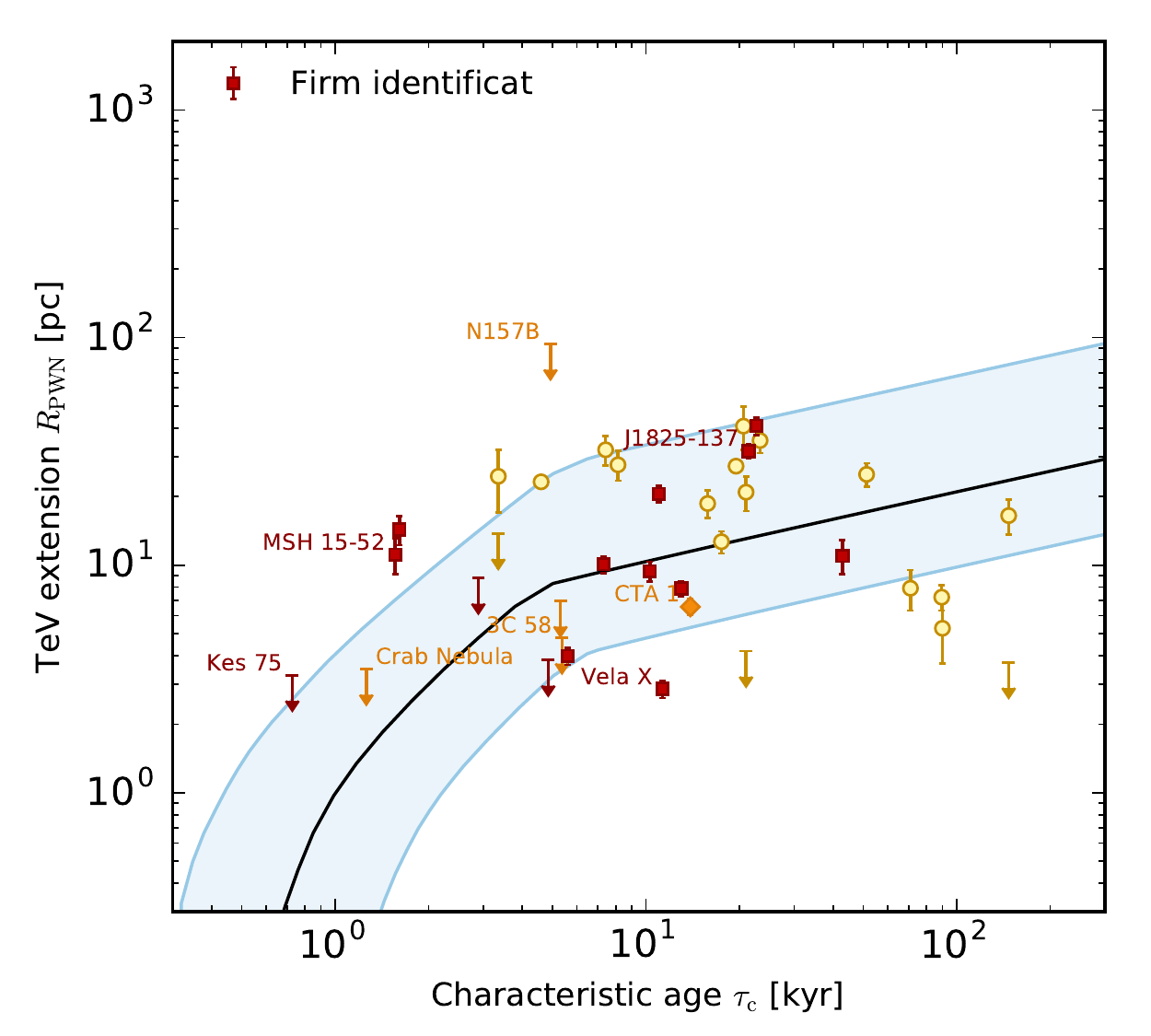} 
\includegraphics[width=0.32 \textwidth]{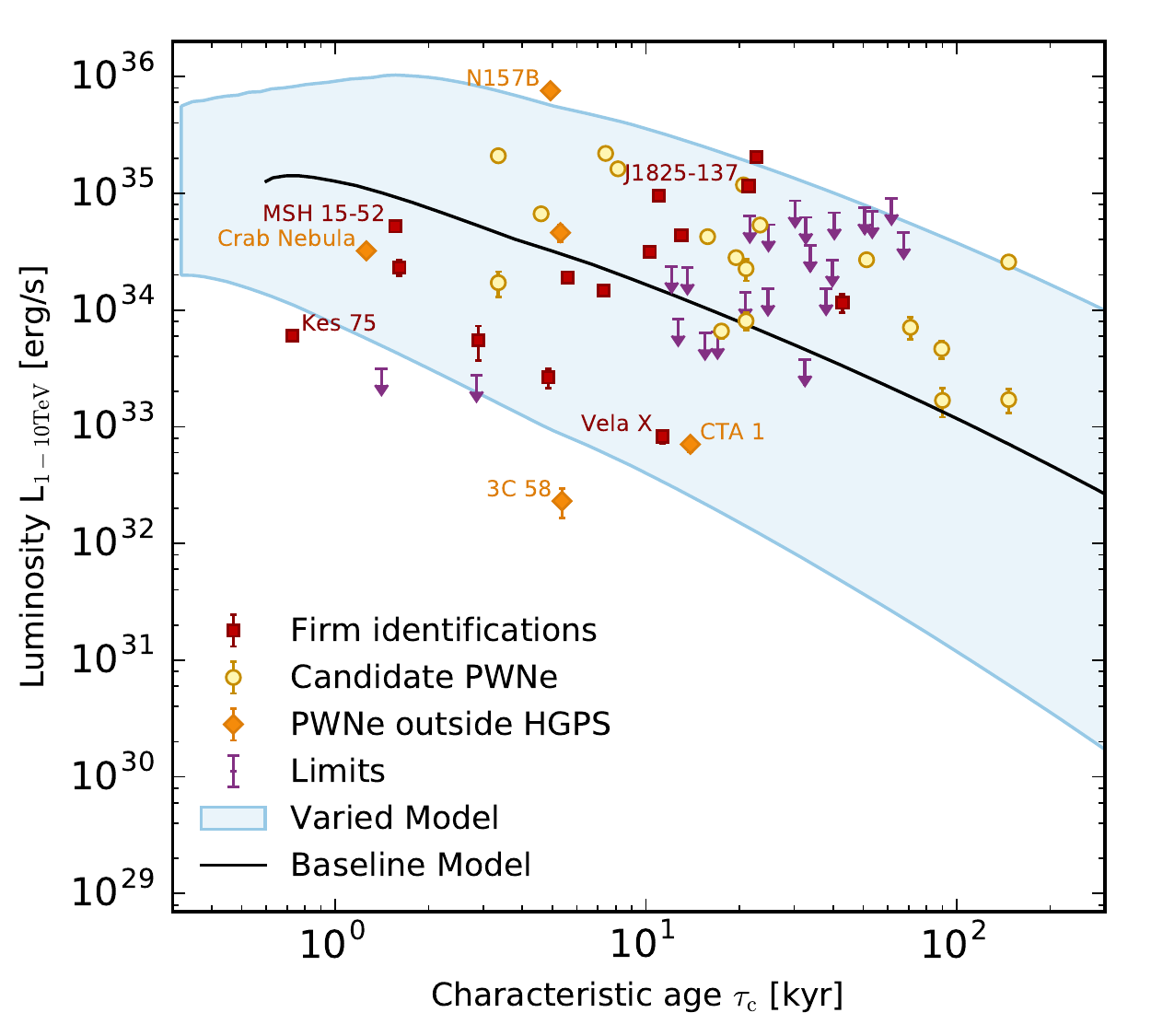}
\caption{\label{fig:PWDPop}Results of PWN population study. 
({\bf Left}) Population of pulsars in the spin-down power $\dot E$ vs. characteristic age plane. Young and energetic
pulsars, clustering at the top-left, are associated with plausible PWN candidates. 
({\bf Middle}) PWN extension evolution with time.
({\bf Right}) Evolution of the PWN TeV luminosity with characteristic age.
Reproduced from \cite{2018HESS-PWNPOP} with permission \copyright\ ESO.} 
\end{figure} 
\vspace{-9pt}
\begin{paracol}{2}
\switchcolumn

\subsubsection{Supernova Remnant Populations}

The second HGPS population study is related to the second galactic population by frequency, namely the supernova
remnants~\cite{2018HESS-SNRPOP}. In this study, upper limits are computed for all SNRs that fall in the HGPS region
and which are not detected at VHE (i.e., not overlapping with a significant excess). A sample of 108 SNRs is constructed
this way, biased towards low flux, since the detected SNRs are excluded from the sample. Using the canonical cosmic ray paradigm,
constraints on the typical ambient density values around SNR shells ($n \leq 7 \mathrm{cm}^{-3}$) and on 
the electron-to-proton energy fraction ($\epsilon_{ep} \leq 5 \times 10^{-3}$) are derived. A shift of $1.01$ (mean)
is observed in the significance distribution, which might be due to the cumulative effect of sub-threshold SNR shells
and the galactic diffuse component. 
Using the SNR shells that are detected in the VHE band, some constraints on the luminosity evolution of SNRs in the
radio and VHE bands are also derived.  The $(L_\mathrm{VHE} / L_\mathrm{radio})$ luminosity ratio exhibits a clear
correlation with source age, which is interpreted as being due to the fact that, in the first several thousand years, 
the radio-synchrotron emission of the SNRs decreases quickly, while the VHE emission decreases slowly. 

Here again, the understanding of SNR evolution will greatly benefit from future, deeper surveys.
\section{Perspectives and Outlook}

After the tremendous success of the third-generation IACTs during the last two decades, driven essentially by \hess, VERITAS and
MAGIC, a new step towards an international facility is currently being taken, merging the efforts of the different
collaborations into a single, world-wide project, named ``The Cherenkov Telescope Array'' (CTA).
Lessons learnt from the various concepts tested in the third generation instruments are being used to design a new array, 
focusing on (i) performance, (ii) reliability and (iii) flexibility, with some major challenges ahead.
Recent developments in survey instruments established the use of particle array survey instruments as a viable and complementary technique to IACTs, 
particularly suited to large-scale surveys and to all-sky monitoring. New technical developments and new upcoming projects 
are expected to further boost the performances and, on a longer timescale, provide a nearly full sky coverage.

\subsection{\label{sec:CTA}CTA, the Next Generation IACT}

The Cherenkov telescope array (CTA) is the next generation array of imaging atmospheric telescopes, currently 
in the prototyping stage. It aims to transform our understanding of the VHE universe. It will consist of two arrays; one
in the Canary Islands, and one in Chile~\citep{2013CTAConcept}, with different telescope layouts for a total of $\sim\,100$
telescopes. 
In order to increase the dynamical range in energy, telescopes of three different sizes will be combined in the same
array: large-sized telescopes (LSTs) with a field of view of $>$$4.5\deg$, and a dense layout will focus on the lowest
energies, medium-sized telescopes (MSTs), with a larger field of view of $>$$7\deg$ on a sparser layout will provide the
sensitivity in the core of the energy domain, and small-sized telescope (SST) with an even larger field of view
($>$$8\deg$) spread over a very large area will explore the highest energies.
These arrays of telescopes of different size have been designed to provide an improvement by a factor of $\sim 10$ in
sensitivity compared to the previous generation, with a substantial improvement in angular and energy resolution,  but at
the cost of a much higher ($\sim$$\times 10$) event rate~\citep{2013MC}, and a huge data volume ($\sim \mathrm{PB}/\mathrm{year}$).

Amongst the key science projects that have been identified for the first years of operation, large surveys
play a particular role in providing unbiased samples of particle accelerators, but also to search for the unexpected.
In the design of the telescopes and the array, a strong focus has been made on the survey capabilities, in particular
through the conception of large field-of-view cameras and the first investigation of an alternate pointing scheme, such as
divergent pointing (Section~\ref{sec:Observations}). 
Three major surveys are currently foreseen~\cite{2013CTA-HGPS}:

\begin{itemize}
  \item An extragalactic survey, covering 1/4 of the sky with a sensitivity of $\sim 0.6\%$ Crab in ~1000 h of
  observations. This will provide an unbiased sample of active galactic nuclei and other possible extragalactic sources,
  and a snapshot of their activity (since AGNs are intrinsically variables at almost all timescales)
  \item A deep galactic plane survey, reaching $\sim$$0.2\%$ Crab sensitivity in the inner regions (and Cygnus region)
  and $\sim$$0.4\%$ in the entire plane region. This will provide a horizon of $\sim$$20 ~\mathrm{kpc}$ (point-like), thus
  covering a large fraction of the Galactic sources. 
  \item A deep survey of the LMC region, aiming at an excellent angular resolution to resolve structures down to $\sim$$20~\mathrm{pc}$, in order to be able to resolve individual objects and map the diffuse emission.
\end{itemize}

The characteristics of the three surveys differ in terms of physics goals. Most likely, the configuration of the array 
will have to be optimised accordingly.
With the density of sources in the extragalactic sky being fairly low, and the sources being (almost) point-like, the
angular resolution and background systematics requirements are not subjects of major concern (except perhaps at
the lowest energy end). 
To quickly cover a large fraction of the sky, and to increase the chances of catching transient
events, one might want to increase the effective field of view by using, for instance, the divergent or skewed pointing mode
(Section~\ref{sec:Observations}). 
In contrast, due to the absorption of VHE \grays\ by pair creation on the extragalactic background light (EBL), one might 
want to achieve the lowest possible energy threshold, which is best obtained in the convergent pointing mode (at high altitude),
because it maximises the collection of light. The use of LSTs is, therefore, being considered to lower the energy
threshold, but they have a smaller field of view than MSTs and SSTs, resulting in (i) a longer time being required to cover
the survey region (ii) a possibly complicated acceptance shape when used in conjunction with the other telescopes. 
Further optimisation (e.g., grid spacing on the sky, run duration, \ldots) is still ongoing.


For the galactic (Figure~\ref{fig:CTAHGPS}) (and the LMC) surveys, the angular resolution is of prime importance to
mitigate the source confusion problem. The angular resolution is optimal in convergent pointing mode (with the maximum telescope multiplicity), and
improves in the core of the energy range ($\sim \mathrm{TeV}$), thus calling for the use of MSTs and SSTs mainly.
Moreover, during recent years, it has been observed that, for many galactic sources (and the PWNs in particular), the extension 
decreases at high energy, thus pushing in favour of the best possible angular resolution. Two points remain large
points of concern for the galactic survey:

\begin{itemize}
  \item The background systematics will most likely be the limiting factor for the sensitivity achievable, most notably for
  the (very) extended sources. Given the foreseen increase of the background rate by $\sim$$\times 10$, the
  state-of-the-art uncertainties in background estimation of 1--2\% will need to be substantially improved by refining the acceptance models.
  Changes in the array layout (telescopes under maintenance, \ldots), inhomogeneities of camera response and/or
  atmospheric effects (Section~\ref{sec:Acceptance}) should be studied carefully and, whenever possible, incorporated in
  the model. In this regard, the simulated acceptance being currently developed might be a promising approach.
  The mitigation techniques recently developed (Section~\ref{sec:Systematics}) can certainly help, but they tend to reduce the
  sensitivity of the array. Further work is clearly needed to take into account the various sources of systematics in
  the calculation of the acceptance.
  \item With the detection of $\sim$$\times 10$ sources in the same field of view, up to several hundreds of sources,
  source confusion and overlap are expected to become a major issue, especially in the context of the unknown shapes of
  the sources and the unknown level of large scale diffuse emission.
  Some preliminary estimates performed with an extrapolation of the current $\log N$--$\log S$ source distribution
  indicate a source confusion lower limit on the order of $\sim$$20\%$ in the core CTA energy range~\citep{2013CTA-HGPS}.
  Template fitting and 3D modelling of the sources (Section~\ref{sec:TemplateFitting}) can help with the separation of  
  superimposed sources with different spectral characteristics. 
\end{itemize}

\end{paracol}
\nointerlineskip
\begin{figure}[H]
\centering
\includegraphics[width=\textwidth]{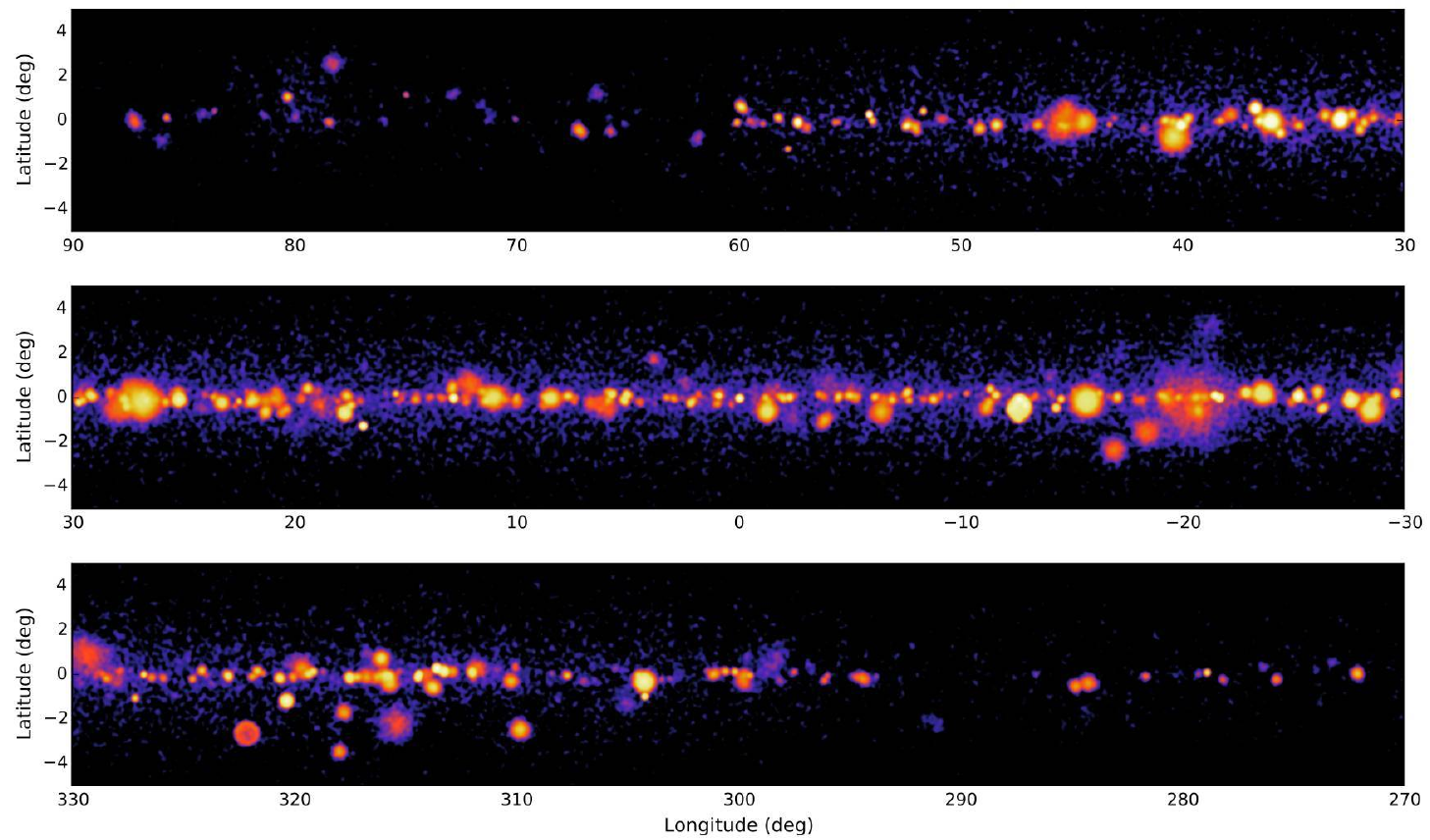}
\caption{\label{fig:CTAHGPS}{Simulated} results from the CTA Galactic Plane Survey in very high-energy $\gamma$ rays for half
of the plane. Reproduced from~\citep{CTAKSP6}.} 
\end{figure}
\vspace{-9pt}
\begin{paracol}{2}
\switchcolumn

\subsection{Next Generation Particle Array Survey Instruments}

Survey instruments are also preparing new upgrades to boost their sensitivity. In 2018, HAWC completed~\cite{2019HAWCOutrigger},
a major upgrade consisting of the addition of a sparse array of 345 small water Cherenkov tanks spread over a large
area. By improving the rejection of showers that are not well contained in the main array, this upgrade allowed 
the core resolution to be improved by a factor of $\sim$$3$ above $1\,\mathrm{TeV}$, and the effectives are to be increased substantially,
particularly at the highest energies~\citep{2019ICRC...36..707J}. Further optimisation of the analysis to include these
data is under way.

The Southern Wide-field Gamma-ray Observatory (SWGO)~\citep{2019SWGO} aims to become the next generation, large scale
survey instrument in the southern hemisphere, covering an energy range from $100\,\mathrm{GeV}$ to hundreds of TeV.
It is similar in concept to HAWC, but $\sim$$\times 4$ larger (for the inner array), and would include a sparser, outer
array of $\sim$$1~\mathrm{km}^2$ to expand the energy range towards the highest-energy frontier.
Planned for installation in South America, it will cover the central regions of the Galaxy with an unprecedented sensitivity,
and will complement the CTA view. SWGO is not yet funded for construction.

LHAASO, currently being deployed at high altitude ($4410~\mathrm{m}$ above see level) in the Sichuan Province, China, is a
novel concept combining three interconnected detectors: an array of underground water Cherenkov detectors, a kilometre
square array made of plastic scintillator and an array of wide field-of-view Cherenkov telescopes.
Early data from LHAASO demonstrate the presence of at least twelve source of petaelectronvolt $\gamma$\ rays in the
Galaxy~\cite{2021LHAASO}, thus boosting the interest for the extremely high energy frontier. It should be noted that LHAASO is a
multi-messenger observatory with unprecedented capabilities in the field of cosmic-ray physics. Its deployment should be
completed by the end of 2021.

Particle array survey instruments are currently becoming invaluable companions to IACTs. They can provide an unbiased view on the
\gray\ emission from the Galactic plane (Section~\ref{sec:SurveysFromSurvey}), whereas IACTs can perform deeper observations,
revealing the details of the cosmic-ray acceleration and \gray\ emission mechanisms. Through their all-sky monitoring capabilities,
Particle array survey instruments can also monitor the long-term activity of variable sources, and alert the community to particular eruptive events that IACTs can
sample with much greater precision. The synergy between targeted, IACT observations and long-term, particle array monitoring
instruments has recently been put under the spotlight with the detailed and anticipated \hess-HAWC
comparison~\cite{2021ApJ...917....6A}. These efforts should gain additional visibility in the coming years.




\section{Conclusions}

Over the course of the last $\sim$$20$ years, the field of VHE astronomy has experienced an incredible and somewhat unexpected  blooming
caused first by (i) the developments in high-speed acquisition techniques, (ii) the advent of third generation
instruments building on the success of the previous instruments, and (iii) the increased capabilities in image
classification and pattern recognition. This evolves into an exponential increase in the number of VHE
\gray\ sources detected with time.
The so-called ``Kifune-plot'' (\mbox{Figure~\ref{fig:Kifune}}), named after T.~Kifune, who first showed a first version of this figure at 
the 1995 ICRC conference in Rome, indicating that the number of X-ray, HE and VHE sources detected has not yet saturated,
and the CTA simulations predict a continuation of this trend.

Moving away from the analysis of single, well-targeted sources, 
scientists have developed new algorithms to map the \gray\ emission of large regions of the sky in varying observational conditions. 
The analysis of very large, heterogeneous data sets comprising observations spread over several years on very diverse
positions has been implemented, leading to major developments in acceptance determination (Section~\ref{sec:Acceptance}) and 
background subtraction techniques (Section~\ref{sec:BackgroundSubtraction}), and has led the way towards the first VHE
source catalogues.
Recent large-scale surveys of unexplored regions were the main driver for the discovery of new sources, 
and made the first population studies finally possible (\mbox{Section~\ref{sec:SourcePopulations}}).

At the same time, many
developments in pattern recognition and image analysis led to elaborate reconstruction and separation techniques, 
which are now rather close to the fundamental limits, thus with only moderate improvements being possible in the future.
Tremendous efforts were made to improve the shower and detector simulation codes, by including subtle instrumental and
atmospheric effects. State-of-the-art, realistic simulations are now able to reproduce the background with sufficient
resolution to open a new paradigm, replacing the classical background subtraction techniques with a modern template-fitting
approach, including a fully simulated background model (Section~\ref{sec:RWSAcceptance}).
 
Particle array survey instruments (Section~\ref{sec:SurveysFromSurvey}) recently demonstrated their maturity and their strong
synergies with IACTs, delivering a complementary and unbiased view on the VHE sky.
The exponential rise in the number of sources shown in Figure~\ref{fig:Kifune} indicates that the number of sources is
currently not limited by their scarcity, but by the sensitivity of the instruments. The next generation of instruments,
and in particular the Cherenkov telescope array, will most likely have to deal with hundreds, if not thousands
of sources.
Major projects, such as deep surveys of the Galactic plane, but also the first survey of a significant fraction
of the extragalactic sky with unprecedented sensitivity (Section~\ref{sec:CTA}), will deliver large and unbiased catalogues
of VHE sources, enabling the statistical analysis of source populations and the clarification of the underlying
evolution models. They will, however, face fundamental challenges caused by the huge amount of acquired data. The
background will need to be understood and modelled with a sub-percent precision to avoid uncontrolled background
systematics, which would limit the sensitivity of the instruments. The proper background estimation will require 
very detailed monitoring of the instrumental
and atmospheric conditions, and use  extensive simulations of the instrument response to varying conditions and
the incorporation of these effects in the acceptance determination algorithms.
Source confusion will most likely become a major issue in regions of the sky with large source density, such as
significant parts of the Galactic plane. Improved angular resolution will be of little help due to the large  size of
most VHE sources. Modern analysis approaches, including template fitting and 3D modelling of the VHE sources, provide
promising paths currently being explored.
 
\end{paracol}
\nointerlineskip
\begin{figure}[H] 
 \widefigure
\includegraphics[width=\textwidth]{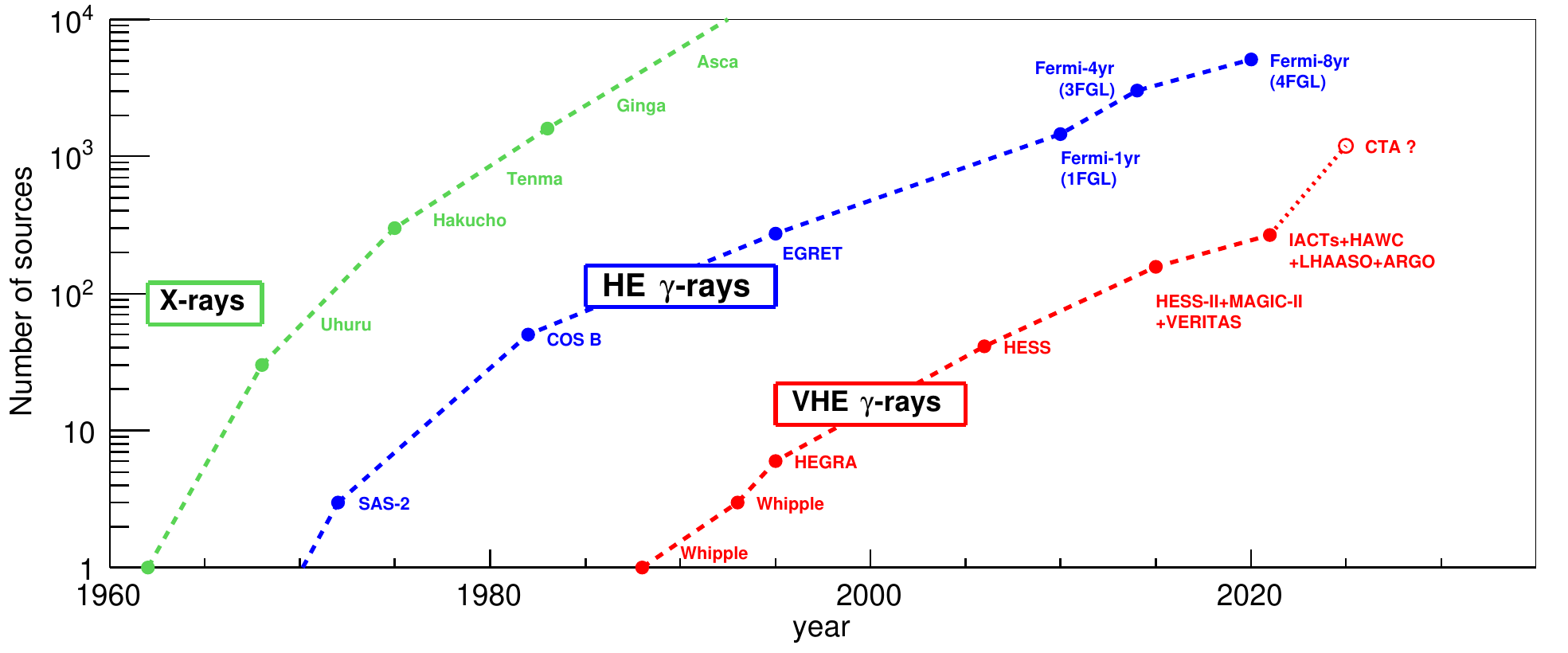} 
\caption{\label{fig:Kifune}Number of established sources as function of time in different energy domains, also dubbed the
``{Kifune Plot}'', in honour of Prof. Tadashi Kifune, who produced the first version of this figure.}
\end{figure}
\begin{paracol}{2}
\switchcolumn

The status of VHE \gray\ astronomy is, in fact, similar to that of X-ray or high-energy $\gamma$ rays:
every time a new astronomical window is opened and a sensitivity threshold is achieved, one can observe 
an exponential rise in the number of sources. From that,
there is little doubt that the field of VHE astronomy can look forward to a very bright future.

\vspace{6pt} 




\funding{This research received no external funding.}

\institutionalreview{Contained in the main text.}

\informedconsent{Contained in the main text.}

\dataavailability{Contained in the main text.} 
\acknowledgments{
We thank S.~Wagner, spokesman of the H.E.S.S. Collaboration and  O.~Reimer,
chairman of the Collaboration board, for allowing us to use H.E.S.S. data in this publication. 
We
are grateful to D.~Horan for carefully reading the manuscript and for providing us with
very useful suggestions. 
}

\conflictsofinterest{The authors declare no conflict of interest.} 


\abbreviations{The following abbreviations are used in this manuscript:\\

\noindent 
\begin{tabular}{@{}ll}
AGN & active galactic nucleus \\
CTA & Cherenkov telescope array \\
EBL & extragalactic background light \\
FoV & field of view \\
HGPS & \hess\ Galactic Plane Survey \\
HE & high energy \\
IACT & imaging atmospheric Cherenkov telescope \\
LMC & large magellanic cloud \\
LST & large-sized telescope \\
MST & medium-sized telescope \\
NSB & night sky background \\
Pdf & probability density function \\
PWN & pulsar wind nebula \\
RA  & right ascension \\
ROI & region of interest \\
RPC & resistive plate chamber \\
SNR & supernova remnant \\
SST & small-sized telescope \\
VHE & very high energy
\end{tabular}}

\end{paracol}
\reftitle{References}



\begin{thebibliography}{999}

\bibitem[{Hess}(1912)]{Hess262750}
{Hess}, V.F.
\newblock {\"Uber Beobachtungen der durchdringenden Strahlung bei sieben
  Freiballonfahrten}.
\newblock {\em Z. Phys.} {\bf 1912}, {\em 13},~1084.

\bibitem[{Cherenkov}(1934)]{Cherenkov1934a}
{Cherenkov}, P.A.
\newblock {Visible emission of clean liquids by action of $\gamma$ radiation}.
\newblock {\em Dokl. Akad. Nauk SSSR} {\bf 1934}, {\em 2},~451.

\bibitem[{Blackett}(1948)]{1948esns.conf...34B}
{Blackett}, P.M.S.
\newblock {A possible contribution to the night sky from the Cerenkov radiation
  emitted by cosmic rays}.
\newblock  In \emph{The Emission Spectra of the Night Sky and Aurorae};  The Physical Society: London, UK, 1948; p.~34. 

\bibitem[{Galbraith} and {Jelley}(1953)]{1953Natur.171..349G}
{Galbraith}, W.; {Jelley}, J.V.
\newblock {Light Pulses from the Night Sky associated with Cosmic Rays}.
\newblock {\em Nature} {\bf 1953}, {\em 171},~349--350, doi:10.1038/171349a0.

\bibitem[{Weekes} and et. al.(1989)]{1989ApJ...342..379W}
Weekes, T.C.;  et al.  
\newblock {Observation of TeV Gamma Rays from the Crab Nebula Using the
  Atmospheric Cerenkov Imaging Technique}.
\newblock {\em \apj} {\bf 1989}, {\em 342},~379, doi:10.1086/167599.

\bibitem[Guy(2003)]{TheseJulienGuy}
Guy, J.
\newblock Premiers Résultats de L'expérience HESS et étude du Potentiel de
  Détection de Matière Noire Supersymétrique.
\newblock Ph.D. Thesis, Université Pierre et Marie Curie-Paris VI,  Paris, France, 2003.

\bibitem[{de Naurois} and {Rolland}(2009)]{denaurois2009}
{de Naurois}, M.; {Rolland}, L.
\newblock {A high performance likelihood reconstruction of {$\gamma$}-rays for
  imaging atmospheric Cherenkov telescopes}.
\newblock {\em \aspp} {\bf 2009}, {\em 32},~231--252, doi:10.1016/j.astropartphys.2009.09.001.

\bibitem[{Szanecki} and et. al.(2015)]{2015APh....67...33S}
{Szanecki}, M.; et al.
\newblock {Monte Carlo simulations of alternative sky observation modes with
  the Cherenkov Telescope Array}.
\newblock {\em Astropart. Phys.} {\bf 2015}, {\em 67},~33--46,
  doi:10.1016/j.astropartphys.2015.01.008.

\bibitem[D'Amico \em{et~al.}(2021)D'Amico, Terzi\ifmmode~\acute{c}\else
  \'{c}\fi{}, Stri\ifmmode \check{s}\else \v{s}\fi{}kovi\ifmmode~\acute{c}\else
  \'{c}\fi{}, Doro, Strzys, and van Scherpenberg]{PhysRevD.103.123001}
D'Amico, G.; Terzi\ifmmode~\acute{c}\else \'{c}\fi{}, T.; Stri\ifmmode
  \check{s}\else \v{s}\fi{}kovi\ifmmode~\acute{c}\else \'{c}\fi{}, J.; Doro,
  M.; Strzys, M.; van Scherpenberg, J.
\newblock Signal estimation in on/off measurements including event-by-event
  variables.
\newblock {\em Phys. Rev. D} {\bf 2021}, {\em 103},~123001, doi:10.1103/PhysRevD.103.123001.

\bibitem[{Mohrmann} and et. al.(2019)]{2019A&A...632A..72M}
{Mohrmann}, L.; et al.
\newblock {Validation of open-source science tools and background model
  construction in {\ensuremath{\gamma}}-ray astronomy}.
\newblock {\em \aap} {\bf 2019}, {\em 632},~A72,
  doi:10.1051/0004-6361/201936452.

\bibitem[{de Naurois}(2012)]{HDRMathieu}
{de Naurois}, M.
\newblock Very High Energy astronomy from H.E.S.S. to CTA. Opening of a new
  astronomical window on the non-thermal Universe.
\newblock Habilitation Thesis, Université Pierre et Marie Curie-Paris VI, Paris, France, 
  2012. 

\bibitem[{Klepser}(2012)]{2012Klepser}
{Klepser}, S.
\newblock {A generalized likelihood ratio test statistic for Cherenkov
  telescope data}.
\newblock {\em Astropart. Phys.} {\bf 2012}, {\em 36},~64--76,
  doi:10.1016/j.astropartphys.2012.04.008.

\bibitem[{Holler} and et. al.(2020)]{2020RWS}
{Holler}, M.; et al.
\newblock {A run-wise simulation and analysis framework for Imaging Atmospheric
  Cherenkov Telescope arrays}.
\newblock {\em Astropart. Phys.} {\bf 2020}, {\em 123},~102491,
  doi:10.1016/j.astropartphys.2020.102491.

\bibitem[{Holler} and et. al.()]{HollerRWS}
{Holler}, M.; et al.
\newblock {Run-Wise Simulations for Analyses with Open-Source Tools in IACT
  Gamma-Ray Astronomy}.
\newblock {In Preparation}.

\bibitem[{Berge} \em{et~al.}(2007){Berge}, {Funk}, and
  {Hinton}]{2007HESS-Backgrounds}
{Berge}, D.; {Funk}, S.; {Hinton}, J.
\newblock {Background modelling in very-high-energy {$\gamma$}-ray astronomy}.
\newblock {\em \aap} {\bf 2007}, {\em 466}, 1219--1229,
  doi:10.1051/0004-6361:20066674.

\bibitem[{Li} and {Ma}(1983)]{liandma}
{Li}, T.P.; {Ma}, Y.Q.
\newblock {Analysis methods for results in gamma-ray astronomy}.
\newblock {\em \apj} {\bf 1983}, {\em 272},~317--324, doi:10.1086/161295.

\bibitem[{Cash}(1979)]{1979ApJ...228..939C}
{Cash}, W.
\newblock {Parameter estimation in astronomy through application of the
  likelihood ratio.}
\newblock {\em \apj} {\bf 1979}, {\em 228},~939--947, doi:10.1086/156922.

\bibitem[{H.E.S.S. Collaboration} \em{et~al.}(2018){H.E.S.S. Collaboration},
  {Abdalla}, and et. al.]{2018HESS-HGPS3}
{H.E.S.S. Collaboration}; {Abdalla}, H.; et al.
\newblock {The H.E.S.S. Galactic plane survey}.
\newblock {\em \aap} {\bf 2018}, {\em 612},~A1,
  doi:10.1051/0004-6361/201732098.

\bibitem[{H.E.S.S. Collaboration} \em{et~al.}(2012){H.E.S.S. Collaboration},
  {Abramowski}, and et. al.]{2012A&A...548A..38A}
{H.E.S.S. Collaboration}; {Abramowski}, A.; et al.
\newblock {Probing the extent of the non-thermal emission from the Vela X
  region at TeV energies with H.E.S.S.}
\newblock {\em \aap} {\bf 2012}, {\em 548},~A38,
  doi:10.1051/0004-6361/201219919.

\bibitem[{Rowell}(2003)]{2003Rowell-Template}
{Rowell}, G.P.
\newblock {A new template background estimate for source searching in TeV gamma
  -ray astronomy}.
\newblock {\em \aap} {\bf 2003}, {\em 410},~389--396,
  doi:10.1051/0004-6361:20031194.

\bibitem[{Fernandes} and et. al(2014)]{2014A&A...568A.117F}
{Fernandes}, M.V.; et. al.
\newblock {A new method of reconstructing VHE {\ensuremath{\gamma}}-ray
  spectra: The Template Background Spectrum}.
\newblock {\em \aap} {\bf 2014}, {\em 568},~A117,
  doi:10.1051/0004-6361/201323156.

\bibitem[{H.E.S.S. Collaboration} \em{et~al.}(2021){H.E.S.S. Collaboration},
  {HAWC Collaboration}, {Abdalla}, and et. al.]{2021ApJ...917....6A}
{H.E.S.S. Collaboration}; {HAWC Collaboration}; {Abdalla}, H.; et al.
\newblock {TeV Emission of Galactic Plane Sources with HAWC and H.E.S.S.}
\newblock {\em \apj} {\bf 2021}, {\em 917},~6,
  doi:10.3847/1538-4357/abf64b.

\bibitem[{Holler}(2001)]{HollerSystematics}
{Holler}, M.
\newblock {Assessing Background Systematics in an Analysis Field of View}.
\newblock {\em \hess\ Internal Note} {\bf 2001}, {\em 21}. 

\bibitem[{Spengler}(2015)]{2015Spengler}
{Spengler}, G.
\newblock {Significance in gamma ray astronomy with systematic errors}.
\newblock {\em Astropart. Phys.} {\bf 2015}, {\em 67},~70--74,  \linebreak 
  doi:10.1016/j.astropartphys.2015.02.002.

\bibitem[{Dickinson} and {Conrad}(2013)]{2013Dickinson}
{Dickinson}, H.; {Conrad}, J.
\newblock {Handling systematic uncertainties and combined source analyses for
  Atmospheric Cherenkov Telescopes}.
\newblock {\em Astropart. Phys.} {\bf 2013}, {\em 41},~17--30,
  \mbox{doi:10.1016/j.astropartphys.2012.10.004}.

\bibitem[{Strong} and {Moskalenko}(1998)]{1998ApJ...509..212S}
{Strong}, A.W.; {Moskalenko}, I.V.
\newblock {Propagation of Cosmic-Ray Nucleons in the Galaxy}.
\newblock {\em \apj} {\bf 1998}, {\em 509},~212--228,
  doi:10.1086/306470.

\bibitem[{Vovk} \em{et~al.}(2018){Vovk}, {Strzys}, and
  {Fruck}]{2018A&A...619A...7V}
{Vovk}, I.; {Strzys}, M.; {Fruck}, C.
\newblock {Spatial likelihood analysis for MAGIC telescope data. From
  instrument response modelling to spectral extraction}.
\newblock {\em \aap} {\bf 2018}, {\em 619},~A7,
  doi:10.1051/0004-6361/201833139.

\bibitem[{Kn{\"o}dlseder} and et. al.(2016)]{2016A&A...593A...1K}
{Kn{\"o}dlseder}, J.; et al.
\newblock {GammaLib and ctools. A software framework for the analysis of
  astronomical gamma-ray data}.
\newblock {\em \aap} {\bf 2016}, {\em 593},~A1,
  doi:10.1051/0004-6361/201628822.

\bibitem[{H.E.S.S. Collaboration} \em{et~al.}(2002){H.E.S.S. Collaboration},
  {Aharonian}, and et. al.]{2002A&A...395..803A}
{H.E.S.S. Collaboration}; {Aharonian}, F.A.; et al.
\newblock {A search for TeV gamma-ray emission from SNRs, pulsars and
  unidentified GeV sources in the Galactic plane in the longitude range between
  $-$2 deg and 85 deg.}
\newblock {\em \aap} {\bf 2002}, {\em 395},~803--811,
  doi:10.1051/0004-6361:20021347.

\bibitem[{H.E.S.S. Collaboration} \em{et~al.}(2005){H.E.S.S. Collaboration},
  {Aharonian}, and et. al.]{2005HESS-HGPS1}
{H.E.S.S. Collaboration}; {Aharonian}, F.; et al.
\newblock {A New Population of Very High Energy Gamma-Ray Sources in the Milky
  Way}.
\newblock {\em Science} {\bf 2005}, {\em 307},~1938--1942,
  doi:10.1126/science.1108643.

\bibitem[{H.E.S.S. Collaboration} \em{et~al.}(2006){H.E.S.S. Collaboration},
  {Aharonian}, and et. al.]{2006HESS-HGPS2}
{H.E.S.S. Collaboration}; {Aharonian}, F.; et al.
\newblock {The H.E.S.S. Survey of the Inner Galaxy in Very High Energy Gamma
  Rays}.
\newblock {\em \apj} {\bf 2006}, {\em 636},~777--797,
  doi:10.1086/498013.

\bibitem[{H.~E.~S.~S. Collaboration} \em{et~al.}(2014){H.~E.~S.~S.
  Collaboration}, {Abramowski}, and et. al.]{2014HESSDiffuse}
{H.~E.~S.~S. Collaboration}; {Abramowski}, A.; et al.
\newblock {Diffuse Galactic gamma-ray emission with H.E.S.S.}
\newblock {\em \prd} {\bf 2014}, {\em 90},~122007,
  doi:10.1103/PhysRevD.90.122007.

\bibitem[{Steppa} and {Egberts}(2020)]{2020A&A...643A.137S}
{Steppa}, C.; {Egberts}, K.
\newblock {Modelling the Galactic very-high-energy {\ensuremath{\gamma}}-ray
  source population}.
\newblock {\em \aap} {\bf 2020}, {\em 643},~A137,
  doi:10.1051/0004-6361/202038172.

\bibitem[{H.E.S.S. Collaboration} \em{et~al.}(2015){H.E.S.S. Collaboration},
  {Abramowski}, and et. al.]{2015HESS-LMC}
{H.E.S.S. Collaboration}; {Abramowski}, A.; et al.
\newblock {The exceptionally powerful TeV {\ensuremath{\gamma}}-ray emitters in
  the Large Magellanic Cloud}.
\newblock {\em Science} {\bf 2015}, {\em 347},~406--412,
  doi:10.1126/science.1261313.

\bibitem[{Abeysekara} and et. al.(2018)]{2018VERITAS-CYGNUS}
{Abeysekara}, A.U.; et al.
\newblock {A Very High Energy {\ensuremath{\gamma}}-Ray Survey toward the
  Cygnus Region of the Galaxy}.
\newblock {\em \apj} {\bf 2018}, {\em 861},~134,
  doi:10.3847/1538-4357/aac4a2.

\bibitem[{ARGO-YBJ Collaboration} \em{et~al.}(2002){ARGO-YBJ Collaboration},
  {Bacci}, and et. al.]{2002ARGO}
{ARGO-YBJ Collaboration}.; {Bacci}, C.; et al.
\newblock {Results from the ARGO-YBJ test experiment}.
\newblock {\em Astropart. Phys.} {\bf 2002}, {\em 17},~151--165.

\bibitem[{Tibet AS {\ensuremath{\gamma}} Collaboration} \em{et~al.}(2019){Tibet
  AS {\ensuremath{\gamma}} Collaboration}, {Amenomori}, and et.
  al.]{2019TibetAS}
{Tibet AS {\ensuremath{\gamma}} Collaboration}.; {Amenomori}, M.; et al.
\newblock {First Detection of Photons with Energy beyond 100 TeV from an
  Astrophysical Source}.
\newblock {\em \prl} {\bf 2019}, {\em 123},~051101,
  doi:10.1103/PhysRevLett.123.051101.

\bibitem[{Atkins} and et. al.(2003)]{2003Milagro}
{Atkins}, R.; et al.
\newblock {Observation of TeV Gamma Rays from the Crab Nebula with Milagro
  Using a New Background Rejection Technique}.
\newblock {\em \apj} {\bf 2003}, {\em 595},~803--811,
  doi:10.1086/377498.

\bibitem[{Abeysekara} and et. al.(2017)]{2017ApJ...841..100A}
{Abeysekara}, A.U.; et al.
\newblock {Daily Monitoring of TeV Gamma-Ray Emission from Mrk 421, Mrk 501,
  and the Crab Nebula with HAWC}.
\newblock {\em \apj} {\bf 2017}, {\em 841},~100,
  doi:10.3847/1538-4357/aa729e.

\bibitem[{Cao} and et. al.(2021)]{2021LHAASO}
{Cao}, Z.; et al.
\newblock {Ultrahigh-energy photons up to 1.4 petaelectronvolts from 12
  {\ensuremath{\gamma}}-ray Galactic sources}.
\newblock {\em \nat} {\bf 2021}, {\em 594},~33--36, doi:10.1038/s41586-021-03498-z.

\bibitem[{Younk} \em{et~al.}(2015){Younk}, {Lauer}, {Vianello}, {Harding},
  {Ayala Solares}, {Zhou}, {Hui}, and {HAWC
  Collaboration}]{2015ICRC...34..948Y}
{Younk}, P.W.; {Lauer}, R.J.; {Vianello}, G.; {Harding}, J.P.; {Ayala Solares},
  H.A.; {Zhou}, H.; {Hui}, M.; {HAWC Collaboration}.
\newblock {A high-level analysis framework for HAWC}.
\newblock  In Proceedings of the 34th International Cosmic Ray Conference (ICRC2015),  The Hague, The Netherlands, 30 July--6 August 2015; Volume~34,
  p. 948.

\bibitem[{Atkins} and et. al.(2004)]{2004Milagro}
{Atkins}, R.; et al.
\newblock {TeV Gamma-Ray Survey of the Northern Hemisphere Sky Using the
  Milagro Observatory}.
\newblock {\em \apj} {\bf 2004}, {\em 608},~680--685, doi:10.1086/420880.

\bibitem[{HAWC Collaboration} \em{et~al.}(2016){HAWC Collaboration},
  {Abeysekara}, and et. al.]{2016HAWC-1HWC}
{HAWC Collaboration}.; {Abeysekara}, A.U.; et al.
\newblock {Search for TeV Gamma-Ray Emission from Point-like Sources in the
  Inner Galactic Plane with a Partial Configuration of the HAWC Observatory}.
\newblock {\em \apj} {\bf 2016}, {\em 817},~3,
  doi:10.3847/0004-637X/817/1/3.

\bibitem[{Abeysekara} and et. al.(2017)]{2017HAWC-2HWC}
{Abeysekara}, A.U.; et al.
\newblock {The 2HWC HAWC Observatory Gamma-Ray Catalog}.
\newblock {\em \apj} {\bf 2017}, {\em 843},~40,
  doi:10.3847/1538-4357/aa7556.

\bibitem[{HAWC Collaboration} \em{et~al.}(2020){HAWC Collaboration}, {Albert},
  and et. al.]{2020HAWC-3HWC}
{HAWC Collaboration}.; {Albert}, A.; et al.
\newblock {3HWC: The Third HAWC Catalog of Very-high-energy Gamma-Ray Sources}.
\newblock {\em \apj} {\bf 2020}, {\em 905},~76,
  doi:10.3847/1538-4357/abc2d8.

\bibitem[{Abeysekara} and at. al.(2017)]{2017Sci...358..911A}
{Abeysekara}, A.U.; et al.
\newblock {Extended gamma-ray sources around pulsars constrain the origin of
  the positron flux at Earth}.
\newblock {\em Science} {\bf 2017}, {\em 358},~911--914,
  doi:10.1126/science.aan4880.
  
  \bibitem[{Patel} \em{et~al.}(2021){Patel}, {Maier}, and
  {Kaaret}]{2021arXiv210806424P}
{Patel}, S.; {Maier}, G.; {Kaaret}, P.
\newblock {VTSCat---The VERITAS Catalog of Gamma Ray Observations}.
\newblock {\em arXiv} {\bf 2021},  arXiv:2108.06424. 

\bibitem[{Wakely} and {Horan}(2008)]{2008TeVCat}
{Wakely}, S.P.; {Horan}, D.
\newblock {TeVCat: An online catalog for Very High Energy Gamma-Ray Astronomy}.
\newblock {\em Int. Cosm. Ray Conf.} {\bf 2008}, {\em
  3},~1341--1344.



\bibitem[{H.E.S.S. Collaboration} \em{et~al.}(2018{\natexlab{a}}){H.E.S.S.
  Collaboration}, {Abdalla}, and et. al.]{2018HESS-PWNPOP}
{H.E.S.S. Collaboration}; {Abdalla}, H.; et al.
\newblock {The population of TeV pulsar wind nebulae in the H.E.S.S. Galactic
  Plane Survey}.
\newblock {\em \aap} {\bf 2018}, {\em 612},~A2,
  doi:10.1051/0004-6361/201629377.

\bibitem[{H.E.S.S. Collaboration} \em{et~al.}(2018{\natexlab{b}}){H.E.S.S.
  Collaboration}, {Abdalla}, and et. al.]{2018HESS-SNRPOP}
{H.E.S.S. Collaboration}; {Abdalla}, H.; et al.
\newblock {Population study of Galactic supernova remnants at very high
  {\ensuremath{\gamma}}-ray energies with H.E.S.S.}
\newblock {\em \aap} {\bf 2018}, {\em 612},~A3,
  doi:10.1051/0004-6361/201732125.

\bibitem[{CTA Consortium} \em{et~al.}(2013{\natexlab{a}}){CTA Consortium},
  {Acharya}, and et. al.]{2013CTAConcept}
{CTA Consortium}; {Acharya}, B.S.; et al.
\newblock {Introducing the CTA concept}.
\newblock {\em Astropart. Phys.} {\bf 2013}, {\em 43},~3--18, doi:10.1016/j.astropartphys. 2013.01.007.

\bibitem[{CTA Consortium} \em{et~al.}(2013{\natexlab{b}}){CTA Consortium},
  {Bernl{\"o}hr}, and et. al.]{2013MC}
{CTA Consortium}; {Bernl{\"o}hr}, K.; et al.
\newblock {Monte Carlo design studies for the Cherenkov Telescope Array}.
\newblock {\em Astropart. Phys.} {\bf 2013}, {\em 43},~171--188,
  doi:10.1016/j.astropartphys.2012.10.002.

\bibitem[{CTA Consortium} \em{et~al.}(2013{\natexlab{c}}){CTA Consortium},
  {Dubus}, and et. al.]{2013CTA-HGPS}
{CTA Consortium}; {Dubus}, G.; et al.
\newblock {Surveys with the Cherenkov Telescope Array}.
\newblock {\em Astropart. Phys.} {\bf 2013}, {\em 43},~317--330,
  doi:10.1016/j.astropartphys.2012.05.020.

\bibitem[Chaves \em{et~al.}(2019)Chaves, Mukherjee, and R.A.]{CTAKSP6}
Chaves, R.; Mukherjee, R.; Ong, R.A. KSP: Galactic Plane Survey.
\newblock In {\em Science with the Cherenkov Telescope Array}; World Scientific
  Publishing Company:  2019; Chapter~6, pp. 101--124, doi:10.1142/9789813270091\_0006.

\bibitem[{Marandon} \em{et~al.}(2019){Marandon}, {Jardin-Blicq}, and
  {Schoorlemmer}]{2019HAWCOutrigger}
{Marandon}, V.; {Jardin-Blicq}, A.; {Schoorlemmer}, H.
\newblock {Latest news from the HAWC outrigger array}.
\newblock  In Proceedings of the 36th International Cosmic Ray Conference (ICRC2019),   Madison, WI, USA,  24 July--1 August 2019; Volume~36,
  p. 736.

\bibitem[{Joshi} and {Schoorlemmer}(2019)]{2019ICRC...36..707J}
{Joshi}, V.; {Schoorlemmer}, H.
\newblock {Air shower reconstruction using HAWC and the Outrigger array}.
\newblock  In Proceedings of the 36th International Cosmic Ray Conference (ICRC2019),  Madison, WI, USA, 24 July--1 August 2019; Volume~36,
  p. 707.

\bibitem[{Albert} and et. al.(2019)]{2019SWGO}
{Albert}, A.; et al.
\newblock {Science Case for a Wide Field-of-View Very-High-Energy Gamma-Ray
  Observatory in the Southern Hemisphere}.
\newblock {\em arXiv} {\bf 2019}, arXiv:1902.08429. 

\end{thebibliography}



\end{document}